\documentstyle[fleqn,11pt]{article}
\def\pj{\hspace{-.26cm}}
\def\fpj{\hspace{-.7cm}}
\def\mum{$\mu_r/M$}

\def\thalf{{\textstyle{\frac{1}{2}}}}
\def\tquar{{\textstyle{\frac{1}{4}}}}

\def\oneth{{\textstyle{\frac{1}{3}}}}
\def\thrhalf{{\textstyle{\frac{3}{2}}}}
\def\twofive{{\textstyle{\frac{2}{5}}}}
\def\thrfive{{\textstyle{\frac{3}{5}}}}
\def\twothr{{\textstyle{\frac{2}{3}}}}
\def\fiveth{{\textstyle{\frac{5}{3}}}}
\def\be{\begin{eqnarray}}
\def\ee{\end{eqnarray}}
\def\pmb#1{\setbox0=\hbox{$#1$}
\kern-.025em\copy0\kern-\wd0
\kern.05em\copy0\kern-\wd0
\kern-.025em\raise.0433em\box0}
\newcommand{\vm}[1]{\mbox{\bf#1}}
\newcommand{\vms}[1]{\mbox{\scriptsize{\bf#1}}}

\def\fpj{\hspace{-.7cm}}

\begin{document}
\begin{titlepage}
\begin{center}
{\Large  {Composition and Structure of Protoneutron Stars }}
\vskip0.7cm
Madappa Prakash$^{a,b}$, Ignazio Bombaci$^{a}$,  \protect\\Manju
Prakash$^{a,b}$, Paul J. Ellis$^{b,c}$ \\ James  M. Lattimer$^d$ and Roland
Knorren$^a$ 
\vskip0.7cm
\begin{em}{\small
$^a$Department of Physics, State University of New York at Stony Brook\\
Stony Brook, NY 11794}\\
\end{em}
\begin{em}{\small
$^b$Institute for Theoretical Physics, University of California\\
Santa Barbara, CA 93106}\\
\end{em}
\begin{em}{\small
$^c$School of Physics and Astronomy, University of Minnesota\\Minneapolis, MN
 55455}\\
\end{em}
\begin{em}{\small
$^d$Department of Earth and Space Sciences, State University of New York at
Stony Brook \\
Stony Brook, NY 11794}\\
\end{em}
\newpage

\vskip 0.2cm
{\bf Abstract}\\
\end{center}
We investigate the structure of neutron stars shortly after they are
born, when the entropy per baryon is of order 1 or 2 and neutrinos are trapped
on dynamical timescales. We find that the structure depends more  sensitively
on the composition of the star than on its entropy, and that the number of
trapped neutrinos play an important role in determining the composition.  
Since the structure is chiefly determined by the pressure of the strongly
interacting constituents and the nature of the strong interactions is poorly
understood at high density, we consider several models of dense matter,
including matter with strangeness-rich hyperons, a kaon condensate and quark
matter.  
                         
In all cases, the thermal effects for an entropy per baryon of order 2 or  less 
are small when considering the maximum neutron star mass.  Neutrino  trapping,
however, significantly changes the maximum mass due to the abundance of
electrons.  When matter is allowed to contain only nucleons and leptons,
trapping decreases the maximum mass by an amount comparable to, but somewhat
larger than,  the increase due to finite entropy.  When matter is allowed to
contain strongly interacting negatively charged  particles, in the form of
strange baryons, a kaon condensate, or quarks, trapping instead results in an
{\it increase} in the maximum mass, which adds to the effects of finite
entropy.  A net increase of order $0.2M_\odot$ occurs.          
                                                    
The presence of
negatively-charged particles has two major implications for the neutrino
signature of gravitational collapse supernovae.   First, the value of the
maximum mass will decrease during the early evolution of a neutron star as it
loses trapped neutrinos, so that if a black hole forms, it either does so
immediately after the bounce (accretion being completed in a 
second or two) or it is delayed for a neutrino diffusion timescale of 
$\sim 10$ s.  The latter case is most likely if the maximum mass of the 
hot star with trapped neutrinos  is near $1.5M_\odot$.  In the absence of
negatively-charged hadrons, black hole formation would be due to accretion and
therefore is likely to occur only immediately
after bounce.  
Second, the appearance of hadronic negative charges results in a general
softening of the equation of state that may be observable in the neutrino
luminosities and average energies. Further, these additional negative charges
decrease the electron fraction and may be observed in the relative excess of
electron neutrinos compared to other neutrinos.

\end{titlepage}

\setcounter{tocdepth}{3}
\tableofcontents
\newpage

\section {A neutron star is born}
\setcounter{section}{1}
\setcounter{subsection}{0}

A protoneutron star is formed in the aftermath of the gravitational collapse
of the core of a massive star. Its evolution proceeds through several
distinct stages~\cite{birth, bur0}, which  may have various
outcomes, as shown schematically in Fig. 1.         
\begin{enumerate}
\item Immediately following core bounce and the passage of a shock through the
outer protoneutron star's mantle, the star contains an unshocked, low entropy
core of $0.5-0.7$ M$_\odot$ in which neutrinos are trapped~\cite{sato,maz}. 
This is surrounded by a low density, high entropy mantle that is both accreting
matter falling through the shock and rapidly losing energy due to beta decays
and neutrino emission. The shock is momentarily stationary prior to an eventual
explosion.                                                               
\item On a timescale of about 0.5 s, accretion becomes much less important as
the supernova explodes and the shock lifts off the stellar envelope. 
Extensive neutrino losses and deleptonization of the mantle lead to the loss of
lepton pressure and to collapse of the mantle on the same timescale.  Neutrino
diffusion times from the core are too long to significantly alter the core
during this stage. If enough accretion occurs, and the initial core were large
enough, the mass of the hot, 
lepton-rich matter could exceed the maximum mass which is stable, in which 
case the star would collapse to form a black
hole.  In this event, neutrino emission would effectively cease, since the event
horizon is believed to form outside the neutrino photosphere~\cite{bur}.
\item This stage is dominated by neutrino diffusion causing deleptonization
and heating of the core.  Neutrino-nucleon absorption reactions set the
diffusion timescale to about 10--15 s. The maximum entropy per baryon reached 
in the core is about 2 (in units of Boltzmann's constant).  
When the core 
deleptonizes, the threshold for the appearance of strangeness, in the form of 
hyperons, a Bose kaon condensate, or quarks, will be 
reduced~\cite{tpl,keiljan,pcl,subside}. 
If one (or
more) of these additional components is present, the equation of state will
soften,  leading to a decrease in the maximum mass. There is, therefore, the
possibility that a black hole could form at this later time.                
\item Following deleptonization, the star has a high entropy, so that 
thermally produced neutrino pairs of all flavors are abundant, and 
thermal diffusion and cooling of 
the hot neutron star takes place. 
Because the entropy is higher at the beginning of cooling than it is at the
beginning of deleptonization, the neutrino mean free paths are smaller and the 
timescales longer. 
In approximately 50 s, as the average neutrino energy decreases, the star
becomes essentially transparent to neutrinos and the core achieves a cold,
catalyzed configuration.  The loss of thermal energy leads to a small increase
in the threshold density for the appearance of strange matter, so that, in the
absence of further accretion, it is unlikely that a black hole could form
during this or later phases.           
\item Following the onset of neutrino transparency, the core continues to 
cool by neutrino emission, but the star's crust cools less because of its lower
neutrino emissivity. The crust acts as an insulating blanket which prevents
the star from coming to complete thermal equilibrium and keeps the surface
relatively warm ($T\approx3\times10^6$ K) for up to 100 years.  This timescale
is primarily sensitive to the neutron star's radius and the thermal
conductivity of the mantle~\cite{lvpp}.
\item Ultimately, the star achieves thermal equilibrium when the energy stored
in the crust is depleted.  The temperature to which the surface now cools is
determined by the rate of neutrino emission in the star's core. If this rate
is large, the surface temperature will become relatively small, and the photon
luminosity may become virtually undetectable from the Earth.  This will be the
case if the direct Urca (beta decay) process can occur, which happens if 
the nuclear symmetry energy is large or if hyperons, a Bose condensate, or 
quarks are present.  Somewhat higher surface temperatures occur if
superfluidity in the core cuts off the direct Urca rate below the superfluid's
critical temperature.  A relatively high surface temperature will persist if
the Urca process can only occur indirectly with the participation of a
spectator nucleon -- the so-called standard cooling scenario. 
\end{enumerate}

Neutrino observations from a galactic supernova will shed much light on the
first four of the above stages.  Observations of X-rays or $\gamma-$rays from
very young neutron stars are crucial for the last stages.  The duration of each
of these stages is essentially determined by neutrino diffusion timescales, and
thus depends both upon the microphysics and the macrophysical  structure of
neutron stars.  Roughly, the diffusion timescale is proportional to
$R^2(c\lambda)^{-1}$, where $R$ is the star's radius and $\lambda$ is the
effective neutrino mean free path.  Thus, important constraints upon the
properties of dense matter can be achieved by looking at this relation as it
applies to each stage.  Generally, the structure (i.e., mass, radius, etc.) of
both hot and cold, and both neutrino-rich and neutrino-poor, stars is fixed by
the equation of state (EOS) and the composition.  Both are also crucial to 
knowledge of the neutrino mean free paths. 

The behavior of the maximum mass as a function of temperature and neutrino
trapping is of practical importance~\cite{browbet}. If the maximum mass of a
cold, neutrino-free neutron star is near the largest measured masses of neutron
stars~\cite{thor}, namely 1.442$\pm 0.001$ M$_\odot$ in the case of PSR
1913+16, the question arises as to whether or not a larger mass can be
stabilized during the preceding evolution of a star from a hot, lepton-rich
state.  If it cannot, then any transition from a neutron star to a black hole
in the stellar collapse process should occur extremely early, either
immediately or during the first few tenths of a second, when the proto-neutron
star is rapidly accreting mass from unejected matter behind the shock. 
If a larger mass can be stabilized, however, the transition from a neutron
star to a black hole,  if it occurs at all,  could happen later, on the
neutrino diffusion or thermal timescale, namely $\sim10$ seconds. 
Burrows~\cite{bur} has demonstrated that the appearance of a black hole should
be accompanied by a dramatic cessation of the neutrino signal (since the event
horizon invariably forms outside the neutrinosphere). 
                                                                             
A newly-formed neutron star should accrete its final baryon mass within a
second or two of its birth, so that the neutrinos will not have had time to 
diffuse from the
stellar core.  Thus, {\it both} the maximum mass for a hot, neutrino-trapped
star and a cold, catalyzed star must be greater than the largest measured
neutron star mass.  Moreover, due to the binding energy released because of
neutrino emission and cooling, the maximum mass for a hot, neutrino-trapped
star must be at least $0.1-0.2~M_\odot$ larger than this limit.  This feature
could provide a more severe constraint upon the equation of state than limits
based solely upon cold, catalyzed matter.  This is especially true if the
neutrino-trapped maximum mass is less than the cold, catalyzed maximum mass for
a given equation of state.

\section{Scope of this work}
\setcounter{section}{2}
\setcounter{subsection}{0}

The evolutionary scenario presented above is based on  dynamical
calculations~\cite{birth} carried out with a schematic equation of state, since
little work has been carried out on the structure of protoneutron stars shortly
after their birth, although, of course, there have been many investigations of
cold neutron  stars. The purpose of this work is to investigate the composition
and structure of these newly born stars. It would then be of interest to  study
the implications for the  dynamical evolution of neutron stars, but this we
defer to the future.  

There are two new effects to be considered for a newborn star. Firstly, 
thermal effects which result in an approximately uniform  entropy/baryon 
of 1--2  across the star~\cite{birth}. Secondly, the fact that 
neutrinos are trapped in the 
star, which means that the neutrino chemical potential is non-zero and this
alters the chemical equilibrium, which leads to compositional changes. Both
effects may result in observable consequences in the neutrino signature from a
supernova and may also play an important role in determining whether or not a
given supernova ultimately produces a cold neutron star or a black hole.  

Since the composition of a neutron star chiefly depends upon the nature of 
the strong interactions, which are not well understood in
dense matter, we shall investigate many of the possible models. After a 
brief discussion of the equilibrium conditions in section 3, we begin section 
4 by discussing non-relativistic potential models and their predictions
for protoneutron stars.  We then turn to relativistic models, which may be more 
appropriate, since the central densities involved are high.
At first, we allow only nucleons to be present in addition to the leptons.
Neutrino trapping increases the electron chemical potential and, therefore,  
the lepton and (due to charge  neutrality) the proton abundances. With more 
protons, the equation of state is softer and the maximum neutron star 
mass is lower. However, it has been recently realized that if additional 
negatively charged particles, such as kaons, hyperons or quarks, are present,  
this situation can be qualitatively changed~\cite{tpl,keiljan,pcl,subside}.
This is  due to the change in the chemical potentials, which delays the
appearance of  these strongly interacting particles that lower the pressure.
Their softening effect is therefore  less in evidence, and a larger maximum mass
is obtained for the young star. This means that the star could become unstable
when the initial neutrino population has departed, as we mentioned in section
1. In the remainder of this section, 
we therefore investigate in some detail the 
effect of hyperons, kaon condensates, and quarks upon the structure of the 
protoneutron star. 

The evolution of the protoneutron star is determined by the time scales
involved. While accurate results require detailed numerical work that 
incorporates neutrino transport, it is  nevertheless possible to use
semi-analytical techniques to gain an  understanding of the times involved and
the general energetics. This is discussed in section 5.           

In section 6, we discuss the implications of this work for delayed black hole 
formation and for neutrino signals from supernovae, in general, and SN1987A
in particular. 
                                                  
\section {Equilibrium conditions} 
\setcounter{section}{3}
\setcounter{subsection}{0}

For stars in which the strongly interacting particles are only baryons, the
composition  is determined by the requirements of charge neutrality and
equilibrium under the weak processes         
\begin{eqnarray}
B_1 \rightarrow B_2 + \ell + {\overline \nu}_\ell \,; \qquad 
B_2 + \ell \rightarrow B_1 + \nu_\ell \,,
\label{bproc}
\end{eqnarray}
where $B_1$ and $B_2$ are baryons, and $\ell$ is a lepton, either an electron or
a muon.  Under conditions when the neutrinos have left the system, these two 
requirements imply that the relations
\begin{eqnarray}
\sum_i \left( n_{B{_i}}^{(+)} + n_{\ell{_i}}^{(+)} \right) &=& 
\sum_i \left( n_{B{_i}}^{(-)} + n_{\ell{_i}}^{(-)} \right) 
\label{charge} \\
\mu_i &=& b_i\mu_n - q_i\mu_\ell \, ,
\label{beta}
\end{eqnarray}
are satisfied.  Above, $n$ denotes the number density and the superscripts
$(\pm)$ on $n$  signify positive or negative charge.  The symbol $\mu_i$ refers
to the chemical potential of baryon $i$, $b_i$ is its baryon number and $q_i$
is its charge.  The chemical potential of the neutron is denoted by $\mu_n$.  

Under conditions when the neutrinos are trapped in the system, the beta
equilibrium condition Eq.~(\ref{beta}) is altered to
\begin{eqnarray}
\mu_i &=& b_i\mu_n - q_i(\mu_\ell -\mu_{\nu_\ell}) \, ,
\label{tbeta}
\end{eqnarray}
where $\mu_{\nu_\ell}$ is the chemical potential of the neutrino $\nu_\ell$.  
Because of trapping, the numbers of leptons per baryon of each flavor of
neutrino, $\ell = e$ and $\mu$, 
\begin{eqnarray}
Y_{L\ell} = Y_\ell + Y_{\nu_\ell} \,,
\label{lnumber}
\end{eqnarray}
are conserved on dynamical time scales. Gravitational collapse calculations of
the white-dwarf core of massive stars indicate that at the onset of trapping,
the electron lepton number $Y_{Le} = Y_e + Y_{\nu_e} \simeq 0.4$, the precise
value depending on the efficiency of electron capture reactions during the
initial collapse stage. Also, because no muons are present when neutrinos
become trapped, the constraint $Y_{L\mu} = Y_\mu + Y_{\nu_\mu} = 0$ can be
imposed. We fix $Y_{L\ell}$ at these values in our calculations for {\it
neutrino trapped} matter.       

For completeness,  we give here the partition function for the leptons. Since
their interactions give negligible contributions~\cite{kapusta}, it is
sufficient to use the  non-interacting form of the partition function:  
\begin{eqnarray}
\ln Z_L&\pj=&\pj \frac{V}{T}\sum\limits_i g_i 
\frac {\mu_i^4}{24\pi^2} 
\left[ 1 + 2\left(\frac {\pi T}{\mu_i} \right)^2 +
\frac{7}{15}\left(\frac {\pi T}{\mu_i} \right)^4 \right] \nonumber \\ 
&&\qquad\qquad + Vg_\mu  \int\frac{d^3k}{(2\pi)^3} \,
\left[ \ln\left(1+{\rm e}^{-\beta(e_{\mu}-\mu_{\mu})}\right) 
      + \ln\left(1+{\rm e}^{-\beta(e_{\mu}+\mu_{\mu})}\right) \right]\,, 
\label{zlept}
\end{eqnarray}
where $V$ is the volume and $\beta=T^{-1}$ is the inverse temperature. The
first term gives the contribution of massless particles and  antiparticles. 
Since $\mu_e\gg m_e$ in all regimes considered here, this term applies to both
electrons and neutrinos. 
The  degeneracies, $g_i$, are 2 and 1 for electrons and neutrinos,
respectively. The second term gives the muon contributions, with 
degeneracy $g_\mu=2$, $e_\mu=\sqrt{k^2+m_{\mu}^2}$ and the muon chemical
potential designated by $\mu_{\mu}$.   The pressure, density and energy density
of the leptons are obtained from Eq.~(\ref{zlept}) in the standard fashion. The
total  partition function, $Z_{\rm total}=Z_HZ_L$, where $Z_H$ is the partition
function of the hadrons discussed below.

\section{Stellar Composition and Structure } 
\setcounter{section}{4}
\setcounter{subsection}{0}

\subsection{Models of hot and dense hadronic matter} 
\setcounter{subsubsection}{0}

The properties of neutron stars can be obtained from the well-known hydrostatic
equilibrium equations of Tolman, Oppenheimer and Volkov~\cite{tov}, once 
the equation of state is specified. At very low densities 
$(n < 0.001$ fm$^{-3})$, we use the Baym-Pethick-Sutherland~\cite{bps} EOS,
while for densities $0.001 < n<0.08$ fm$^{-3}$, we employ the  EOS
of Negele and Vautherin~\cite{nv}. At higher densities the EOS
depends on the nature of the strong interactions.  These are not yet known 
with certainty,
although several intriguing possibilities are currently being investigated.  In
view of this, we will investigate the influence of thermal effects and
neutrino-trapping on the structure of neutron stars by considering widely
differing models of dense matter and trying to identify the common features 
shared by these models. 
                                            
The models considered include (a) a generalization of a schematic potential 
model based on the work of Prakash, Ainsworth and Lattimer~\cite{pal}, which
reproduces the results of more microscopic calculations~\cite{wff} of dense
matter, (b) a relativistic field theoretical model~\cite{pehr}, based on the
archetypal Walecka model~\cite{sew}, in which baryons interact via the exchange
of $\sigma$-, $\rho$- and $\omega$-mesons,  (c) a model
based on the chiral Lagrangian of Kaplan and Nelson~\cite{kapnel} and a
model based on a meson exchange picture, in which a
kaon condensate occurs at about $4n_0$, and finally, (d) a model which 
allows both quark and hadron phases to be present. The composition of the star 
differs among these models due to differences in the nuclear interactions at 
high density, and, also whether or not strange baryons or strange mesons are
included in the description of matter.  For example, when only nucleons are
included in models (a) and (b), the proton fraction in matter is determined by
the density dependence of the symmetry  energy; the more rapidly the symmetry
energy increases with density, the greater is the proton fraction.  However,
neutrons remain the most abundant species in such stars.  In model (b), the
inclusion of hyperons, which carry strangeness, has the effect of substantially
softening the EOS at high density.  In particular, the  presence of a
substantial number of negatively charged particles, such as  the $\Sigma^-$
hyperon, raises the proton concentration in neutrino free  matter and reduces
the lepton concentrations.  In model (c), $K^-$ mesons in
the condensate effectively replace electrons to achieve charge neutrality, with
the result that nearly as many protons as neutrons are found in the dense
interior regions of the ``nucleon'' star.  In model (d), additional negative 
charge is provided by $d$ and $s$ quarks.

The composition of matter is  significantly  altered when neutrinos are
trapped.  This is due to the fact that at the onset of trapping,  $Y_{Le} \sim
0.4$, and the electron fraction $Y_e$ is significantly larger than that found
in a cold catalyzed star.   The corresponding changes in the structure are 
quantitatively  different among the different models.  However, the effects
of neutrino trapping in matter containing negatively charged, strongly
interacting particles, either $\Sigma^-$s in hyperonic matter,  $K^-$s in Bose
condensed matter, or quarks in matter that has undergone a phase
transition at high density, are qualitatively similar.   Since the size of
thermal effects depends on the relative concentrations, which determine the
degree to which each constituent is  degenerate ($T/T_{F_i} \ll 1$, where
$T_{F_i}$  is the Fermi temperature of species $i$), the structural changes are
expected to be accordingly different for the different models.

\subsection{Potential models}
\setcounter{subsubsection}{0}

Based on a two-body potential fitted to nucleon-nucleon scattering, and a
three body term whose form is suggested by theory and whose parameters are
determined by the binding of few body-nuclei and the saturation properties of
nuclear matter, Wiringa {\em et al.}~\cite{wff} have performed microscopic
calculations of neutron star matter at zero temperature.  At high density,
there are uncertainties in the three-body interactions, which are
reflected in the density dependence of the symmetry energy.  
Calculations at finite temperature to encompass 
an entropy/baryon in the range $S=1-2$
are not yet available.   In this section, we therefore outline a schematic
potential model which is designed  to reproduce the results of the more
microscopic calculations (see, for example, Refs. ~\cite{wff,bl}) of both 
nuclear and neutron-rich matter at zero temperature, and which can be 
extended to finite temperature~\cite{bomb}. 

We begin with the energy density               
\begin{eqnarray}
\varepsilon = \varepsilon_n^{(kin)} + \varepsilon_p^{(kin)} + V(n_n,n_p,T) \,,
\label{expal1}
\end{eqnarray}
where $n_n$ ($n_p$) is the neutron (proton) density and the total density
$n=n_n+n_p$. The contributions arising from the kinetic parts are
\begin{eqnarray} 
\varepsilon_n^{(kin)} + \varepsilon_p^{(kin)} = 
2\int\frac {d^3k}{(2\pi)^3} \, \frac {\hbar^2k^2}{2m} \left( f_n + f_p\right)\;,
\end{eqnarray}
where the factor 2 denotes the spin degeneracy and $f_i~$ for $i=n,p$ are 
the usual Fermi-Dirac distribution functions and $m$ is the nucleon mass.
It is common to employ local contact interactions to model the nuclear
potential. Such forces lead to power law density-dependent terms in $V(n)$.
Since 
repulsive contributions that vary faster than linearly give rise to acausal
behavior at high densities, care must be taken to screen such repulsive
interactions \cite{pal}.  Including the effect of finite-range forces between 
nucleons, we parametrize the potential contribution as
\begin{eqnarray}
V(n_n,n_p,T) &=& 
\oneth An_0\left[\thrhalf -\left(\thalf + x_0\right)(1-2x)^2\right]u^2
\nonumber \\ 
 && +\frac {\frac  23 Bn_0\left[\frac 32-\left(\frac 12 + x_3\right)
(1-2x)^2\right]u^{\sigma+1}  }
{1 + \frac  23 B^\prime n_0\left[\frac 32-\left(\frac 12 + x_3\right)
(1-2x)^2\right]u^{\sigma-1}  } \nonumber \\
&& +\twofive u \sum_{i=1,2} \left\{ (2C_i+4Z_i) 
2 \int \frac {d^3k}{(2\pi)^3} \, g(k,\Lambda_i) \left( f_n + f_p\right) 
\right.\nonumber \\ 
&& +(C_i-8Z_i) 
\left. 2 \int \frac {d^3k}{(2\pi)^3} \, g(k,\Lambda_i) 
[ f_n (1-x) + f_p x] \right\} \;,
\label{expal2}
\end{eqnarray}
where $x=n_p/n$ and $u=n/n_0$, with $n_0$ denoting equilibrium nuclear 
matter density.  The function $g(k,\Lambda_i)$ is suitably
chosen to simulate finite range effects.  The constants $A,~B,\sigma
,~C_1,~C_2$, and $~B^\prime$, which enter in the  description of symmetric
nuclear matter, and the additional constants $x_0,~x_3, ~Z_1$, and $Z_2$, which
determine the properties of asymmetric nuclear matter, are treated as
parameters that are constrained by empirical knowledge. 

Various limits of the energy density are of interest, and are listed below.
Setting $x=1/2$ and $f_n=f_p$, the energy density of symmetric nuclear
matter is 
\begin{eqnarray} 
\varepsilon_{nm} &=& 4 \int \frac {d^3k}{(2\pi)^3} \, \frac {\hbar^2k^2}{2m}\,
f_n 
+ \thalf A n_0 u^2 + \frac {Bn_0 u^{\sigma+1}}{1+B^\prime u^{\sigma -1}}
\nonumber \\
&& +u \sum_{i=1,2} C_i~4 \int \frac {d^3k}{(2\pi)^3} \, g(k,\Lambda_i)
f_n \;,
\label{PAL1}
\end{eqnarray}
and, with $x=0$, the corresponding result for pure neutron matter is
\begin{eqnarray}
\varepsilon_{nem} &=& 2 \int \frac {d^3k}{(2\pi)^3}\,\frac {\hbar^2k^2}{2m}f_n  
+ \oneth An_0 (1-x_0) u^2 + \frac {{2\over 3} Bn_0 (1-x_3) u^{\sigma+1}}
{1+{2\over 3}B^\prime (1-x_3)u^{\sigma -1}} \nonumber \\
&& +\twofive u \sum_{i=1,2} (3C_i-4Z_i)~2 
\int \frac {d^3k}{(2\pi)^3} \, g(k,\Lambda_i) f_n \;.
\label{enem}
\end{eqnarray}
At zero temperature, $f_i = \theta (k_{F_i} - k)$, where $k_{F_i}$ is the
Fermi momentum of particle $i$.  Thus the kinetic energy densities are
\begin{eqnarray}
\varepsilon^{(kin)}_{nm} &=& \thrfive E_F^{(0)}n_0  u^{5/3} \qquad \qquad 
{\rm for ~~nuclear ~~matter} \nonumber \\
\varepsilon^{(kin)}_{nem} &=& 2^{2/3}\left(\thrfive E_F^{(0)}n_0u^{5/3} \right) 
\, \qquad {\rm for ~~neutron ~~matter} \;,
\end{eqnarray}
where $E_F^{(0)}=(\hbar k_F^{(0)})^2/2m$ is the Fermi energy of nuclear matter
at the equilibrium density.  To simulate finite range effects, we investigate 
two commonly used  functional forms for $g(k,\Lambda_i)$, which lead to closed
form expressions at zero temperature.  

(i) $g(k,\Lambda_i) =  [1+(k/\Lambda_i)^2]^{-1}$:  In this case, the
finite range terms{\footnote{In dynamical situations, such as heavy-ion
collisions, it is more appropriate to model matter using  momentum dependent
Yukawa interactions,  as pointed out in Ref.~\cite{gwpld}.  For static matter,
both cold and hot, the simpler form chosen here is adequate insofar as
identical physical  properties may be recovered with suitable choices of the
parameters entering the description of the EOS.}} at $T=0$ in Eq.~(\ref{PAL1})
and Eq.~(\ref{enem}) may be written as                         
\begin{eqnarray}                                     
V_{nm}^{(fr)} &=& 3n_0u \sum_{i=1,2} C_i R_i^3
\left( \frac {u^{1/3}}{R_i} - 
\arctan \frac {u^{1/3}}{R_i}  \right) 
\label{PAL2} \\
V_{nem}^{(fr)} &=& \thrfive  n_0u \sum_{i=1,2} (3C_i-4Z_i) R_i^3
\left(\frac {(2u)^{1/3}}{R_i} - 
\arctan \frac {(2u)^{1/3}}{R_i} \right)  \,,
\label{PAL3}
\end{eqnarray}
where $R_i =  \Lambda_i/(\hbar k_F^{(0)})$. 
Note that the potential energy density for symmetric nuclear matter  in
Eq.~(\ref{PAL1}) and Eq.~(\ref{PAL2}) is the same as that employed in
Ref.~\cite{pal}.  

(ii) $g(k,\Lambda_i) = 1-(k/\Lambda_i)^2$: Here, the finite range
interactions   are approximated by effective local interactions by retaining
only the quadratic momentum dependence. The energy densities in Eq.~(\ref{PAL1})
and Eq.~(\ref{enem}) then take the form of Skyrme's effective 
interactions~\cite{vb}.  The  finite range terms, again at $T=0$, now read
\begin{eqnarray}
V_{nm}^{(fr)} &=& n_0u^2 \sum_{i=1,2} C_i 
\left[ 1 - \frac 35 \frac {u^{2/3}}{R_i^2} \right] \nonumber \\
V_{nem}^{(fr)} &=& \twofive n_0u^2 \sum_{i=1,2} (3C_i-4Z_i) 
\left[ 1 - \frac 35 \frac {(2u)^{2/3}}{R_i^2} \right]\;.
\label{sknem}                                        
\end{eqnarray}
Note that, at high density, the quadratic momentum dependence inherent in
Skyrme-like (SL) interactions will  eventually lead to an acausal behavior due
to the $u^{8/3}$ dependence of the energy densities.  This situation does not
occur in case (i). 

The parameters $A,~B,\sigma ,~C_1,~C_2$, and $~B^\prime$,  a small parameter
introduced to maintain causality, are determined from constraints provided by
the empirical properties of symmetric nuclear matter  at the equilibrium
density $n_0=0.16$ fm$^{-3}$.  With appropriate choices of the parameters, it
is possible to parametrically vary the nuclear incompressibility $K_0$ so that
the dependence on the stiffness of the EOS may be explored.  Numerical values
of the parameters, appropriate for symmetric nuclear matter, are given in 
Table~1.   EOSs based on Eqs.~(\ref{expal1}) and (\ref{expal2}) are hereafter 
referred to as the BPAL EOSs if the function $g(k,\Lambda_i) =  
[1+(k/\Lambda_i)^2]^{-1}$, and  the SL EOSs if $g(k,\Lambda_i) 
= 1-(k/\Lambda_i)^2$. 
                                                  
In the same vein,  by suitably choosing the parameters $x_0,~x_3,~Z_1$, and
$Z_2$, it is possible to obtain  different forms for the density dependence  of
the symmetry energy $S(n)$ defined by the relation 
\begin{eqnarray}
E(n,x) = \varepsilon (n,x)/n = E(n,1/2) + S(n) (1-2x)^2 + \cdots \;,
\end{eqnarray}
where $E$ is the energy per particle, and $x=n_p/n$ is the proton
fraction.   Inasmuch as the density dependent terms associated with
powers higher than $(1-2x)^2$ are generally small, even down to $x=0$, 
$S(n)$ adequately describes the properties of asymmetric
matter.   The need to explore different forms of $S(n)$ stems from the
uncertain behavior at high density and has been amply detailed in earlier
publications~\cite{pal,lpph}.  We have chosen to study three cases, in which
the potential part of the symmetry energy varies approximately as ${\sqrt
u},~u$, and $2u^2/(1+u)$, respectively, as was done in Ref.~\cite{pal}. 
Numerical values of the parameters that generate these functional forms are
given in Table~2 for the BPAL and SL EOSs.  The notation BPAL$n_1n_2$ and
SL$n_1n_2$ is used to denote different EOSs; $n_1$ refers to different values
of $K_0$, and $n_2=1,2$ and $3$ indicate, respectively,  ${\sqrt u},~u$ and
$2u^2/(1+u)$ for the dependence of the nuclear symmetry potential energy on the
density.            
                            
The main advantage of casting the schematic parametrization of Ref.~\cite{pal}
in the form of Eq.~(\ref{expal1}) through Eq.~(\ref{expal2})  is that it is now
possible to study asymmetric matter at finite temperature.  As an illustration
of the calculational procedure at finite temperature, consider first the case
of pure neutron matter.  The evaluation of the baryon density           
\begin{eqnarray}
n = n_n = 2 \int \frac {d^3k}{(2\pi)^3} \, \left[1 + 
\exp \left(\frac {e_k - \mu_n }{T} \right) \right]^{-1} 
\label{dens}
\end{eqnarray}
requires a knowledge of the single particle spectrum 
\begin{eqnarray}
e_k = \frac {\hbar^2 k^2}{2m} + U(n,k;T) \,,
\label{spectrum}
\end{eqnarray}
where the single particle potential $U(n,k;T)$, which is explicitly momentum
dependent, is obtained by a functional differentiation of the potential energy
density in Eq.~(\ref{enem}), with respect to the distribution function $f_n$. 
Explicitly, 
\begin{eqnarray}
U(n,k;T) = {\tilde U}(n;T) + \twofive u \sum_{i=1,2} (3C_i-4Z_i) 
\left[1+ \left(\frac {k}{\Lambda_i} \right)^{\!2}\:\right]^{-2} \,,
\end{eqnarray}
for the BPAL EOS, where the explicit momentum dependence is contained in the
last term.  The momentum-independent part is given by 
\begin{eqnarray}
{\tilde U}(n;T) &=& \twothr A(1-x_0)u \nonumber \\
&+& \frac {{2\over 3} B (1-x_3) u^\sigma}
{\left[{1+{2\over 3}B^\prime (1-x_3)u^{\sigma -1}}\right]^2} \cdot 
\left[ (\sigma + 1) + \frac 43 B^\prime (1-x_3)u^{\sigma-1}\right] \nonumber \\
&+& \frac {2}{5n_0} \sum_{i=1,2} (3C_i-4Z_i) ~2 \int \frac
{d^3k^\prime}{(2\pi)^3} 
\left[1+ \left(\frac {k^\prime}{\Lambda_i} \right)^{\!2}\:\right]^{-2} 
f_n(k^\prime) \,.
\label{utilde}
\end{eqnarray}
For a fixed baryon density $n$ and temperature $T$, Eq.~(\ref{dens}) may be
solved iteratively for the as yet unknown variable                        
\begin{eqnarray}
\eta = \frac {\mu_n-{\tilde U}}{T} \,.
\end{eqnarray}
The knowledge of $\eta$ allows the last term in Eq.~(\ref{utilde}) to be 
evaluated, yielding ${\tilde U}$, which may then be used to infer the
chemical potential from
\begin{eqnarray}
\mu_n = T \eta  - {\tilde U} \,,
\end{eqnarray}
which is required in the calculation of the single particle  spectrum
$e_k$ in Eq.~(\ref{spectrum}).  With this $e_k$, the energy
density in Eq.~(\ref{enem}) is readily evaluated.   The entropy density has the
same functional form as that of a non-interacting system: 
\begin{eqnarray}
s = - 2 \int \frac {d^3k}{(2\pi)^3} 
\left [f_n \ln ~f_n + (1-f_n) \ln ~(1-f_n) \right] \,,
\end{eqnarray}
from which the pressure is obtained using
\begin{eqnarray}
P = sT + n\mu_n - \varepsilon       \,.
\label{pres}
\end{eqnarray}

The above procedure is also applicable, with obvious modifications, to a system
containing unequal numbers of neutrons and protons, which is generally the case
for charge-neutral matter in beta equilibrium.  \\

\subsubsection{Pure neutron matter } 

The influence of finite entropy on the structure of a star is most easily
studied when the star is idealized to be composed of neutrons only. 
The top panels in Fig.~2
show the density dependence of the Landau effective mass,
$m^*=k_F/(\partial e_k/\partial k)|_{k_F}$,
for the BPAL22 and SL22 EOSs.  
For the BPAL22 model considered, the
effective mass is weakly dependent on density and has a value $\simeq 0.6m$,
where $m$ is the bare mass. On the other hand, for the SL22 EOS the
variation of the effective mass with density (top panels) is significantly more
rapid. This qualitative difference follows from the explicit expressions for the
effective mass:
\begin{eqnarray}
\left(\frac {m^{*}}{m}\right)_{nem} = 
\left[ 1 + \sum_{i=1,2}\alpha_i~u\left(
1 + \frac {(2u)^{2/3}}{R_i^2} \right)^{\!-2} \:\right]^{-1}\quad; \quad 
\alpha_i = \frac 25 \frac {(4Z_i-3C_i)}{E_F^{(0)}R_i^2}
\label{emasbpals}
\end{eqnarray}   
for the BPAL EOS, and  
\begin{eqnarray}
\left(\frac {m^{*}}{m}\right)_{nem} = \left[ 1 + \alpha~u \right]^{-1}; \qquad 
\alpha = \alpha_1+\alpha_2
\label{skmstar}
\end{eqnarray}
for the SL EOS. 

Turning to the pressure (center panels of Fig. 2), 
it is possible to analytically establish
the quadratic dependence on the entropy per baryon, and, perhaps more
importantly,  understand the order of magnitude of the increase in the maximum
mass if we analyze the
thermal contributions using the methods of Fermi liquid theory.   Following the
analysis in Ref.~\cite{pabw}, the thermal pressure of interacting neutrons may
be cast in the form       
\begin{eqnarray}
P_{th} = \left[ nT \frac {\pi^2}{4} \frac {T}{T_F} \right] \cdot 
\frac 23 {\left[1 - \frac 32 \frac {d\ln m^*}{d \ln n} \right]}  \,,
\label{ptht}
\end{eqnarray}
where the Fermi temperature, $T_F = \hbar^2k_F^2/2m^*$, sets the scale of the
temperature dependence of the thermodynamical functions. 
In the dense central regions of the star, $T/T_F \ll 1$, so the
neutrons, which are the only constituents
considered now, are in a highly degenerate configuration.
Since, in the degenerate limit, the entropy per particle $S=(\pi^2/2)(T/T_F)$,
the ratio of the thermal pressure to that of the zero temperature pressure
$P_0$ may
be expressed as                       
\begin{eqnarray} 
\frac {P_{th}}{P_0} = \left [ \frac {5}{3\pi^2} S^2 + \cdots \right] 
{\left[1 - \frac 32 \frac {d\ln m^*}{d \ln n} \right]} 
{\left[1 + \frac {P_{pot}}{P_{kin}}\right]^{-1} }  \,,
\label{pth}
\end{eqnarray}                              
where $P_{kin}$ and $P_{pot}$, the kinetic and potential pressures, as well as
the Landau effective  mass, $m^*$, in this relation refer
to zero temperature matter.  This equation adequately reproduces the 
exact thermal pressure at the entropies likely to be relevant in the
evolution of a neutron star. The differences of Eq. (\ref{pth}) from the ideal 
gas result
are obvious.   First, the density dependence of the effective mass introduces a
correction; this is clearly the origin of the larger thermal pressure for
the SL22 case compared to the BPAL22 case for a given entropy (see Fig.~2). 
Second, the pressure arising from potential interactions, which is
generally larger than the kinetic pressure due to the predominance of repulsive
interactions at high density, produces a significant reduction in the ratio
$P_{th}/P_0$. 

Since the thermal pressure increases quadratically with the entropy/baryon,
$S=s/n$, the neutron star
masses should show a similar behavior, as we see from the bottom panels of 
Fig.~2. Qualitatively similar
results are obtained for our other EOSs  with different behavior for the 
symmetry energy.
Quantitative results for the basic physical attributes of a maximum-mass star 
are shown in 
Table~3 for the EOSs termed BPAL22 and SL22, respectively.  It is  clear that
the  maximum mass at finite entropy is well approximated by 
\begin{eqnarray}
M_{max}(S) = M_{max}(0) \left[ 1 + \lambda S^2 + \cdots \right] \,,
\label{lambda}
\end{eqnarray}
where the coefficient $\lambda $ is EOS dependent.  The values of $\lambda$
given in Table 3 are quite small, $\sim 10^{-2}$.  

These results highlight the point that the magnitude of the increase in the
maximum mass is chiefly governed by the  magnitude of the thermal pressures at
fixed entropy.  In the simple models considered above, the thermal pressures
depend sensitively on the momentum dependence of the nuclear interactions. 
Insofar as one can establish this momentum dependence, either from experiment
or theory, the thermal pressures may be constrained.  For example,  the energy
dependence of the real part of the optical model potential required to explain 
proton-nucleus scattering experiments provides a stringent constraint  on the
momentum dependence of the single particle potential at nuclear density, but
for values of $x\sim0.5$.   For higher densities in the range $(2-3)n_0$
attained in GeV/particle heavy-ion collisions,  the momentum dependence largely
governs the flow of matter, momentum and energy~\cite{gwpld}.   While the BPAL
single particle potentials  are consistent with the observed behavior, the
SL potentials, with their quadratic momentum dependence, are known to be
inconsistent with data at high momentum~\cite{pkd}.  Also, at high density, the
quadratic momentum dependence has the disadvantage that it leads to an acausal
behavior, due to the $u^{8/3}$ dependence of the energy densities in
Eq.~(\ref{sknem}). \\

\subsubsection{Neutrino-free matter in beta equilibrium } 
                 
We turn now to the more realistic case in which matter consists of neutrons,
protons, electrons, and muons, with their relative concentrations determined
from the conditions of charge neutrality, Eq.~(\ref{charge}), and equilibrium
under  beta decay processes in the absence of neutrino trapping, 
Eq.~(\ref{beta}).   The EOS of strongly interacting
matter above nuclear density is given by Eq.~(\ref{expal1}) and
Eq.~(\ref{expal2}).  The EOS of leptons is obtained from Eq.~(\ref{zlept}).  

The relative concentrations, the electron chemical potential, and the 
individual entropies per baryon (for $S=1$) are shown in Fig.~3 for the BPAL22
and SL22 EOSs, both of which are characterized by a symmetry energy whose
potential part varies linearly with density.  Note that at high density, the 
proton  concentration lies in the
range $(20-30)\%$ (for these EOSs), which is balanced by an equal amount of
negatively charged leptons to maintain charge neutrality.  Also, the
lepton contribution to the total entropy per baryon is comparable to that 
of the degenerate nucleons at high density, thus lowering the nucleon 
contribution at fixed $S$. This results in a smaller increase in the maximum 
mass than when leptons are not present, as is evident from Fig.~4, where the 
nucleon effective masses,  
isentropic pressures, and the mass curves for the corresponding EOSs are  shown.
The increase in the maximum mass is again
quadratic with entropy.
This is due to the fact that both nucleonic and leptonic pressures stem  mostly
from quadratic terms in the entropy, terms involving higher powers of entropy
giving rather small contributions.  When the nucleons are degenerate, the
generalization of Eq. (\ref{pth}) is 
\begin{eqnarray}
\frac {P_{th}}{P_0} = \left[ \frac {5}{3\pi^2}S^2 + \cdots \right]
 \frac {\sum_i\frac {Y_i}{T_{F_i}} 
\left(1 - \frac 32 \frac {d\ln m_i^*}{d\ln n_i}\right)}
{\left(\sum_i \frac{Y_i}{T_{F_i}}\right)^2 \left(\sum_i Y_i T_{F_i} \right)}
\left [ 1 + \frac {P_{pot}}{P_{kin}} \right]^{-1}\,, \quad i=n,p
\end{eqnarray}
where $Y_i=n_i/n$ and $T_{F_i}=(\hbar k_{F_i})^2/(2m_i^*)$ denote the number
concentration and the Fermi temperature of species $i$, respectively.   
Further, $P_{kin}$, $P_{pot}$, and $P_0=P_{kin}+P_{pot}$ are the kinetic, 
potential, and the total pressures at zero temperature and include the
contributions from both neutrons and protons.    
                                                                            
Quantitative results for the physical attributes of the maximum mass star  are
summarized in Table~4 for the different BPAL  EOSs which have varying
stiffnesses and different density dependence of the symmetry energy, the latter
determining  the proton and lepton concentrations in the star.  Results for the
SL  interactions with a linear potential contribution to the symmetry energy,
but for EOSs with different stiffness are shown in Table~5.  The results in
these tables reflect the influence of entropy on the gross properties of stars. 
With few exceptions, the temperature at  the core remains below 100 MeV. 
Although the central density and pressure  of the maximum mass star are
significantly reduced, $\lambda$ is uniformly $\sim10^{-2}$, so the increase in 
the maximum mass amounts to only a few percent of  the cold catalyzed star,
even up to $S=2$.  In all cases studied, the increase in mass is quadratic with
entropy.             

The moments of inertia, $I$, displayed in Tables 4 and 5 
refer to maximum mass stars, so that the effect of cooling a given star from 
entropy per baryon, $S$, of 2 to 0 cannot be directly assessed.
We therefore show in Fig.~5 the moment of inertia as
a function of the central density and of the baryonic mass, $M_B$. (The full 
dots on the various curves represent the maximum mass configurations.) Since 
the baryonic mass is proportional to the number of nucleons, it will remain 
fixed as the star cools, in the absence of accretion. 
Fig.~5 shows that the moment of inertia decreases as the star cools from 
$S=2$ to $S=0$.  The magnitude is dependent on the mass in
question, but a  typical value is $\sim20$\%. Since angular momentum is 
conserved, again in the absence of accretion, this implies an increase of the 
angular velocity of similar 
order. Thus a significant spin-up of the neutron star is expected to occur
during the cooling process. \\

\subsubsection{Neutrino-trapped matter in beta equilibrium } 

We turn now to the case in which neutrinos are trapped in matter.  As mentioned
earlier, we fix the electron lepton number at $Y_{Le} = 0.4$ and the muon
lepton number at $Y_{L\mu} = 0$. Fig.~6 shows the effects of neutrino trapping
on the relative abundance, the chemical potentials, and the partial entropies
for an entropy per baryon $S=1$. The major effect of trapping is to keep the
electron concentration high so that matter is more proton rich in comparison to
the case in which neutrinos are not trapped (cf. Fig.~3).  The EOS
with trapped neutrinos is less stiff than that without neutrinos, since the
decrease in pressure due to the nuclear symmetry energy is greater than the
extra leptonic pressure.  This is reflected in the limiting masses
obtained with neutrino trapped matter (see Table~6), which are lower than those
without neutrinos (cf. Table~4).  Results for the maximum mass stars are
summarized in Table~6 for the BPAL models with different stiffnesses.  
The decrease in the maximum mass due to the
effect of trapping is generally larger than the increase due to thermal
effects for $S=1$. For $S=2$ the thermal contributions are larger, which 
results in a delicate balance between the two competing effects.

It is instructive to contrast the effects of thermal contributions on white
dwarf and neutron star structures.  
Unlike the case of a white dwarf, it is not
possible to analytically predict the effects of finite entropy on the structure
of a neutron star.  While both configurations are degenerate, and thus one
expects the dominant effects of finite entropy to enter quadratically, the role
of interactions in the two cases are quite different.  In contrast to the case
of a white dwarf, in which the equation of state is highly 
ideal~\cite{kapusta}, the equation of state in a neutron star is strongly
non-ideal.  This results in a white dwarf having a much greater sensitivity to
entropy than a neutron star.  For a white dwarf, in which the electron pressure
dominates, the thermal correction to the Chandrasekhar mass is about 10\% at an
entropy per baryon of 1~\cite{hf}.  In neutron stars, the pressure support is
largely provided by the strongly interacting baryons, which have relatively
smaller thermal contributions to the pressure and, therefore, a smaller
increase in the maximum neutron star mass.  As a result, the compositional 
variables of the EOS play a more important role than the temperature for the 
structure of neutron stars.

\subsection{Field theoretical models} 
\setcounter{subsubsection}{0}

From the results of the previous section, we notice that the central density
of the maximum mass star typically exceeds $(4-5)~n_0$.  At such densities, the
Fermi momentum and effective nucleon mass are both expected to be on the order
of $500~$MeV. Thus a relativistic description is preferred. 
Relativistic local
quantum field theory models (see, for example, Ref.~\cite{sew}) of finite
nuclei and infinite nuclear, and neutron star, matter have had some success,
albeit with rather more schematic interactions and with less sophisticated
approximations than their nonrelativistic counterparts (see, for example,
Refs.~\cite{wff,bl}).  It is our purpose here to examine the effects of finite
entropy and neutrino-trapping 
in a relativistic description and to contrast the results with those of
potential models.

Specifically,  we employ a  relativistic field theory model in which baryons,
$B$, interact via the exchange of $\sigma$-, $\rho$-, and  $\omega$-mesons. In
the case that only nucleons are considered, $B=n,p$; this is the well-known 
Walecka model~\cite{sew}, which we evaluate in the Hartree approximation (or,
equivalently, at the one-loop level). It has been shown, however, that hyperons
significantly soften the zero-temperature equation of
state~\cite{glenhyp,kaphyp}. Therefore, we shall also consider the case where
the hyperons, $\Lambda$, $\Sigma$, and $\Xi$, are included in the set of
baryons $B$. (The inclusion of the  spin-$\frac 32~~\Delta$ quartet and the
$\Omega^-$ hyperon is straightforward, but does not quantitatively alter the
results, since they appear at densities much higher than found in the cores of
stars.)   Specifically, our Lagrangian is
\begin{eqnarray}
{\cal L}_H&\pj=&\pj\sum_B\bar B\left(i\gamma^{\mu}\partial_{\mu}-g_{\omega B}
\gamma^{\mu}\omega_{\mu}-g_{\rho B}\gamma^{\mu}{\bf b}_{\mu}\cdot
{\bf t} -M_B+g_{\sigma B}\sigma\right)B\nonumber\\
&&+\thalf\partial_{\mu}\sigma\partial^{\mu}\sigma
-\thalf m^2_{\sigma}\sigma^2-U(\sigma)\nonumber\\
&&-\tquar F_{\mu\nu}F^{\mu\nu}+\thalf m^2_{\omega}\omega_{\mu}\omega^{\mu}
\nonumber\\
&&-\tquar {\bf B}_{\mu\nu}\cdot{\bf B}^{\mu\nu}+\thalf m^2_{\rho}{\bf b}_{\mu}
\cdot{\bf b}^{\mu} \; .\label{hyp1}
\end{eqnarray}
Here the $\rho$-meson field is denoted by ${\bf b}_{\mu}$,
the quantity ${\bf t}$ denotes the isospin operator which acts on the baryons, 
and the field strength tensors  for the vector mesons are given by the usual
expressions:--
$F_{\mu\nu}=\partial_{\mu}\omega_{\nu}-\partial_{\nu}\omega_{\mu}$,
${\bf B}_{\mu\nu}=\partial_{\mu}{\bf b}_{\nu}-\partial_{\nu}{\bf b}_{\mu}$.

It is straightforward to obtain the partition function for
the hadronic degrees of freedom, denoted by $Z_H$, 
\begin{eqnarray}
\ln Z_H&\pj=&\pj\beta V\left[\thalf m_{\omega}^2\omega_0^2+\thalf 
m_{\rho}^2b_0^2 - \thalf m_{\sigma}^2\sigma^2-U(\sigma)
-\sum_B\Delta E(M^*_B)\right]\nonumber\\
&&\qquad\qquad+ 2V\sum_B \int\frac{d^3k}{(2\pi)^3} \,\ln\left(1+{\rm e}
^{-\beta(E^*_B-\nu_B)}\right)\;. 
\label{hyp2}
\end{eqnarray}
We shall consider the relativistic Hartree approximation, for which we take
$U(\sigma)=0$. We shall also consider the mean field approximation, in which 
the negative energy sea contributions $\Delta E(M_B^*)$ is neglected and 
additional scalar self-couplings are included with 
$U(\sigma) = (bM/3)(g_{\sigma N}\sigma)^3 + (c/4)(g_{\sigma N}\sigma)^4$.
The contribution of antibaryons
are not significant for the thermodynamics of interest here, and is therefore
not included in Eq.~(\ref{hyp2}). Here, the effective baryon masses 
$M_B^*=M_B-g_{\sigma B}\sigma$ and
$E^*_B=\sqrt{k^2+M^{*2}_B}$. The chemical potentials are given by
\begin{equation}
\mu_B=\nu_B+g_{\omega B}\omega_0+g_{\rho B}t_{3B}b_0\;,\label{hyp3}
\end{equation}
where $t_{3B}$ is the third component of isospin for the baryon. Note
that particles with $t_{3B}=0$, such as the $\Lambda$ and $\Sigma^0$   
do not couple to the $\rho$. When hyperons are included, the negative energy sea
contribution from all baryons, inclusive of the hyperons, is considered 
as indicated by the notation $\Delta E(M^*_B)$.         

The shift in the energy density of the negative energy baryon states
is evaluated in the one loop Hartree approximation.
After removing divergences,  
$\Delta E(M^*_B)$ can be written~\cite{hr} in the form
\begin{eqnarray}
\Delta E(M^*_B)&\pj=&\pj-\frac{1}{8\pi^2}\biggl[4\left(1-\frac{\mu_r}{M}+
\ln\frac{\mu_r}{M}\right)M_B(M_B-M^*_B)^3-\ln\frac{\mu_r}{M}\;(M_B-M^*_B)^4
\nonumber\\
&&\qquad+M^{*4}_B
\ln\frac{M^*_B}{M_B}+M^3_B(M_B-M^*_B)-{\textstyle\frac{7}{2}}M^2_B(M_B-M^*_B)^2
\nonumber\\
&&\qquad\qquad\qquad+{\textstyle \frac{13}{3}}
M_B(M_B-M^*_B)^3-{\textstyle \frac{25}{12}}(M_B-M^*_B)^4\biggr]\;,
\label{walvac}
\end{eqnarray}
where $M=939$ MeV is the nucleon mass.
Here, the necessary renormalization introduces a scale parameter, $\mu_r$. For 
the standard choice~\cite{sew,chin} of \mum=1 (termed RHA), the first two terms 
in Eq.~(\ref{walvac}) vanish. This will not be the case for other choices of 
\mum, which introduce explicit $\sigma^3$ and $\sigma^4$ contributions.
At the phenomenological level, the
$\sigma^3$ and $\sigma^4$ couplings, generated from the baryon-loop diagrams,
modify the density dependence of the energy, which makes it possible~\cite{hr}
to obtain nuclear matter compression moduli that are significantly lower than
in the standard RHA.  We shall exploit this freedom to vary \mum\, and we call
this approach the  modified  relativistic Hartree approximation (MRHA). In
previous work~\cite{pehr} without hyperons, we found that while neutron star 
masses do not
significantly constrain  \mum, finite nuclei favor a value of 0.79. However, our
interest here is to explore the dependence on the varying stiffness of the
equation of state, which is permitted by modest variations in \mum,  for the
purpose of studying the impact on the structure of the star at finite
entropy.

Using $Z_H$, the thermodynamic quantities can be obtained in the standard way. 
The pressure $P_H=TV^{-1}\ln Z_H$, the number density for species $B$, and the
energy density $\varepsilon_H$ are given by
\begin{eqnarray}
n_B&\pj=&\pj2\int\frac{d^3k}{(2\pi)^3} 
\left({\rm e}^{\beta(E^*_B-\nu_B)}+1\right)^{\!-1}\;,\nonumber\\
\varepsilon_H&\pj=&\pj\thalf m_{\sigma}^2\sigma^2+U(\sigma)+
\thalf m_{\omega}^2\omega_0^2+\thalf m_{\rho}^2 b_0^2 
+\sum_B\Delta E(M^*_B)\nonumber\\
&&\qquad\qquad +2\sum_B
\int\frac{d^3k}{(2\pi)^3} 
E_B^*\left({\rm e}^{\beta(E^*_B-\nu_B)}+1\right)^{\!-1}\;.\label{hyp4}
\end{eqnarray}
The entropy density is then given by
$s_H=\beta(\varepsilon_H+P_H-\sum_B\mu_Bn_B)$.

The meson fields are obtained by extremization of the partition function, 
which yields the equations
\begin{eqnarray}
&&\fpj m_{\omega}^2\omega_0=\sum_Bg_{\omega B} n_B\quad;\quad
m_{\rho}^2b_0=\sum_Bg_{\rho B}t_{3B}n_B\;,\nonumber\\
&&\fpj m_{\sigma}^2\sigma=-\frac{dU(\sigma)}{d\sigma}
+\sum_B g_{\sigma B}\left\{2\hspace{-1mm}\int\!\frac{d^3k}{(2\pi)^3} 
\frac{M_B^*}{E_B^*}\left({\rm e}^{\beta(E^*_B-\nu_B)}+1\right)^{\!-1}
\hspace{-2mm}+\frac{\partial}{\partial M_B^*}\left[\Delta E(M^*_B) 
\right]\!\right\}.\label{hyp5}
\end{eqnarray}

The additional conditions needed to obtain a solution are provided by the
charge neutrality requirement, Eq. (\ref{charge}), and, when neutrinos are not 
trapped, the set of chemical 
potential relations provided by Eq. (\ref{beta}). For example, when $\ell=e^-$,
this implies the equalities  
\begin{eqnarray}
&&\fpj \mu_{\Lambda}= \mu_{\Sigma^0} = \mu_{\Xi^0} = \mu_n \,, \nonumber \\
&&\fpj \mu_{\Sigma^-}= \mu_{\Xi^-} = \mu_n+\mu_e \,, \nonumber \\
&&\fpj \mu_p = \mu_{\Sigma^+} = \mu_n - \mu_e \,. 
\label{murel}
\end{eqnarray}

In the case that the neutrinos are trapped, Eq. (\ref{beta}) is replaced by
Eq. (\ref{tbeta}).  The new equalities are then obtained by the replacement
$\mu_e \rightarrow \mu_e - \mu_{\nu_e}$ in the above equations.  The
introduction of additional variables, the neutrino  chemical potentials,
requires additional constraints, which we supply by fixing the lepton fractions,
$Y_{L\ell}$, as noted above. \\

\subsubsection{Neutrino-free, non-strange baryonic matter } 

For the MRHA calculations, the nucleon coupling constants were fitted to the 
equilibrium nuclear matter 
properties: a binding energy per particle of 16 MeV, an equilibrium 
density of 0.16 fm$^{-3}$, and a symmetry energy of 30 MeV (at the equilibrium 
density), as in Ref.~\cite{pehr}. Defining $C_i^2=(g_{iN}M/m_i)^2$, where
$N$ represents $n$ or $p$,
the parameters $C^2_{\omega}$, $C^2_{\sigma}$, 
and $C^2_{\rho}$ 
are reproduced in Table~7 for a range of values of
\mum. We see that they encompass a fairly wide range of values for the
nucleon effective mass, $M^*_N$,  and the compression modulus, $K_0$, 
at saturation. Correspondingly, the stiffness of the equation of 
state shows significant
variations. Note that $K_0$ alone is not always a reliable indicator of the
stiffness at high density; for example, the equation of state for \mum=0.73 is
noticeably stiffer than for \mum=1.25, yet the compression modulus is 
smaller, see Ref.~\cite{pehr}.  

In the mean field approximation, we use one of the parameter sets of
Glendenning and Moszkowski \cite{glenmos}, so that comparisons may be made. 
Specifically, we use the couplings 
\begin{eqnarray}
b &=& 0.008659  \qquad c = -0.002421 \nonumber \\
C_\omega^2 &=& 109.14, \qquad C_\sigma^2 = 224.78 \qquad {\rm and} \qquad
C_\rho^2 = 108.49 \,.
\end{eqnarray}
With these constants, the equilibrium density of nuclear matter is 
$n_0=0.153~{\rm fm}^{-3}$ with the Landau effective mass 
$M_{N{\rm sat}}^*/M = 0.827$ and compression modulus $K_0=240$ MeV.
This model is referred to as the GM model.

Table~8 shows the basic properties of stars in beta equilibrium for the 
MRHA model with various values of \mum\ and for the GM model. The structural 
changes at finite entropy compared to  the zero
entropy case are qualitatively similar to those given by the BPAL potential
model in the previous section. In most cases, $\lambda$ is a little 
smaller in magnitude, and this is probably due to the strong repulsion of the 
$\omega$-meson at high density, which increases the $P_{pot}$ term in
Eq. (\ref{pth}). Again,
a nearly quadratic increase of the maximum mass  with entropy is obtained.
The central temperatures, $T_c$, are lower here than in the non-relativistic 
case. Since it is the entropy per baryon that is constant, the temperature 
will vary with density. The temperature profile as a function of density
is shown by the full curves in the upper panel of Fig.~7 for \mum=1.25.
The temperature is a maximum at the center of the star (here the density 
ratio is $\sim7$ for a maximum mass star, see Table 8) and decreases with 
decreasing density, the fall off 
being particularly marked at low density. The density of a neutron 
star is approximately constant in the interior and drops to zero over a
radial distance $\Delta R/R\sim0.1$. Thus the interior of the star will 
have a constant temperature, and this will fall off rapidly in the surface
region.

The top panel in Fig.~8 shows the composition of the star for the model with
\mum=1.25. The middle panel shows the electron (or muon) chemical potential,
while the bottom panel shows the individual contributions to the total 
entropy.  Note that the lepton contribution to the entropy is smaller than in 
the non-relativistic case. In Fig.~9, we show in the top panel the behavior 
of the Landau effective masses for neutrons and protons; these are defined by 
$m^*_n\equiv E^*_{Fn}=\sqrt{k^2_{Fn}+M^{*2}_n}$, 
where $k^3_{Fn}=3\pi^2n_n$ for neutrons, and similarly for protons.
The isentropic pressures are shown in the middle panel of Fig.~9, and the
bottom panel indicates the star mass versus central density ratio for
fixed entropies.  These results are quite similar to those shown in Figs.~3 
and 4 for BPAL22, although the maximum masses obtained here are 
somewhat larger. 

As with the potential models, the nucleonic contributions to the thermal
pressure may again be simply estimated in the degenerate limit, $T/T_F \ll 1$.
For one nucleon species,
the Fermi temperature $T_F = k_F^2/(2E_F^*$), and 
the thermal pressure for this relativistic  model
is given by an expression of the form 
(see Ref.~\cite{pabw} for more details) 
\begin{eqnarray}
P_{th} = nT \frac {\pi^2}{4} \frac {T}{T_F} \cdot 
\frac 13 \left[ 1 + \left(\frac {M^*}{E_F^*}\right)^2 
\left (1-3 \frac {d\ln M^*}{d\ln n} \right)\right] + \cdots \,,
\end{eqnarray}
where the term containing the logarithmic derivative arises from the density 
dependence of the effective mass.  
The  entropy per particle in the degenerate limit is again 
given by $S=(\pi^2/2)(T/T_F)$,  but now with the Fermi temperature appropriate
for a relativistic spectrum.  Generalizing to two nucleon species, the total 
thermal pressure of the nucleons, up to quadratic terms in $S$, may be 
written as 
\begin{eqnarray}
P_{th} = n \left[ \frac {1}{3\pi^2}S^2 + \cdots \right] 
 \frac {\sum_i\frac {Y_i}{T_{F_i}} 
\left[ 1 + \left(\frac {M^*_i}{E_{F_i}^*}\right)^2 
\left (1-3 \frac {d\ln M^*_i}{d\ln n_i} \right)\right]  }
{\left(\sum_i \frac{Y_i}{T_{F_i}}\right)^2 }
\,, \quad i=n,p \,,
\label{thp}
\end{eqnarray}
where $Y_i=n_i/n$ with $n$ the total nucleon density. This provides an 
accurate approximation to the exact results. \\

\subsubsection{Neutrino-trapped, non-strange matter } 

With $Y_{Le}=0.4$ and $Y_{L\mu}=0$, Fig.~10 shows the effects of neutrino
trapping on the relative abundance, the chemical potentials, and the partial
entropies for an entropy per baryon $S=1$. The temperature profile of the 
star, indicated by the dotted curves in Fig.~7, differs rather little 
from the untrapped case. Results for the maximum mass
stars are summarized in Table~9 for the MRHA and GM models.
Note that the effect of trapping is similar to that found
earlier with the potential model, especially for the abundance of 
protons due to the large concentration of electrons.  
The amount by which the maximum decreases, $\sim 0.07M_\odot$ 
when neutrinos are trapped, is also similar to the results of the potential
models. Since thermal effects are smaller in the relativistic models,
the maximum mass of the $S=2$, neutrino-trapped star is always less than 
that of the cold $S=0$, neutrino-free star.

Together with the findings of the earlier section, we conclude
that in matter in which the only baryons are neutrons and protons, neutrino 
trapping usually decreases the maximum mass by a larger amount than thermal
effects increase it.    \\    
                                                      
\subsubsection{Neutrino-free, strangeness-rich baryonic matter } 

When hyperons are included, their coupling constants are needed; however, 
these are largely unknown. We shall assume that all the hyperon coupling 
constants are the same as those of the $\Lambda$, for which we can
take some guidance from hypernuclei. Following Glendenning
and Moszkwoski~\cite{glenmos}, the binding energy
of the lowest $\Lambda$ level in nuclear matter at saturation yields
\begin{eqnarray}
B_\Lambda &=& \mu_\Lambda - M_\Lambda 
= x_\omega g_{\omega n}\omega_0 + M^*_\Lambda - M_\Lambda \,, 
\quad {\rm or} \nonumber\\ 
-28 {\rm~MeV}&=& x_\omega g_{\omega n}\omega_0
-x_\sigma g_{\sigma n}\sigma\;,\label{hyp6}
\end{eqnarray}
where $x_\sigma={g_{\sigma\Lambda}}/{g_{\sigma n}}$ and 
$x_\omega={g_{\omega\Lambda}}/{g_{\omega n}}$.  Further, as suggested in 
Ref.~\cite{glenmos} on the basis of  fits to hypernuclear levels and neutron
star properties, we take  $x_\sigma=0.6$.  The value of $x_\omega$ may then be
determined from Eq.~(\ref{hyp6}).  For the $\rho$ meson, we take  $x_\rho=
g_{\rho\Lambda}/g_{\rho n}=x_{\sigma}$. The alternative choice, 
$x_\rho=x_\omega$, is found to produce essentially similar results.  
                                     
In the MRHA model, the negative energy sea
contributions $\Delta E(M^*_B)$  from all baryons, inclusive of the hyperons, 
contribute even when the positive energy states of the hyperons are empty. 
This entails a redetermination of the constants for the case 
in which hyperons are included. 
In Table 10, the coupling constants that reproduce nuclear matter saturation
properties and the binding energy of the lowest $\Lambda$ level in nuclear
matter at saturation  (with $x_\sigma=0.6$) 
are given for different choices of \mum.  Also shown
are the nucleon effective masses and the compression moduli.  Since the
negative energy sea contributions of all the baryons are positive, the values
of $C_\omega^2$ are somewhat smaller than those in Table 7 for the case in
which only nucleons are considered.  Consequently, the values of $K_0$ are also
smaller than those in Table 7.

For the MRHA model, with \mum=1.25, we show in Fig.~11 the relative fractions 
(top panel) of the baryons and leptons in
beta equilibrium,  and the electron chemical potential (bottom panel) as a
function of baryon density at zero temperature.   One expects that $\Lambda$,
with a mass of 1116 MeV, and the $\Sigma^-$, with a mass of 1197 MeV, first
appear at roughly the same density, because the somewhat higher mass of the
$\Sigma^-$ is compensated by the the presence of the electron chemical
potential in the equilibrium condition (see Eq.~(\ref{murel})) of the
$\Sigma^-$.  More massive, and more positively charged, particles than these
appear at higher densities.  Notice that with the
appearance of the negatively charged $\Sigma^-$  hyperon, which competes with
the leptons in maintaining charge neutrality, the lepton concentrations begin
to fall.  This is also reflected, for example, in the magnitude of the 
electron chemical potential which saturates at around 200 MeV and begins to
fall once the $\Sigma^-$ population begins to rise rapidly.  (That the
negatively charged $\Sigma^-$ is the cause for $\mu_e$ to fall with density may
be verified by allowing only the neutral particles to appear.  In this case,
$\mu_e$ continues to rise with density, albeit with a reduced slope compared to
the case in which no neutral hyperons are present.)   The rapid build up of the
other hyperons with increasing density has two major consequences. First, the
system is strangeness-rich at high density, with  nearly as many protons as
neutrons. Second, since at a given total baryon density the system contains 
many more baryon species with sizeable concentrations, the EOS is  considerably
softer  than when no hyperons are present. This causes the  maximum mass to be
reduced \cite{glenhyp,kaphyp}.

Fig.~12 contains the corresponding results at an entropy per baryon $S=1$.  
As expected, at
finite temperature the  hyperons attain significant fractions at lower baryon
densities than at zero temperature. The order of appearance of the various
hyperons follows from the chemical potential relations, Eq. (\ref{murel}), and
their differing masses. The bottom panel of Fig.~12 shows that the baryons 
carry most of the entropy, since the lepton populations remain low at high
density, due to the magnitude of the electron chemical potential. 

Table 11 summarizes the gross features of the maximum mass star populated with
hyperons.  Compared to the case in which only nucleons are present, the 
addition of hyperons causes the central temperatures to be reduced.
This is evident by comparison of the bottom and the top panels in 
Fig.~7. This figure also shows that with hyperons present the temperature 
changes rather little with density until $u<2$, so that a constant
temperature would be achieved over much of the star.
The softening introduced in the MRHA EOS by hyperons is evident in 
the maximum masses in Table 11. These are about $0.4-0.9M_\odot$, smaller than 
the results of Table 8, for which only nucleons are allowed in matter. Notice 
that in some cases, the maximum mass falls below $1.44M_\odot$, and the
pressure  support of finite entropy is not adequate to raise the maximum mass
above  $1.44M_\odot$.  We also give here results for the mean field GM model
\cite{glenmos} (with $x_\sigma = x_\rho = 0.6$ and $x_\omega = 0.659$).   This
model shows a similar reduction in the maximum mass of $\sim 0.5~M_\odot$,  due
to the softening induced by hyperons.   Qualitatively similar results, albeit
with somewhat  larger limiting masses, are obtained for the other choices of
couplings in Ref.~\cite{glenmos}.

In contrast to the maximum masses of stars containing nucleons and leptons
only,  the maximum masses of the hyperon populated stars do not show a regular
behavior with increasing entropy. In those cases for which the maximum mass
increases with entropy, a quadratic increase is observed. However, with
\mum=1, a decreasing trend with entropy is found.  We have verified that
this surprising behavior does not violate any laws of thermodynamics. It can be
traced back to the softening of the EOS caused by the appearance of hyperons at
relatively smaller densities than are found at zero temperature.  \\

\subsubsection{Neutrino-trapped, strangeness-rich matter } 

Fig.~13 is the counterpart of Fig.~11 for the case in which neutrinos are 
trapped at zero temperature. Trapped
neutrinos have a large influence on the charged hyperon thresholds. For
example, the appearance of the $\Sigma^-$ hyperon is governed by 
$\mu_{\Sigma^-}=\mu_n+\mu$. Here, $\mu=\mu_e-\mu_{\nu_e}$ is much smaller than
in the untrapped case for which $\mu=\mu_e$, so the appearance of the
$\Sigma^-$ occurs at a higher density.  With the appearance of hyperons, the
neutrino population begins to increase with density, in contrast to the
monotonic decrease exhibited in the hyperon free case. The finite chemical
potential of the  neutrinos requires the electron chemical potential to be at a
higher value than in the neutrino free case  to maintain chemical equilibrium.
Thus, unlike the neutrino free case, the electron chemical potential increases
with density.  Notice that muons play little role here.  The preponderance
of negatively charged particles, both leptons and baryons,  now has the
consequence that the system is proton rich over an extended region  of density. 
These qualitative features are retained also at finite entropy with baryons
carrying most of the entropy, as shown in Fig.~14.

The physical properties of the maximum mass stars with trapped neutrinos  are
listed in Table 12.   The changes due to entropy alone are small and not always
in the direction of increasing the maximum mass. Notice, however, that for each
entropy shown,  the  maximum masses are all about $0.2M_\odot$ {\it
larger} than those in Table 11 for neutrino free stars.  Since the star has to
cool down from an $S\sim 1$ configuration, with neutrinos trapped, to a
configuration of $S\sim 0$ without neutrinos, the maximum stable mass
decreases.  This may be contrasted with the case in which 
no hyperons are present,
in which case neutrino trapping and finite entropy effects are opposed to each
other and effectively cancel.  This is a general result that stems from the
softening induced by the presence of negatively charged hadrons in the star, as 
demonstrated in the next section, where kaon condensation is allowed to occur in
dense matter.                               

\subsubsection{Metastability of neutron stars with strangeness-rich baryons} 

The most striking conclusion of the above discussion is the possibility of
metastable neutron stars if matter contains hyperons.  Metastable stars  occur
within a range of masses near the  maximum mass of the initial configuration
and remain  stable only for several seconds after formation.    In contrast, in
matter with only nucleons, any star that is below  the mass limit of the
initial configuration will be stable during the  subsequent evolution (in the
absence of mass accretion).      

In an idealized picture, the two features that govern the evolution of 
the maximum mass are the lepton content and finite entropy. The latter plays 
a minor role, but effects of the high lepton content are very important. 
The  combined effects of lepton fraction and finite entropy are shown in 
Fig.~15. Here, the abscissa is the baryonic mass which is proportional to the 
number of baryons in the system and is constant during the evolution of the
star (in the absence of accretion of matter, again, an idealization). The 
ordinate is the gravitational mass, hitherto referred to simply as the mass,
which includes interactions and, therefore, changes as the star evolves. If 
hyperons
are present (lines ending with a dot), then deleptonization, that is the
transition from $Y_{L_e}=0.4$  to $Y_\nu=0$, accompanied by heating from $S=1$
to $S=2$, lowers the range of baryonic masses that can be supported by the
equation of state  (of model GM here) from about  $1.95M_\odot$ to about
$1.73M_\odot$. The window in the baryonic mass in which  neutron stars are
metastable is thus about $0.22M_\odot$ wide. On the other hand, if hyperons 
are absent (lines ending with a star), the baryonic mass increases during 
deleptonization, and no metastability occurs. Similar results apply to the
gravitational masses, which can be obtained from the baryonic masses with the
help of the right panel.

Fig.~16 illustrates the evolution of the maximum mass during 
deleptonization.  For clarity, the small 
effects of finite temperature are not considered
here. When neutrinos diffuse out, the neutrino fraction decreases  from an
initial maximum value of about $Y_{\nu_e}=0.08$,  corresponding to
$Y_{L_e}=0.4$, to zero. The  evolution of the neutrino fraction with time can
only be obtained by solving the neutrino and energy transport equations, as was
done, for example, by Burrows and Lattimer~\cite{birth}. This is beyond the
scope of the  present work. However, if one assumes that the neutrino fraction
decreases approximately uniformly with time, then one obtains a good picture of
how the maximum mass will evolve. In Fig.~16, the maximum mass is shown as a
function of the electron neutrino number  per baryon, $Y_{\nu_e}$, for the GM
model. It is evident that different compositions lead to different trends. In
nucleonic matter, the maximum mass slowly {\em increases} as neutrinos diffuse
out; in hyperonic matter,  maximum mass {\em decreases}.  It is also clear that
the rate of change is largest shortly before the neutrino fraction drops to
zero. Similar qualitative behavior is obtained for other  choices of mean field
parameters from Ref.~\cite{glenmos}.

\subsection{Matter with kaon condensation}
\setcounter{subsubsection}{0}

The idea that, above some critical density, the ground state of baryonic 
matter might contain a Bose-Einstein condensate of kaons is due to Kaplan and  
Nelson~\cite{kapnel}. The formulation, in terms of chiral perturbation theory, 
was subsequently discussed by Politzer and Wise~\cite{pol} and Brown  {\it et 
al.}~\cite{bro2}.  The composition and structure of kaon condensed stars and 
also some evolutionary aspects were considered by Thorsson, Prakash and
Lattimer~\cite{tpl}.  Further related calculations at the mean-field level  
may be found in Refs.~\cite{muto,maruy}. Most recently, loop contributions have  
also been investigated~\cite{bro3}.   Depending on the parameters employed, 
particularly the strangeness content of the proton, kaon condensation is 
typically found to occur at
about 4 times the equilibrium nuclear matter density. This has the
effect of softening the equation of state and lowering the maximum  mass of 
the neutron star. 

In chiral models the kaon-nucleon interaction  occurs directly via four point 
vertices; however, it can also be modelled as an indirect interaction 
that arises from the exchange of mesons~\cite{jul}. The latter approach has 
the virtue that it is more consistent with the meson exchange picture that is 
usually employed for the baryon interactions, and it is of interest to compare 
the predictions with those of the chiral model. A further question is the 
role that hyperons might play in addition to kaons,
since we have seen that they appear at a density similar to, or somewhat 
lower than, the condensate threshold. This raises the issue of 
the interplay between the strange baryons and mesons and the net  strangeness
content of neutron stars.  These questions have been the subject of recent 
work~\cite{ekp,kpe,schaffner}, and we shall address them in the present context.
                                                         
Our purpose here is to determine the impact of a kaon condensate on 
protoneutron stars. In order to keep the number of calculations within 
bounds, we will focus on the mean field case with the GM parameters. 
We will study the effect of neutrino trapping in chiral and meson 
exchange models, both with and without hyperons. As regards finite entropy 
effects, we have seen that they are less significant than neutrino trapping.
It is not clear how to develop a consistent finite temperature formalism 
for the chiral case, while it is reasonably straightforward for the 
meson exchange model. We therefore study the effects of finite entropy in 
the latter approach and restrict our chiral formalism to $T=0$. \\

\subsubsection{Chiral formalism}

The Kaplan-Nelson $SU(3)\times SU(3)$ chiral Lagrangian for
the kaons and the $s$-wave kaon-baryon interactions takes the form
\begin{eqnarray}
{\cal L}_K&\pj=&\pj\tquar f^2 {\rm Tr}\partial_{\mu}U\partial^{\mu}U\!+
C{\rm Tr}\,m_q(U+U^{\dagger}\!-2)+i{\rm Tr}\bar{B}\gamma^{\mu}[V_{\mu},B]
+a_1{\rm Tr}\bar{B}(\xi m_q\xi+\!h.c.)B\nonumber\\
&&\pj+a_2{\rm Tr}\bar{B}B(\xi m_q\xi+h.c.)+a_3\left\{{\rm Tr}\bar{B}B\right
\}
\left\{{\rm Tr}(m_qU+h.c.)\right\}\;.\label{pje0}
\end{eqnarray}
Here, $U=\xi^2$ is the non-linear field involving the pseudoscalar meson octet, 
from which we retain only the $K^{\pm}$ contributions--
\begin{equation}
U=\exp\left(\frac{\sqrt{2}i}{f}M\right)\quad;\quad
M=\left(\matrix{0&0&K^+\cr
0&0&0\cr K^-&0&0\cr}\right)\;.
\end{equation}
The baryon octet -- nucleons plus hyperons -- is given by
\begin{equation}
B=\left(\matrix{\sqrt{\frac{1}{2}}\Sigma^0+\sqrt{\frac{1}{6}}\Lambda&
\Sigma^+&p\cr
\Sigma^-&-\sqrt{\frac{1}{2}}\Sigma^0+\sqrt{\frac{1}{6}}\Lambda&n\cr
\Xi^-&\Xi^0&-\sqrt{\frac{2}{3}}\Lambda\cr}\right)\;.
\end{equation}
In Eq. (\ref{pje0}), the quark mass matrix
$m_q={\rm diag}(0,0,m_s)$;  {\it i.e.}, only the mass of the strange quark 
is taken to be non-zero. For the mesonic vector current, $V_{\mu}$,
only the time component  survives in an infinite system with
$V_0=\thalf(\xi^{\dagger}\partial_0\xi+\xi\partial_0\xi^{\dagger})$. Also, the
pion decay constant $f=93$ MeV, and $C,a_1,a_2$ and $a_3$ are constants.
After some algebra, the relevant part of ${\cal L}_K$ takes the form
\begin{eqnarray}
{\cal L}_K&\pj=&\pj\left(\frac{\sin\chi}{\chi}\right)^{\!2}\Biggl\{
\partial_{\mu}K^+\partial^{\mu}K^-
+\frac{i}{4f^2}\frac{(K^+\partial_0K^--K^-\partial_0K^+)}{\cos^2\thalf\chi}
\sum_B(Y_B+q_B)B^{\dagger}B
\nonumber\\
&&\fpj-\biggl(m_K^2+\frac{m_s}{2f^2}\sum_B\Bigl[(a_1+a_2)(1+Y_Bq_B)
+(a_1-a_2)(q_B-Y_B)+4a_3\Bigr]\bar{B}B\nonumber\\
&&\fpj+\frac{m_s}{6f^2}(a_1+a_2)(2\bar{\Lambda}\Lambda
+\sqrt{3}[\bar{\Sigma}^0\Lambda+\bar{\Lambda}\Sigma^0])
\biggr)\frac{K^+K^-}{\cos^2\thalf\chi}\biggr\}\;,\label{tapp1}
\end{eqnarray}
where $q_B$ and $Y_B$ are the baryon charge and
hypercharge, respectively, and $q_B=\thalf Y_B+t_{3B}$. (The hypercharge is 
given in terms of the baryon number and strangeness by $Y_B=b_B+S_B$.)
In Eq. (\ref{tapp1}), we have defined $\chi^2=2K^+K^-/f^2$ and taken 
the kaon mass to be given by $m_K^2=2Cm_s/f^2$.
We have not included in Eq. (\ref{tapp1}) terms which simply give a constant
shift to the baryon masses; they indicate that $a_1m_s=-67$ MeV and
$a_2m_s=134$ MeV, using the hyperon-nucleon mass differences. The remaining
constant $a_3m_s$ is not accurately known, and we shall use values in the
range $-134$ to $-310$ MeV corresponding to 0 to 20\% strangeness content
for the proton. The corresponding range for the kaon-nucleon sigma term,
\begin{equation}
\Sigma^{KN}=-\thalf(a_1+2a_2+4a_3)m_s \,,
\end{equation}
is 167--520 MeV. Some 
guidance is provided by recent lattice gauge 
simulations~\cite{dong}, which find that the strange quark condensate in the 
nucleon is large,
{\it i.e.}, $\langle N|\overline ss|N\rangle = 1.16\pm0.54$. From the relation
$m_s\langle\overline ss\rangle_p = - 2(a_2+a_3)m_s$ and using $m_s=150$ MeV,
we obtain $a_3m_s = -(220 \pm40)$ MeV, which is in the middle of our range
of values.

The pure kaon part of Eq. (\ref{tapp1}) gives a contribution to the grand 
potential, $\Omega_K=-T\ln Z_K$, which can be
evaluated in the mean field approximation by writing the time dependence of 
the fields $K^{\pm}=\frac{1}{\sqrt{2}}f\theta e^{\pm i\mu t}$~\cite{tpl}.
Here, $\theta$ gives the condensate amplitude, and $\mu$ is the kaon chemical
potential. One easily finds
\begin{equation}
\Omega_K=
Vf^2(2m_K^2\sin^2\!\thalf\theta-\thalf\mu^2\sin^2\!\theta)\;.
\end{equation}

The kaon-baryon interactions in Eq. (\ref{tapp1}) can be incorporated in the 
baryon Lagrangian of Eq. (\ref{hyp1}) by suitable definitions of the 
effective masses and chemical potentials.
Notice first that the $\Lambda-\Sigma^0$ mass matrix needs to be diagonalized.  
We write it in the form
\begin{eqnarray}
&&\fpj
\left( \begin{array}{cc}
2\alpha&2\beta\\
2\beta&2\gamma
\end{array} \right) \,,
\qquad {\rm where}\nonumber\\
&&\fpj2\alpha=M_{\Lambda}-g_{\sigma\Lambda}\sigma
+(\fiveth a_1+\fiveth a_2+4a_3)m_s\sin^2\!\thalf\theta\nonumber\\
&&\fpj2\beta=3^{-\frac{1}{2}}(a_1+a_2)m_s\sin^2\!\thalf\theta\nonumber\\
&&\fpj2\gamma=M_{\Sigma}-g_{\sigma\Sigma}\sigma
+(a_1+a_2+4a_3)m_s\sin^2\!\thalf\theta\;.
\end{eqnarray}
Then it is straightforward to obtain the eigenstates and the masses
\begin{eqnarray}
&&\fpj H_1=\frac{\Sigma^0-\delta\Lambda}{(1+\delta^2)^{\frac{1}{2}}}\quad;\quad
H_2=\frac{\Lambda+\delta\Sigma^0}{(1+\delta^2)^{\frac{1}{2}}}\quad{\rm with}
\quad2\beta\delta=\gamma-\alpha-\sqrt{(\gamma-\alpha)^2+4\beta^2}
\nonumber\\
&&\fpj M^*_{H_1}=\gamma+\alpha+\sqrt{(\gamma-\alpha)^2+4\beta^2}\quad;\quad
M^*_{H_2}=\gamma+\alpha-\sqrt{(\gamma-\alpha)^2+4\beta^2}
\;.\label{tmix}
\end{eqnarray}
Since $\gamma>\alpha$ here, in the limit of no mixing ($\beta=0$) $\delta$ is
zero.  Henceforth, the sum over baryon states, $B$, includes ${H_1}$ and 
${H_2}$ along with the $n,\,p,\,\Sigma^{+,-},\,\Xi^{0,-}$. For the latter 
cases, the masses are given by
\begin{equation}
M^*_B=M_B-g_{\sigma B}\sigma+[(a_1+a_2)(1+Y_Bq_B)+(a_1-a_2)(q_B-Y_B)
+4a_3]m_s\sin^2\!\thalf\theta\;.
\label{temasses}
\end{equation}
The chemical potentials $\mu_B$ are given in terms of the effective chemical
potentials, $\nu_B$, by
\begin{equation}
\mu_B=\nu_B+g_{\omega B}\omega_0+g_{\rho B}t_{3B}b_0
-(Y_B+q_B)\mu\sin^2\!\thalf\theta\;.
\label{tapp3}
\end{equation}
Then, the zero-temperature limit of Eq. (\ref{hyp2}) yields the baryon
grand potential $\Omega_H=-T\ln Z_H$. (We consider only the mean field case 
here, so $\Delta E=0$.)

The total hadron grand potential $\Omega_{\rm tot}=\Omega_H+\Omega_K=-PV$, 
where $P$ is the total pressure. The energy density is given by 
$\varepsilon=-P+\sum_B\mu_Bn_B+\mu n_K$, where the baryon number density, 
$n_B$, is given by the zero-temperature limit of Eq. (\ref{hyp4}) and the 
kaon number density is
\begin{equation}
n_K=-\frac{1}{V}\frac{\partial\Omega_{\rm tot}}{\partial\mu}=
f^2(\mu\sin^2\!\theta+4b\sin^2\!\thalf\theta)\quad{\rm with}\quad
b=\sum_B(Y_B+q_B)n_B/(4f^2)\;.
\end{equation}

The meson fields are obtained by extremizing $\Omega_{\rm tot}$, 
yielding Eq. (\ref{hyp5}) with $T=0$. The condensate amplitude, $\theta$, 
is also found by extremizing $\Omega_{\rm tot}$.
This yields the solutions $\theta=0$ (no condensate), or, if a condensate
exists, the equation
\begin{eqnarray}
&&\fpj\mu^2\cos\theta+2\mu b-m_K^2-d_1-d_2=0\,\qquad{\rm where}\;\nonumber\\
&&\fpj2f^2d_1=\sum_{B\ne H_1,H_2}\Bigl[(a_1+a_2)(1+Y_Bq_B)
+(a_1-a_2)(q_B-Y_B)+4a_3\Bigr]m_sn^s_B\;,\nonumber\\
&&\fpj2f^2 d_2 \sin^2\!\thalf\theta =\sum_{B=H_1,H_2}(M^*_B
+g_{\sigma\Lambda}\sigma)n^s_B-M_{\Lambda}n^s_{H_1}
-M_{\Sigma}n^s_{H_2}\nonumber\\
&&\hspace{6cm}+\frac{1}{1+\delta^2}
(M_{\Lambda}-M_{\Sigma})(n^s_{H_1}-n^s_{H_2})\;.
\label{tthresh1}
\end{eqnarray}
Here, we have taken $g_{\sigma\Lambda}=g_{\sigma\Sigma}$ for simplicity, and
the baryon scalar density is
\begin{equation}
n^s_B=\frac{1}{\pi^2}\int\limits_0^{k_{FB}}dk\,k^2\frac{M^*_B}{(k^2
+M^{*2}_B)^{\frac{1}{2}}}\;,
\end{equation}
with $k_{FB}$ denoting the baryon Fermi momentum.
Equation (\ref{tthresh1}) is equivalent to the requirement that $\mu$ be 
equal to the energy of the $K^-$ zero-momentum state.

Finally, we need to satisfy the charge neutrality condition of 
Eq. (\ref{charge}), which reads 
\begin{equation}
\sum_B q_Bn_B-n_K-n_e-n_{\mu}=0\,,
\end{equation}
and the 
chemical equilibrium conditions of Eq. (\ref{murel}). In the latter, we replace
$\mu_{\Sigma^0}$ and $\mu_{\Lambda}$ by $\mu_{H_1}$ and $\mu_{H_2}$; thus, 
the first of Eqs. (\ref{murel}) becomes 
$\mu_{H_1} = \mu_{H_2} = \mu_{\Xi^0} = \mu_n $. Also, chemical
equilibrium in the reaction $n\leftrightarrow p+K^-$ requires that the kaon
chemical potential satisfy $\mu=\mu_n-\mu_p$.

We shall also need the optical potential  for a kaon in nuclear matter
for comparison with the meson exchange approach.
We can use Eq. (\ref{tapp1}) with equal numbers of neutrons and 
protons, no hyperons, and no condensate ($\chi=0$). Lagrange's equations for 
an $s$-wave $K^-$, with $K^-=k^-(\vm{x})e^{-iEt}$ and $E=\sqrt{p^2+m_K^2}$,
can be written
\begin{eqnarray}
[\pmb\nabla^2+E^2-m_K^2]k^-(\vm{x})&\pj=&\pj\left[-\frac{3nE}{4f^2}-
\frac{\Sigma^{KN}n^s}{f^2}\right]k^-(\vm{x})\nonumber\\
&\pj=&\pj2m_KU^K_{opt}\:k^-(\vm{x})\;,\label{opt}
\end{eqnarray}
where $n$ is the density of nuclear matter and $n^s$ is the scalar density.
For a zero-momentum $K^-$ meson, the optical potential reduces to
\begin{equation}
U^K_{opt}=S^K_{opt}+V^K_{opt}\quad;\quad S^K_{opt}=-\frac{\Sigma^{KN}n^s}
{2m_Kf^2}\quad;\quad V^K_{opt}=-\frac{3n}{8f^2}\;.\label{chopt}
\end{equation}
Equilibrium nuclear matter density fixes $V^K_{opt}=-51$ MeV, and, for our
range of values of the sigma term, $S^K_{opt}=-$(22-69) MeV. Thus, 
$U^K_{opt}\sim-100$ MeV. This is about half of the favored ``deep solution" 
obtained by Friedman, Gal and Batty~\cite{batty} in their analysis of the 
kaonic atom  data, although it is comparable to the value obtained in their
$t_{eff}\rho$ approximation. There are, of course, uncertainties in the 
analysis and also in simply expropriating the real part of a complex 
potential. \\

\subsubsection{Meson exchange formalism} 

In this approach, we take a Lagrangian for the kaon sector, which contains 
the usual kinetic energy and mass terms, along with the meson interactions,
\begin{eqnarray}
{\cal L}_K&\pj=&\pj\partial_{\mu}K^+\partial^{\mu}K^-
-(m_K^2-g_{\sigma K}m_K\sigma)K^+K^-\nonumber\\
&&\hspace{2cm}+i\left[g_{\omega K}\omega^{\mu}+g_{\rho K}b^{\mu}\right]
(K^+\partial_{\mu}K^--K^-\partial_{\mu}K^+)\;.
\label{kaonlag}
\end{eqnarray}
Here, $b^{\mu}$ denotes the $\rho^0$ field. 
(The vacuum kaon mass, $m_K$, is present in the third term to render
$g_{\sigma K}$ dimensionless.) Schaffner and Mishustin~\cite{schaffner} have
included an additional four-point interaction in their Lagrangian, so that 
Lagrange's equations yield $\partial_{\mu}\omega^{\mu}=0$ as required for a 
particle of spin 1~\cite{og}. We would argue that at the mean field level
the vector fields are constants, so the divergence is necessarily zero, and 
Eq. (\ref{kaonlag}) is sufficient. We can simplify notation by
introducing an effective kaon mass defined by
\begin{equation}
        {m_K^*}^2 = m_K^2 - g_{\sigma K}m_K \sigma\;.
\end{equation}
Since only the time components of the vector fields survive, it is also 
useful to define
\begin{equation}
X=g_{\omega K}\omega_0+g_{\rho K}b_0.
\end{equation}

In order to determine the kaon partition function at finite temperature, 
we generalize the procedure outlined in Kapusta~\cite{kapusta}. First, by
studying  the invariance of the Lagrangian under the transformation 
$K^{\pm}\rightarrow K^{\pm}e^{\pm i\alpha}$, the conserved current density 
can be identified. The zero component, {\it i.e.} the charge density, is
\begin{equation}
J_0=i(K^+\partial_0K^- - K^-\partial_0K^+) + 2XK^+K^-\;.
\end{equation}
Next, we transform to real fields $\phi_1$ and $\phi_2$,
\begin{equation}
K^{\pm}=(\phi_1\pm\phi_2)/\sqrt{2}\;,
\end{equation}
and determine the conjugate momenta
\begin{equation}
\pi_1=\partial_0\phi_1-X\phi_2\qquad;\qquad\pi_2=\partial_0\phi_2+X\phi_1\;.
\end{equation}
The Hamiltonian density is
${\cal H}_K=\pi_1\partial_0\phi_1+\pi_2\partial_0\phi_2
-{\cal L}_K$, and the partition function of the grand canonical ensemble can 
then be written as the functional integral
\begin{eqnarray}
Z_K&\pj=&\pj\int[d\pi_1][d\pi_2]\int_{periodic}[d\phi_1][d\phi_2]\nonumber\\
&&\times\exp\left\{\int\limits_0^{\beta}d\tau\int d^3x\left(
i\pi_1\frac{\partial\phi_1}{\partial\tau}+
i\pi_2\frac{\partial\phi_2}{\partial\tau}-{\cal H}_K(\phi_i,\pi_i)
+\mu J_0(\phi_i,\pi_i)\right)\right\}\;.\label{zint}
\end{eqnarray}
Here, $\mu$ is the chemical potential associated with the conserved charge 
density, and the fields obey periodic boundary conditions in the imaginary time
$\tau=it$, namely $\phi_i(\vm{x},0)=\phi_i(\vm{x},\beta)$.

The Gaussian integral over momenta in Eq. (\ref{zint}) is easily performed.
Next the fields are Fourier decomposed according to
\begin{eqnarray}
\phi_1&\pj=&\pj f\theta\cos\alpha+\sqrt{\frac{\beta}{V}}\sum_{n,\vms{p}}
e^{i(\vms{p}\cdot\vms{x}+\omega_n\tau)}\phi_{1,n}(\vm{p})\nonumber\\
\phi_2&\pj=&\pj f\theta\sin\alpha+\sqrt{\frac{\beta}{V}}\sum_{n,\vms{p}}
e^{i(\vms{p}\cdot\vms{x}+\omega_n\tau)}\phi_{2,n}(\vm{p})\;,
\end{eqnarray}
where the first term describes the condensate, so that in the second term
$\phi_{1,0}(\vm{p}=0)=\phi_{2,0}(\vm{p}=0)=0$. The Matsubara frequency
$\omega_n=2\pi nT$. The partition function can then be written
\begin{eqnarray}
Z_K&\pj=&\pj N\int \prod_{n,\vms{p}}[d\phi_{1,n}(\vm{p})][d\phi_{2,n}(\vm{p})]
e^S\;,\qquad{\rm where}\nonumber\\
S&\pj=&\pj\thalf \beta Vf^2\theta^2(\mu^2+2\mu X-m_K^{*2})
-\thalf\sum_{n,\vms{p}}\Bigl(\phi_{1,-n}(-\vm{p}),\phi_{2,-n}(-\vm{p})\Bigr)
\vm{D}\left(\matrix{\phi_{1,n}(\vm{p})\cr\phi_{2,n}(\vm{p})\cr}\right)
\;,\nonumber\\
\vm{D}&\pj=&\pj\beta^2\left(\matrix{\omega_n^2+p^2+m_K^{*2}-2\mu X-\mu^2&
2(\mu+X)\omega_n\cr-2(\mu+X)\omega_n&\omega_n^2+p^2+m_K^{*2}-2\mu X-\mu^2\cr}
\right).
\end{eqnarray}
We define the $K^{\pm}$ energies according to 
\begin{equation}
\omega^{\pm}(p)=\sqrt{p^2+m_K^{*2}+X^2}\pm X\;,
\end{equation}
and, in most cases, we shall suppress the explicit dependence of $\omega^{\pm}$ 
on $p$. With this definition, the determinant of $\vm{D}$ is
\begin{equation}
{\rm det}\,\vm{D}=\beta^4\left[\omega_n^2+(\omega^--\mu)^2\right]
\left[\omega_n^2+(\omega^++\mu)^2\right]\;.
\end{equation}
Then
\begin{equation}
\ln Z_K=\thalf\beta Vf^2\theta^2(\mu^2+2\mu X-m_K^{*2})-\thalf
\sum_{n,\vms{p}}\ln{\rm det} \vm{D}\;.
\end{equation}
Performing the sum over $n$ and neglecting the zero-point contribution, 
which is not appropriate to a mean field approach, we obtain
\begin{eqnarray}
\ln Z_K&\pj=&\pj\thalf\beta Vf^2\theta^2(\mu^2+2\mu X-m_K^{*2})\nonumber\\
&&-V\int\limits_0^{\infty}\frac{d^3p}{(2\pi)^3}\left[
\ln(1-e^{-\beta(\omega^--\mu)})+
\ln(1-e^{-\beta(\omega^++\mu)})\right]\;.\label{zkexch}
\end{eqnarray}

The partition function for the baryon sector takes precisely the form given 
in Eq. (\ref{hyp2}), and the total partition function is 
$Z_{\rm total}=Z_HZ_K$. Extremization of $Z_{\rm total}$ yields the fields. 
It is useful to define the thermal quantities
\begin{eqnarray}
n_K^{TH}&\pj=&\pj\int\frac{d^3p}{(2\pi)^3}[f_B(\omega^--\mu)
-f_B(\omega^++\mu)]\;,\nonumber\\
A_K^{TH}&\pj=&\pj\int\frac{d^3p}{(2\pi)^3}\left(p^2+m_K^{*2}+X^2\right)^
{-\frac{1}{2}}[f_B(\omega^--\mu)+f_B(\omega^++\mu)]\;,
\end{eqnarray}
where the Bose occupation probability $f_B(x)=(e^{\beta x}-1)^{-1}$.
Then the field equations are
\begin{eqnarray}
m_{\omega}^2\omega_0 &\pj=&\pj \sum_Bg_{\omega B} n_B -g_{\omega K}
\left[(f\theta)^2\mu +n_K^{TH}-XA_K^{TH}\right]\nonumber\\
m_{\rho}^2b_0 &\pj=&\pj \sum_Bg_{\rho B}t_{3B}n_B -g_{\rho K}
\left[(f\theta)^2\mu +n_K^{TH}-XA_K^{TH}\right]\nonumber\\
\nonumber\\
m_{\sigma}^2\sigma &\pj=&\pj-\frac{dU(\sigma)}{d\sigma} 
+2\sum_B g_{\sigma B}\int\frac{d^3k}{(2\pi)^3}
\frac{M_B^*}{E_B^*}\left(e^{\beta(E_B^*-\nu_B)}+1\right)^{-1}\nonumber\\
&&+{\textstyle{\frac{1}{2}}}g_{\sigma K}m_K\left[(f\theta)^2+A_K^{TH}\right]
\;.\label{hhyp5}
\end{eqnarray}
Notice that the condensate contributes directly to the equations of motion
(\ref{hhyp5}), whereas in chiral models the contribution appears in the
effective chemical potentials and effective masses. 
The condensate amplitude, $\theta$, is also found by extremization of
$Z_{\rm total}$. This yields the solutions $\theta=0$ (no condensate), or, 
if a condensate exists, the equation
\begin{equation}
\mu^2+2\mu X-m_K^{*^2}=[\mu-\omega^-(0)][\mu+\omega^+(0)] =0\;.
\label{tthresh2}
\end{equation}
Since $\mu$ is positive here, we only have the possibility of a $K^-$ 
condensate when $\mu=\omega^-(0)$. We illustrate  the behavior of
the kaon energies with density in Fig.~17. As the chemical potential $\mu$
increases, a condensate forms when $\mu$ becomes equal to $\omega^-(0)$.
At higher densities, a condensate is still present so that $\mu$
remains equal to $\omega^-(0)$. Whether the $\omega^+(0)$ energy
increases rapidly with density, as in Fig.~17, or remains more nearly 
constant, depends on the strength of the couplings employed.
Notice that, utilizing Eq. (\ref{hhyp5})
in Eq. (\ref{tthresh2}), one obtains a threshold ($\theta$ infinitesimal)
equation resembling Eq. (\ref{tthresh1}) of the chiral case;  although the 
weightings of the various baryons are different, and the $dU/d\sigma$ term 
does not play a role in the chiral case (see the further discussion below). 
Above threshold, $\theta$ enters in different ways in the two models.

The baryon thermodynamic variables are given by Eq. (\ref{hyp4}).
For the kaons, the partition function, Eq. (\ref{zkexch}),  gives the 
pressure, $P_K=TV^{-1}\ln Z_K$; notice that Eq. (\ref{tthresh2}) indicates 
that the condensate gives zero contribution to the pressure. The kaon 
number density is easily found to be
\begin{equation}
n_K=f^2\theta^2(\mu+X)+n_K^{TH}\;.
\end{equation}
If we write the baryon energy density in the form of Eq. (\ref{hyp4}),
then, after using the equations of motion Eqs.~(\ref{hhyp5}) and 
(\ref{tthresh2}), the remaining part of the energy density, which arises from
the kaons,  can be written   
\begin{equation}
\varepsilon_K=(f\theta m_K^*)^2
+\int\!\frac{d^3p}{(2\pi)^3}[\omega^-(p)f_B(\omega^--\mu)
+\omega^+(p)f_B(\omega^++\mu)]+X[n_K^{TH}-XA_K^{TH}].
\end{equation}
The total entropy density can be obtained from the standard thermodynamic 
identity 
\begin{equation}
s=\beta(\varepsilon+P-\mu n_K-\sum_B\mu_Bn_B)\,.
\end{equation}
The above equations can be applied at zero temperature, in which case the
thermal Bose occupation probabilities are zero, and the Fermi occupation 
probabilities become step  functions cut off at the Fermi momentum.

As before, the neutron star must be charge neutral,
{\it i.e.} 
\begin{equation}
\sum_B q_Bn_B-n_K-n_e-n_{\mu}=0\,,
\end{equation}
and in chemical equilibrium, with Eq. (\ref{murel}) satisfied and
$\mu=\mu_n-\mu_p$.
                            
Finally, as in the chiral model, we can determine the value of the optical 
potential felt by a single kaon at zero momentum. The analogue of
Eq. (\ref{opt}) gives
\begin{equation}
 U_{opt}^K \equiv {\cal S}^K_{opt} + {\cal V}^K_{opt}  \quad;\quad
{\cal S}^K_{opt}=  - {\textstyle{\frac{1}{2}}}
g_{\sigma K} \sigma \quad;\quad {\cal V}^K_{opt}=-g_{\omega K} \omega_0\;.
\label{uoptmeson}
\end{equation}
Since we want to compare the chiral and meson exchange approaches, we will
demand that they yield the same optical potential in nuclear matter. This
should be a reasonable way of ensuring that the parameterizations are
compatible. Thus, we choose ${\cal S}^K_{opt}$ and ${\cal V}^K_{opt}$ to be 
equal to $S_{opt}^K$ and $V_{opt}^K$ from Eq. (\ref{chopt}).  For the as yet 
undetermined kaon-rho meson coupling, we take $g_{\rho K}/g_{\rho N} = 1/3$, 
as suggested by naive quark counting. \\

\subsubsection{Kaon condensation in non-strange baryonic matter} 

Let us begin by studying the critical, or threshold, density
at which kaons start to condense. The equations obtained in both the chiral 
and the meson-exchange models can be written in the form
\begin{equation}
\mu^2 + 2\mu \alpha - m_K^{*^2} = 0 \,.
\end{equation}
For the present, the only baryons we consider are nucleons, in which case the 
chiral model expressions for $\alpha$ and $m_K^{*^2}$ are
\begin{eqnarray}
\alpha &\pj=&\pj \frac{2n_p+n_n}{2f^2} \nonumber \\ 
m_K^{*2} &\pj=&\pj m_K^{2} + \left[ 2a_1n_p^s + (2a_2+4a_3)(n_p^s + 
n_n^s)\right]
\frac {m_s}{2f^2} \,, \label{chicrit}
\end{eqnarray}
respectively, in nucleons-only matter.  In the meson-exchange model, at 
zero temperature, we have
\begin{eqnarray}
\alpha &\pj=&\pj (G^\omega_{KN} - \thalf G^\rho_{KN}) n_n + 
(G^\omega_{KN} +  \thalf G^\rho_{KN}) n_p \nonumber \\
m_K^{*2} &\pj=&\pj m_K^{2} + G^\sigma_{KN} m_K \left[\frac{1}{g_{\sigma N}}
\frac{dU(\sigma)}{d\sigma}- n_n^s - n_p^s \right]\,, \label{mescrit} 
\end{eqnarray}
where we have used the definitions $G^i_{KN} = g_{iN}g_{iK}/m_i^2$, with $i =
\sigma,\ \omega$ and $\rho$.  If we specialize to isospin symmetric nuclear
matter with $\mu = \mu_n-\mu_p = \mu_e = 0$, Eqs.~(\ref{chicrit}) and
({\ref{mescrit}) take the form 
\begin{eqnarray}
m_K^2 f^2 &\pj=&\pj n^s \Sigma^{KN} \qquad {\rm and} \nonumber \\ 
m_K^2 &\pj=&\pj G^\sigma_{KN} \left[n^s - \frac{1}{g_{\sigma N}}
\frac{dU(\sigma)}{d\sigma}\right] \,,
\end{eqnarray}
respectively. These results give the critical density for
condensation in symmetric nuclear matter to be $u \geq 6.5$,  
well in excess of the
values of $u_{crit}$ shown in Table~13 for neutrino-free 
matter in beta equilibrium.  
Schaffner et al.~\cite{schaff} have recently emphasized this and shown that, 
depending on the chosen parameters, condensation may not occur at all in 
{\it nuclear} matter.  

The presence of leptons in stellar 
matter lowers the critical density for condensation by a significant amount. 
In Table~13, we list the critical density for kaon condensation at $T=0$   for
three choices  of the constant $a_3m_s$, and for matter without and with
hyperons.  (The latter case will be discussed in the next subsection.)  Results
shown include the neutrino-free and trapped neutrino cases.  In all cases shown
here, baryons are described using the GM model, but kaons are described using 
both the chiral and meson exchange models. 

Consider first the zero temperature case, in which neutrinos are absent.   The
critical densities show a marked reduction as the magnitude of  $a_3m_s$
increases, since this enhances the interactions; but they are not very
sensitive to the choice of the model used to describe kaonic interactions 
({\em i.e.}, chiral versus the meson exchange model) as long as compatability of
the kaon optical potentials is required.  The mean field model, GM, yields
somewhat lower values for the critical density than the MRHA models (see
Ref.~\cite{ekp}), due to the more rapid increase of the scalar fields that
enter the interaction terms of Eqs.~(\ref{chicrit}) and (\ref{mescrit}).

We remark that if we were to replace the scalar density in these calculations 
by the number density, the critical density ratio,  $u_{crit}$, for
condensation at $T=0$ drops by approximately 1 unit in the density ratio $u$.
(The effect appears to be of smaller magnitude in the recent work of  Maruyama
et al.~\cite{maruy} employing mean field theory.)

For neutrino-free matter, the effects of finite temperature on the onset of
kaon condensation are  shown in Fig.~18. (The meson exchange formalism is 
used for the finite temperature results in this subsection.)  The results 
shown are for the GM 
model and for three different values of $\Sigma^{KN}$.   The proton and
kaon concentrations in the top two panels, and the chemical potential
$\mu$  in the third panel, refer to the values at the critical density ratio
$u_{crit}$ shown in the bottom panel.  Thermal effects give rise to a 
non-negligible net negative kaon concentration, which  results in larger proton
concentrations than those for the zero entropy case.  This hinders the onset
of condensation.  However, the compensating  changes in the chemical potential
$\mu$ with temperature (in the range that supports an entropy per baryon up to
2)  result in only small net changes in  $u_{crit}$ for condensation.  Similar
results are obtained in the MRHA models.  Thus, the effects of condensation 
remain  significant, even at finite  entropy.                

In Fig.~19, the relative concentrations, the electron chemical potential, and
the hadronic and leptonic contributions to the entropy are shown for $S=1$ for
the case of neutrino-free matter. The results are for the GM model with
$a_3m_s= -222$ MeV.  At finite temperatures and at densities below that
for condensation, the kaon concentration remains smaller than that of the
other particles; consequently,  the changes induced in the total pressure and
energy density are small.  As in the case of zero temperature,  the kaon
concentration builds up rapidly for densities above the condensation density, 
which results in a softening of the equation of state. For an entropy per 
baryon up to 2, the relative concentrations of the various particles 
essentially retain the zero temperature behavior.  Thus, thermal effects do 
not change drastically the softening induced by the condensation.  The changes 
in the structure of the star, in particular, the changes in the maximum mass, 
are therefore at the few per cent level, as in the case of stars without
condensation.             
     
We turn now to the effects of trapped neutrinos.  The results for this case are
contained in Table 13 and Figs.~20 and 21.  Due to the behavior of
the chemical potential  (which is similar to that shown in Figs.~6 and 10), 
the critical density for kaon condensation is much higher when neutrinos are 
trapped than in the case of neutrinos having left the star. (Compare the
critical densities with the central densities in Table 9.)  Thus, in the trapped
case, the hadronic pressures are relatively larger for a substantial range of
densities than in the neutrino-free case. 

In  Table 14, we give the stellar properties for the neutrino-free and
neutrino-trapped cases.  Comparison of the neutrino-free results 
with Table 8 shows rather small effects, except for the largest magnitudes 
for the parameter $a_3m_s$. It is only for these cases that the central 
densities significantly exceed  the critical densities (Table 13) and allow a 
sizeable core of condensed kaons to appear. This also occurs in the MRHA 
models  for the larger magnitudes of $a_3m_s$. Turning to the trapped case, 
we see that the maximum mass is generally larger. This can be 
contrasted  with Tables 8 and 9, where $K^-$  particles were absent; 
and the maximum mass was a little less in the trapped case. This 
qualitative change engendered by kaons is similar to that previously noted 
for hyperons. \\

\subsubsection{Kaon condensation in strangeness-rich baryonic matter } 

We have affirmed the importance of hyperons in neutron stars, so it
is interesting to see how they affect the phenomenon of kaon 
condensation~\cite{ekp}. The necessary formalism has been outlined above. 
Since finite temperature effects are not too large, we 
focus on the zero temperature case. 
For the chiral model with $a_3m_s = -222$ MeV in conjunction with
the mean field GM description of the baryons, the results are shown in
Fig.~22.  The critical densities are given in Table 13.  
These critical densities are higher than those in the case in which 
hyperons were absent. The reason is clear. Once a significant
number of negatively charged hyperons are present (panel (a) of Fig.~20),   
the electron
chemical potential, $\mu$, begins to decrease with density (panel (b)). Since
kaon condensation occurs when this same chemical potential equals the energy of
the zero-momentum $K^-$ state, it is necessary to go to higher density where the
interactions are able reduce the energy further. For the MRHA, in almost all
cases, the critical density is beyond the central densities in Table 11,
 so that condensation does not occur. 
On the other hand, it can take place in  the mean
field model, GM, provided the magnitude of $a_3m_s$ is not too  small. This
follows from the fact that the scalar densities increase more rapidly in mean
field models than in the MRHA, which affects the interaction  terms $d_1$ and
$d_2$ in Eqs.~(\ref{tthresh1}).  Some insight into the role of the scalar
densities may be gained by examining the threshold condition in the chiral
approach when only $\Sigma^-$ and $\Lambda$ hyperons are present:        
\begin{eqnarray}
\mu^2 + \frac {(2n_p+n_n-n_{\Sigma^-})}{2f^2}  \mu - m_K^2
- \left [2a_1 n_p^s + (2a_2+4a_3) (n_p^s+n_n^s+n^s_{\Sigma^-}) \right. 
\nonumber \\
\hspace{2cm} \left. + \left( \fiveth (a_1+a_2)+4a_3 \right) n^s_\Lambda 
\right] \frac {m_s}{2f^2}= 0\;.
\label{tthresh3}
\end{eqnarray}
The first two terms in this equation are smaller than in 
the nucleons-only case, 
and this has to be compensated by the last term, which requires a higher
density.     

The presence of hyperons causes the condensate amplitude to increase 
rapidly with density (panel (d)); so rapidly, in fact, that large changes are
induced in the scalar densities (panel(c)) and the Dirac effective masses of
all the particles (panel(b)). The nucleon effective masses 
(see Eq. (\ref{temasses})) are given by 
\begin{eqnarray}
M_p^* &=& M - g_{\sigma n}\sigma + (2a_1+2a_2+4a_3) m_s \sin^2\!\thalf\theta
\;, \nonumber \\
M_n^* &=& M - g_{\sigma n}\sigma + (2a_2+4a_3) m_s \sin^2\!\thalf\theta\;,
\end{eqnarray}
and go to zero before the central stellar density is reached. This indicates
the need to consider improvements, in particular, an exact evaluation of the
zero-point energy with the non-linear Kaplan-Nelson Lagrangian, 
and this is under investigation.

Summarizing, we find that while the effect of non-zero temperature upon
the onset of kaon condensation is small, both the presence of hyperons and
neutrino trapping inhibit condensation, although the latter is a 
transient effect. It appears that in the MRHA, at least as formulated here,
kaons play a much smaller role than suggested by non-relativistic 
treatments~\cite{tpl}. Mean field approximations give a larger effect, but
we can not yet treat them satisfactorily when hyperons are present.

\subsubsection{Sensitivity of kaon condensation to hyperon couplings}

In the calculations above, we assumed that the couplings of the $\Sigma$ and 
$\Xi$ were equal to those of the $\Lambda$ hyperon. Here, we relax this 
assumption and explore the sensitivity to unequal couplings  of the different 
hyperons using the meson exchange formalism at zero temperature.
Of the many possibilities, we pick three for study. These are listed in
Table 15, in terms of the ratio to the nucleon couplings as defined earlier. 
For the $\Lambda$, we use the values discussed previously.
For the $\Sigma$, we use two sets of values which gave satisfactory fits to the 
$\Sigma^-$ atom  data in the work of Mare\v{s} et al. \cite{sigmacou}.
This was based on a mean field description of
nuclear matter using the nucleon couplings of Horowitz and Serot~\cite{hs}, 
who did not include non-linear terms ($U(\sigma)=0$). The parameters for this
model, termed `HS81' here, are:
\begin{equation}
\frac {g_{\sigma N}}{m_\sigma} = 3.974~{\rm fm} \,, \quad 
\frac {g_{\omega N}}{m_\omega} = 3.477~{\rm fm} \,, \quad {\rm and} \quad
\frac {g_{\rho N}}{m_\rho} = 2.069~{\rm fm} \,.
\end{equation}
Partly for consistency and partly because this 
model is often used as a baseline in the literature,
we will adopt these parameters. (Qualitatively similar results are obtained 
for  other values of the nucleon
couplings, which yield more realistic values of the compression modulus.)
Finally, we need the couplings of the $\Xi$. Since there is little 
information, we take the couplings to be equal to those of either the
$\Lambda$ or the $\Sigma$. Note that case 1 in this table is close to the set 
that we have been using in the previous discussion.

In Fig.~23, the particle fractions shown in the upper, center and lower 
panels refer, respectively, to  hyperon
coupling cases 1, 2 and 3 of Table 15.  The upper panel is
similar to results already  discussed; note that kaons do not condense up to
the maximum density displayed, $u=4.5$. In discussing the other cases, we first
mention the  seeming paradox that increasing the coupling constants of a
hyperon species delays its appearance to a higher density. The explanation 
\cite{glenhyp,kaphyp} is
that the threshold equation receives contributions from  the $\sigma,\ \omega$
and $\rho$ mesons, the net result being positive due to  the $\omega$. Thus, if
all the couplings are scaled up, the positive contribution becomes larger, and
the appearance of the particle is delayed to a higher  density. With this in
mind, consider the center panel of Fig.~23, which corresponds to case
2 of Table 15. The  $\Sigma$ couplings are larger than in 
case 1 (upper panel), so
the $\Sigma^-$ no  longer appears, thus allowing the chemical potential $\mu$
to continue   rising with density.    This allows the $\Xi^-$ to appear at
$u=2.2$,  essentially substituting for the $\Sigma^-$. Of course, were we to
reduce the  $\Xi$ couplings on the grounds that this hyperon contains two
strange quarks, the $\Xi^-$ would appear at an even lower density. Turning to
the lower panel of this figure, we recall that this  corresponds  to case 3 of
Table 15,  for which both the $\Sigma$ and $\Xi$ couplings are increased.
Neither of them now appear, and since the chemical potential, $\mu$, continues
to increase with density, it becomes favorable for kaons to condense at 
$u=3.6$; the fraction $Y_{K^-}$, however, remains rather small.

Clearly, the lesson to be drawn from this is that the thresholds for the
strange particles, hyperons and kaons, are sensitive to coupling  constants
that are poorly known. In matter where hyperons are allowed to be present, 
generally the effects of kaons are small. In fact, Schaffner and Mishustin 
\cite{schaffner} find that, with their choice of coupling constants, kaons do 
not condense. On the other hand, should it turn out that the coupling 
constants of the $\Sigma$ and $\Xi$ are larger than the ones adopted here 
as the standard choice, these hyperons might not be present at all, and 
consequently kaons would play a more important role. 
Thus, while strangeness plays a significant role in
determining the constitution and physical properties of a neutron star, the
detailed behavior cannot be tied down at the present time.

\subsubsection{Metastability of neutron stars with kaon condensates} 

Fig.~24 shows the window of metastability in the baryonic mass,
which, in the absence of mass accretion, is  unchanged during the evolution 
of the star. Here the baryons are nucleons described in the mean field GM model
without and with kaons, for which we use the chiral model with $a_3m_s = -222$ 
MeV. When kaons are present, the range  $M_B = 2.09 - 2.15 M_\odot$ can
be supported by the initial EOS of  lepton-rich matter, but not by the later
EOS of lepton-poor matter (lines ending in dots). This range of metastability 
corresponds to gravitational masses of $M_G = 1.81 - 1.91 M_\odot$.  In the 
absence of kaons, metastability does not occur, since the maximum mass 
decreases when the neutrinos leave (lines ending in stars). Note that the 
qualitative features here are similar to the case of matter with 
strangeness-bearing hyperons, see Fig.~15.

Fig.~25 shows the corresponding evolution of the maximum gravitational mass in
matter with and without kaons  in the deleptonization stage. As we have 
previously remarked,  the initial state with $Y_{L_e}=0.4$ corresponds to 
an electron-neutrino fraction of 
$Y_{\nu_e} \simeq 0.08$ and, of course, in the final state $Y_{\nu_e}=0$. 
As in Fig.~16, the decrease (increase) in the maximum mass when kaons are 
present (absent) is most pronounced shortly before the neutrino fraction
drops to zero. \\
        
\subsection{Matter with quarks}
\setcounter{subsubsection}{0}

We now examine another type of softening of high density matter by
allowing for a hadron to quark phase transition in the interior of the
star~\cite{pcl}.  
Glendenning~\cite{gle1} has suggested that a mixed phase of baryon and quark 
matter exists over a wide range of densities in the case of neutrino-free 
matter. Our primary interest here is in  protoneutron stars and the effects of 
trapped neutrinos~\cite{pcl}, which, conceivably, could lead to  observable
consequences. We focus on zero temperature, since, as we have seen, changes in
the maximum mass due to neutrino trapping are larger than those due to finite
temperature. The influence of complicated finite size structures due to Coulomb
and surface effects~\cite{hps} does not qualitatively affect our conclusions
and will be taken up elsewhere. 

We shall follow fairly closely the treatment of Glendenning~\cite{gle1}. 
Thus, for the pure phase in which the strongly interacting particles 
are baryons, 
we employ the mean field GM model, for which the formalism has been 
discussed in earlier sections.   For the pure quark phase (in a
uniform background of leptons), we use the bag model for which the pressure 
is
\begin{eqnarray}
P_Q = -B + 
\frac 13 \sum_{f=u,d,s} g_f \int\limits_0^{k_{Ff}} \frac {d^3k}{(2\pi)^3}\, 
\frac {k^2}{(k^2+m_f^2)^{1/2}}  + 
\frac 13 \sum_\ell g_\ell \int\limits_0^{k_{F\ell}} \frac {d^3k}{(2\pi)^3}\, 
\frac {k^2}{(k^2+m_\ell^2)^{1/2}}  \,.
\end{eqnarray} 
The first
term accounts for the cavity pressure, and the second and third terms give the
Fermi degeneracy pressures of quarks and leptons, respectively.   The constant
$B$ has a simple interpretation as the thermodynamic potential of the vacuum,
and will be regarded as a phenomenological parameter in the range
$(100-250)~{\rm MeV~fm}^{-3}$.  The lower limit here is dictated by the
requirement that, at low density, hadronic matter is the preferred phase.  For
$B$ much larger than the upper limit, a transition to matter with quarks 
never occurs.  The degeneracy factor for quarks is $g_f = 2\times 3$, accounting
for the spin and color degrees of freedom.  
The chemical potential of  free quarks in the
cavity is $\mu_f = \sqrt{k_{Ff}^2 + m_f^2}$,  where $k_{Ff}$ is the Fermi
momentum of quarks of flavor $f$. For numerical calculations, we take the $u$
and $d$ quarks as massless, and $m_s=150$ MeV. The baryon density and the energy
density are 
\begin{eqnarray}
n_Q &\pj=&\pj \oneth \sum_{f=u,d,s} n_f\,,  \qquad 
n_f = \frac {k_{Ff}^3}{3\pi^2} \nonumber \\
\varepsilon_Q &\pj=&\pj -P_Q+\sum_f n_f\mu_f + \sum_\ell n_\ell\mu_\ell \,.  
\end{eqnarray}
The relevant weak decay processes in the pure quark phase are similar to 
Eq.~(\ref{bproc}), but with $B_i$ replaced by
$q_f$, where $f$ runs over the quark flavors $u~,d$, and $s$.  In neutrino-free
matter, charge neutrality and chemical equilibrium under the weak processes 
imply 
\begin{eqnarray} 
&&\fpj \sum_f q_f n_f + \sum_{\ell=e,\mu} q_{\ell} n_{\ell} = 0
\label{charge2} \\
&&\fpj \mu_d = \mu_u + \mu_\ell = \mu_s\, . 
\label{beta2}
\end{eqnarray}
When neutrinos are trapped, the new chemical equilibrium relation is
obtained by the replacement $\mu_\ell \rightarrow  \mu_\ell - \mu_{\nu_\ell}$
in Eq.~(\ref{beta2}).  

In the mixed phase of hadrons and quarks, it is necessary to
satisfy Gibbs' phase rules: 
\begin{eqnarray}
P_H=P_Q \qquad {\rm and} \qquad \mu_n=\mu_u+2\mu_d \,.  
\end{eqnarray}
Further, following
Glendenning~\cite{gle1}, we require {\em global}, but not local, 
 charge neutrality of
bulk matter, for both separately conserved charges: baryon number and
electric charge. Denoting by $f$ the fraction of volume occupied by the
hadronic phase, we have  
\begin{eqnarray}
&&\fpj f\sum_Bq_Bn_B + (1-f)\sum_{f=u,d,s}q_fn_f +\sum_{\ell=e,\mu}q_{\ell}
n_{\ell}= 0 \label{gcharge} \\
&&\fpj n = f \sum_Bn_{B} + (1-f) n_Q\,,  
\end{eqnarray}
where $q_B$ is the electric charge of each hadron. 
An important consequence of global neutrality is that baryonic
and quark matter coexist for a much larger range of pressures than for the case
of local charge neutrality~\cite{gle1}. 
The total energy density is $\varepsilon =f\varepsilon_H + (1-f)
\varepsilon_Q$. \\

\subsubsection{Metastability of neutron stars with quarks}

Fig.~26 shows a comparison of the compositions  of neutrino-free matter (top
panel) and neutrino-trapped matter (bottom panel).  
In the case of neutrino-free matter, quarks make their appearance at 
around $4n_0$ for $B=200~{\rm MeV~fm^{-3}}$. After this, the neutral and 
negative particle
abundances begin to fall, since quarks furnish both negative charge and baryon
number.  The bottom panel of Fig.~26 shows the influence of trapped neutrinos 
(with $Y_{Le}=0.4$) on the relative fractions.  The primary role of trapped
neutrinos is to increase the proton and electron abundances, which strongly
influences the threshold for the appearance of hyperons.  The $\Lambda$ and the
$\Sigma$'s  now appear at densities higher than those found in the absence of
neutrinos.  In addition, the transition to a mixed phase with quarks is delayed
to about  $10n_0$. Qualitatively similar trends are observed for other values
of the bag constant $B$.     

In Fig.~27, we show the phase boundaries as a function of the bag pressure $B$. 
The onset of the phase transition is at density $n_1=u_1n_0$, and  a pure quark 
phase begins at density $n_2=u_2n_0$.   Also shown are the central densities
$n_c=u_cn_0$ of the maximum mass stars.   Trapping shifts the onset of the
phase transition to higher baryon densities  and also reduces the extent of 
the mixed phase in comparison to the case of neutrino-free matter.  
The existence of a mixed phase inside the star depends on whether or not
hyperons are present.  In the absence of hyperons (top panel),
a mixed phase is present for the entire range of bag constants. 
When hyperons are
present (bottom panel), the mixed phase is present only when $B$ 
is sufficiently low ($B \leq 165~{\rm MeV~fm}^{-3})$ and occurs over
a smaller range in density than that found in the absence of hyperons.   The
abrupt change in the onset of the transition around  $B=140~{\rm MeV~fm}^{-3}$
is caused by the appearance of hyperons prior to that of quarks.  

The dashed lines in these figures correspond to the case in which neutrinos
have left the star.  Whether or not hyperons are present,  the mixed phase is
now present over a wide range of density inside the star.   Note also that,
since the central density of the star $u_c < u_2$ for all cases considered, the
presence of a pure quark phase is precluded.  Finally, note that  quarks are
more likely to appear the fewer the neutrinos remaining in the star. 

Table~16 shows the maximum masses of stars  as a function of the composition of
the matter.    With only nucleons and leptons (last row),  neutrino trapping
generally {\em reduces} the maximum mass from the case of neutrino-free 
matter.  This is caused by the smaller pressure support of lepton-rich matter, 
in which  the gain in the negative  symmetry pressure exceeds the increase in
leptonic pressure.  However, the introduction of quarks, which soften the EOS,
causes the maximum mass for the trapped case to be {\em larger} than that for
neutrino-free matter.  This reversal in behavior is due to the fact that  the
first appearance of quarks occurs at a higher density when neutrinos are
trapped. When hyperons are present, the maximum mass remains larger for  the
trapped case. This is also due to the appearance of hyperons at a higher
density when neutrinos are trapped. It is an  example of the general result
that when matter contains non-leptonic negative charges, the maximum mass of
the neutrino-trapped star is larger than that of the neutrino-free star.  
This result has important ramifications for the evolution of proto-neutron
stars and for the formation of black holes, as we discuss in Sec. 6. \\

\subsection{Global energetics} 
\setcounter{subsubsection}{0}

While the softening of the EOS due to the presence of negatively-charged
particles has a large effect upon the maximum mass, it has surprisingly little
effect upon the binding energy versus mass relationship for neutron stars. The
binding energy is the difference between baryonic and gravitational masses of
the final neutron star configuration.  It is an important observational
parameter, because at least 99\% of it appears as radiated neutrino
energy. In Fig.~28, we display the binding energy as a function of the baryonic
mass $M_B$ for stars with and without the presence of strangeness-bearing
components, such as kaons or hyperons.  The various curves refer to different
EOSs and terminate at the maximum mass supported by their respective EOS.   \\
                                
A few striking results are evident from this figure. 
\begin{enumerate}
\item  The largest binding energy occurs for the EOS that supports the largest 
maximum mass. 
\item For each EOS, the binding energy displays a nearly quadratic behavior up 
to the maximum mass. 
\item There exists a rather narrow band of possible binding energies for a
given mass, implying the following universal relationship for the binding 
energy as a function of mass:   
\begin{eqnarray}
B.E. = (M_B - M_G)c^2 \cong (0.065\pm 0.01)
\left({M_B\over M_\odot}\right)^2M_\odot \,,
\label{bind}
\end{eqnarray}
where the numerical coefficient represents an update of the value quoted
earlier by Lattimer and Yahil~\cite{ly}.  The universality is not altered
by the presence of significant softening in the high density EOS due to the 
appearance of quarks, kaons, or hyperons. 
\item Only near the terminations at the maximum masses do the
binding energies slightly deviate from the lower envelope of the curves.  This
effect is slightly more pronounced for  softer equations of state.    
\end{enumerate}

We have found that the lower envelope of the binding energy--mass relation is
equivalent to that found for the stiffest plausible equation of state, namely
one that is limited by causality.  Denoting by $n_t$ a transition density above
which the EOS is assumed to be causal, the EOS above $n_t$ is given
by~\cite{lpmy}:
\begin{eqnarray}
P &=& {1\over2}
\left[P_t-\epsilon_t+(P_t+\epsilon_t)\left({n\over n_t}\right)^2\right]\,; 
\nonumber \\ 
\epsilon &=& \epsilon_t + P - P_t\,, 
\end{eqnarray}
where the quantities $P_t$ and $\epsilon_t$ are the pressure and energy density
at $n_t$ and thus depend both upon it and upon the equation
of state employed.  However, if $n_t$ is in the range $n_0-2n_0$, both $P_t$
and $\epsilon_t$ are somewhat insensitive to the EOS, and the binding
energy is relatively
insensitive to these quantities.  Fig.~28 includes the binding energy as a
function of baryon mass for an EOS which is causal above $n_t=0.3$
fm$^{-3}$ and matched to the GM 
equation of state below $n_t$.  
Varying the values of $n_t$ and the
EOS below $n_t$ merely alters the termination point (maximum mass) of
this curve, without otherwise noticeably changing it. 

We note that an analytic representation of the binding energy--mass relation
can be determined using the techniques developed by Nauenberg and
Chapline~\cite{nc}, who assumed that the pressure and energy density are
constant within the neutron star as an alternative to an explicit integration
of the Tolman-Oppenheimer-Volkov equation \cite{tov}.  
Defining the parametric variable $\chi$ by 
\begin{eqnarray}
\sin^2\chi={2GM\over Rc^2}\,,
\end{eqnarray}
the mass is  
\begin{eqnarray}
M={\sqrt{3c^6(1-\xi)\over 32\pi G^3(\epsilon_t-P_t)}} \sin^3\chi \,; \quad 
\xi = {6\cos\chi\over 9\cos\chi-Q} -1 \,,
\end{eqnarray}
where $Q = 2\sin^3\chi/(\chi-\sin\chi\cos\chi)$.  
The binding energy can then be written as
\begin{eqnarray}
B.E.=M\left[{{3n_tm_Bc^2\over\sqrt{\epsilon_t^2-P_t^2}}}
{{\sqrt {1-\xi^2}}\over Q} - 1\right]\,.
\end{eqnarray}
Utilizing $Q\cong3(1-3\chi^2/10+\cdots )$ and $\xi \cong \chi^2/10+\cdots$,
as appropriate for low mass stars, one recovers the Newtonian 
result that $B.E. = (3/5)GM^2/Rc^2 + \cdots$, which displays the quadratic
dependence on the mass of the star.                      

Since nearly all of the binding energy is released in the form of neutrinos,  
it appears that an accurate measurement of the total radiated neutrino energy
will lead to a good estimate of the remnant mass.  However, as the results of
Fig.~28 and Eq.~(\ref{bind}) show, it will not be possible to distinguish the
various equations of state from the total binding energy alone.

\section{Evolution timescales } 
\setcounter{section}{5}
\setcounter{subsection}{0}

We have seen that the structure of a neutron star is strongly influenced  by
the presence of trapped neutrinos and, to a lesser extent, by the non-zero
entropy/baryon. The evolution of the star will be governed by the timescale
for release of the trapped neutrinos, referred to as deleptonization, and 
the timescale for thermal cooling, which reduces the entropy to a small value.
We discussed these timescales in Sec. 1 on the basis of detailed numerical 
calculations of the evolution dynamics~\cite{birth}. Since the timescales are of
some  significance, we will show in this section how these can be estimated
from analytical considerations and illustrate the dependence on the equation of
state and the opacities. 

The equations that describe the physical state and evolution of
the nascent neutron star are simply those of standard stellar structure theory,
modified for the effects of general relativity and augmented to include lepton
and neutrino transport. (Photon transport is completely suppressed at the high
densities of the core.)  All the known neutrino species and their antiparticles
carry energy.  Most of the time, in most of the star, the neutrinos
are in thermal equilibrium with the matter and have Fermi distributions.  This
is true because neutrino processes, such as the nucleon Urca, pair production,
and inverse nucleon brehmsstrahlung, are sufficiently rapid.  The calculations
of Maxwell~\cite{max}, for example, illustrate that the timescale for $\nu_e$
($\nu_\mu$) equilibration will be less than 1 s, at nuclear densities, for
temperatures $T\geq2 (3)$ MeV.  This is an overestimate for the equilibration
time for $\nu_e$, since Maxwell did not consider the direct Urca process.  If
that process is included, the timescale decreases by an order of magnitude.

It is a good approximation to lump together
$\nu_\mu$ and $\nu_\tau$ transport together into ``$\mu$'' transport, and
thereby assume them to have zero chemical potential and equal opacities.
Further simplification can be made if one assumes the electron neutrino and
antineutrino opacities are also equal.  Then we need only define one $e$-type
chemical potential, denoted by $\mu_{\nu}$.  This is exact deep in the opaque 
interior, but breaks down
in the transparent regime above the neutrinosphere.  It can be joined smoothly
to a free-streaming approximation here, however.  The relevant equations,
in spherical symmetry, have been given in Ref.~\cite{birth}, to which we 
refer the reader for more elaborate discussion. They are
\begin{eqnarray} 
{dP\over dr} &=& -{G(M+4\pi r^3P)(\rho+P/c^2)\over r(r-2GM/c^2)} 
\label{jim98}\\
{dM\over dr} &=& 4\pi r^2\rho \label{jim99}\\
{dN\over dr} &=& {4\pi r^2n\over\sqrt{1-2GM/rc^2}} \label{jim100}\\
{dY_\nu\over d\tau} &=& -e^{-\phi}{\partial (4\pi r^2F_\nu e^\phi)\over
\partial N} + S_\nu 
\label{ynudot}      \\ 
{dY_e\over d\tau} &=& -S_\nu 
\label{yedot}                \\ 
{dU\over d\tau} &=& -P{d(1/n)\over d\tau }-e^{-2\phi}{\partial L_\nu
e^{2\phi}\over\partial N} \,.
\label{basic}
\end{eqnarray} 
Here Eq. (\ref{jim98}) is the general relativistic equation for hydrostatic 
equilibrium, in which $M(r)$ is the enclosed gravitational mass. The enclosed 
baryon mass, $N(r)$, obeys Eq. (\ref{jim100}). Eqs. (\ref{ynudot}) and
(\ref{yedot}) give the rate of change of the electron neutrino and 
electron concentrations, with $F_\nu$ the number flux of electron neutrinos and 
$S_{\nu}$ the electron neutrino source term. Finally, Eq. (\ref{basic})
gives the rate of change of $U$, the internal energy per baryon, where
$L_\nu$ is the total neutrino luminosity (including all species).  
The term $e^\phi=\sqrt{-g_{00}}$
relates time at infinity $\tau $ with the coordinate time $t$, and one can 
show~\cite{birth} that $d\phi/dP= -(P+\rho c^2)^{-1}$.

In the diffusion approximation, fluxes are driven by density gradients.  In 
our context, this translates into expressions of the form
\begin{eqnarray} 
F_\nu &=& -\int_0^\infty{c\lambda_\nu\over3}
{\partial n_\nu(E_\nu)\over
\partial r}dE_\nu\,;  \label{fflux}\\ 
L_\nu &=& -\int_0^\infty4\pi r^2\sum_i{c\lambda_E^i\over3} 
{\partial \epsilon_i(E_\nu)\over \partial r}dE_\nu\,, 
\label{flux}
\end{eqnarray}
where the sum is over neutrino species.  The $\lambda_\nu$ and $\lambda^i_E$'s
are mean free paths for number and energy transport, respectively, and are
functions of neutrino energy $E_\nu$.  Also, $n_\nu(E_\nu)$ is the number of
electron neutrinos with energy $E_\nu$, and $\epsilon_i(E_\nu)$
is the energy density of species $i=e,\mu$.
The general relativistic corrections have been dropped for
clarity, although they are straightforward to incorporate.

We can combine Eqs.~(\ref{ynudot}) and (\ref{yedot}) to obtain the rate of
change of the total lepton number, and Eq.~(\ref{basic}) and the first law of
thermodynamics to obtain the rate of change of the entropy:
\begin{eqnarray}
n{dY_L\over dt} &=& n\left({dY_e\over dt}+{dY_\nu\over dt}\right)= - {1\over
r^2}{\partial\over\partial r}r^2F_\nu \\ 
nT{ds\over dt}&=& -{1\over4\pi r^2}{\partial L_\nu\over\partial r}-n
\sum_{i=n,p,e,\nu}\mu_i{dY_i\over dt}\,. 
\label{rates}
\end{eqnarray} 

To proceed, we have to understand the energy dependence of the opacities.
There are two main sources of opacity:

\begin{enumerate}
\item Neutrino-nucleon absorption.  This affects $\nu_e$ and $\bar\nu_e$ only,
except at very high densities if muons are present, which is not possible
until relatively late in the cooling.  The absorption mean free path, assuming
nondegenerate nucleons, is
\begin{eqnarray}
\lambda_{abs} = \lambda_{abs}^o\left({E_{\nu o} \over E_\nu}\right)^2 {\rm~cm}
\,, \label{labs} 
\end{eqnarray}
where $\lambda_{abs}^o$ is the fiducial absorption mean free path at the 
fiducial $\nu_e$ energy $E_{\nu o}\simeq 260$ MeV, the typical $\nu_e$ chemical
potential at the beginning of deleptonization.  From Ref.~\cite{iwamo}, we find
\begin{eqnarray}
\lambda_{abs}^o={4\over n_n\sigma_o(1+3g_A^2)}
\left({m_ec^2\over E_{\nu o}}\right)^2 \,,
\end{eqnarray}
where $n_n$ is the neutron number density, $\sigma_o=1.76\times10^{-44}$
cm$^{2}$, and $g_A\simeq1.257$.  
Using $n_n=(8/3)n_0$, 
appropriate for
$Y_e=1/3$ and a total baryon density $n=4n_0$ at the beginning of
deleptonization, we find $\lambda_{abs}^o\simeq0.36$ cm.  Since nucleons will,
in fact, be at least partially degenerate, and because of fermi liquid effects,
the true absorption mean free path will be about 3--10 times larger than this
value.
\item Neutrino-nucleon scattering.  This elastic
scattering affects all $\nu$-types.  For nondegenerate nucleons~\cite{iwamo}, 
\begin{eqnarray}
\lambda_{e s}^o={4\over n\sigma_o}
\left({m_ec^2\over E_{\nu o}}\right)^2\,,\\
\lambda_{\mu s}^o={4\over n\sigma_o}
\left({m_ec^2\over E_{\nu_\mu o}}\right)^2\,,
\end{eqnarray}
for $\nu_e$ and $\nu_\mu$, respectively.  For our reference density, $4n_0$, we
obtain $\lambda_{e s}^o\simeq1.37$ cm for $E_{\nu o}=260$ MeV, and
$\lambda_{\mu s}^o\simeq1.75$ cm for $E_{\nu_\mu o}=230$ MeV; the latter is the
appropriate value for the mean $\nu_\mu$ energy, with $s=2$, at the beginning of
the cooling era.  Corrections to the scattering mean free paths for degeneracy 
and interactions should be similar to those for absorption.
\end{enumerate}
Thus, during deleptonization, $\lambda_{abs}^o < \lambda_s^o$, and $\nu_e$
absorption dominates both energy and lepton number transport.  However, during
thermal cooling, energy transport is effected mostly by $\mu-$ and $\tau-$
neutrinos, since $\nu_e$'s are more tightly coupled to the matter.

The opacities imply three kinds of fluxes:
\begin{enumerate}
\item a number flux $F_\nu$ of $\nu_e$'s, dominated by absorption.  Utilizing 
Eq.~(\ref{labs}), we have
\begin{eqnarray}
- F_\nu=\int_0^\infty{c\lambda_{abs}\over3}
{\partial n_\nu(E_\nu)\over\partial r}
dE_\nu={c\lambda_{abs}^oE_{\nu o}^2\over6\pi^2(\hbar c)^3}{\partial
\mu_\nu\over\partial r}\equiv a{\partial\mu_\nu\over\partial r}.
\label{fnu}
\end{eqnarray} 
\item an energy flux $L^e_\nu$ of $\nu_e$'s, also dominated by absorption:
\begin{eqnarray}
- L^e_\nu=\int_0^\infty4\pi r^2{c\lambda_{abs}\over3} 
{\partial\epsilon_\nu(E_\nu) \over\partial r}dE_\nu = 
4\pi r^2 a{\partial\over\partial r}\left({\pi^2T^2\over6}+{\mu_\nu^2 \over 2}
\right)\,.
\label{le}
\end{eqnarray} 
\item an energy flux $L^\mu_\nu$ of $\nu_\mu$'s and $\nu_\tau$'s, dominated by
scattering:
\begin{eqnarray}
- L^\mu_\nu=2\int_0^\infty4\pi r^2{c\lambda_{\mu
s}^o\over3}{\partial\epsilon_{\nu_\mu}(E_\nu)\over\partial r}dE_\nu &=&
4 \pi r^2{c\lambda_{\mu s}^o(E_{\nu_\mu o})^2\over3\pi^2(\hbar c)^3}
{\partial\over\partial r}\left({\pi^2 T^2\over6}\right) \nonumber \\
&\equiv & 4\pi r^2 b{\partial\over\partial r}\left({\pi^2 T^2\over6}\right)\,,
\label{lmu}
\end{eqnarray} 
where we used the fact that $\nu_\mu$'s and $\nu_\tau$'s have zero chemical 
potential.
\end{enumerate}

\subsection{Deleptonization Era}
\setcounter{subsubsection}{0}

We can now appreciate the separate stages of deleptonization and cooling.
Deleptonization is dominated by number transport, and the controlling
equation is
\begin{eqnarray}
n{dY_L\over dt}=n{\partial Y_L\over\partial Y_\nu}{dY_\nu\over dt}
={a\over r^2}{\partial\over\partial r}
\left(r^2{\partial\mu_\nu\over\partial r}\right)\,.
\label{ndldt}
\end{eqnarray}
During deleptonization, the electron neutrinos are degenerate, and to lowest
order $nY_\nu\simeq(\mu_\nu/\hbar c)^3/6\pi^2$.  The neutrinos and electrons are
in beta equilibrium, and it can be shown that $\partial Y_L/\partial Y_\nu=
(\partial Y_L/\partial Y_\nu)_o (E_{\nu o}/\mu_\nu)$, 
where $(\partial
Y_L/\partial Y_\nu)_o\simeq3$ for $Y_{\nu o}\simeq 0.06$.  Thus,
\begin{eqnarray}
3E_{\nu o} \left(\partial Y_L\over\partial Y_\nu\right)_{\!o}
\mu_\nu{\partial\mu_\nu\over\partial t}={a\over r^2}{\partial
\over\partial r}\left(r^2{\partial\mu_\nu\over\partial r}\right)\,.
\label{dyldt2}
\end{eqnarray} 
We now seek separable solutions of the form $\mu_\nu=E_{\nu o}\phi(t)\psi(r)$.
We obtain
\begin{eqnarray}
3 \left(\partial Y_L\over\partial Y_\nu\right)_{\!o} {E_{\nu o}^2\over a}
{\partial\phi\over\partial t}={1\over r^2\psi^2}
{\partial\over\partial r}\left(r^2{\partial\psi\over\partial r}\right)
=-\alpha\,,
\label{dyldt3}
\end{eqnarray}
where $\alpha$ is a separation constant.  One sees that
\begin{eqnarray}
\phi=1-t/\tau_d\ \ ;\qquad\qquad \tau_d={3\over c\lambda^o_{abs}\alpha}
\left({\partial Y_L\over\partial Y_\nu}\right)_{\!o}\,,
\label{phidep}
\end{eqnarray} 
where $\tau_d$ is the diffusion time.  Note that the decay of the neutrino
chemical potential is approximately linear with time.  This result is borne out
in more detailed numerical calculations.  The radial equation is
\begin{eqnarray}
-\alpha R^2\psi^2={1\over x^2}{\partial\over\partial x}
\left(x^2{\partial\psi\over\partial x}\right)\,,
\label{raddep}
\end{eqnarray} 
which is just that of the Lane-Emden polytrope~\cite{chandra} of index 2.  
In the above, $R$ is the radius of the star.  The solution of
this equation satisfies $\alpha R^2\simeq19$; and, therefore,
\begin{eqnarray}
\tau_d\simeq {9R^2\over19c\lambda^o_{abs}}\simeq 44.3 
\left({R\over10{\rm~km}}\right)^2{\rm~s}.
\label{raddep1}
\end{eqnarray}
Recalling that degeneracy and fermi liquid corrections will increase the mean
free path by a factor of 3--10, a deleptonization time of 5--15 s is indicated.
This is the correct magnitude for the deleptonization time, and shows
clearly how it depends upon the equation of state through the radius of the
star, $R$, and upon the opacity.

Because of the positive temperature gradient and the chemical potential
gradient, the deleptonization is accompanied by heating in the core.  The
entropy/baryon  rises to the value of about 2 before it decreases during the 
cooling era.  The onset of cooling does not begin until deleptonization is 
complete. Rewriting the last term of Eq.~(\ref{rates}) as
\begin{eqnarray}
-n\sum_{n,p,e,\nu}\mu_i{dY_i\over dt}=n(\mu_n-\mu_p-\mu_e+\mu_\nu){dY_e\over dt}
-\mu_\nu{dY_L\over dt}\,, 
\end{eqnarray}
we can combine Eqs. (\ref{basic}) and (\ref{fflux}) to find 
\begin{eqnarray}
nT{ds\over dt}={a+b\over r^2}{\partial\over\partial r}
\left(r^2{\partial(\pi^2T^2/6)\over\partial r}\right) + 
a\left({\partial\mu_\nu\over\partial r}\right)^2\,.
\label{fullcool}
\end{eqnarray} 
The terms proportional to $a$ are due to electron neutrinos, and 
the term proportional to $b$ is due to the other neutrinos.  There
is heating or cooling depending on the direction of the temperature gradient,
but the chemical potential gradients always lead to heating.  When
$\mu_\nu\gg T$, the $(\partial\mu_\nu/\partial r)^2$ term dominates, and we have
heating.  When $\mu_\nu\simeq0$, and the temperature decreases with radius,
cooling occurs.
                                                             
\subsection{Thermal Cooling Era}
\setcounter{subsubsection}{0}

We now turn to the thermal cooling of the protoneutron star, which continues
beyond the deleptonization era.  While the initial entropy
per baryon $s$ in the star's interior is about 1, after the deleptonization
heating is
finished the entropy reaches the value of about 2.  
The entropy is dominated by baryons for temperatures less than about 100 MeV.
Thus, we may write~\cite{pabw}
\begin{eqnarray}
s\approx2a_{\ell d}T\ \ ;\qquad a_{\ell d}=
{1\over15}{m^*\over m}\left({n_0\over n}\right)^{2/3}
{\rm~MeV}^{-1}\,, \label{entro}
\end{eqnarray} 
where $m^*$ is the effective nucleon mass. 
Note that the maximum value of
the central temperature is
\begin{eqnarray}
T_{max}={s_{max}\over2a_{\ell d}}\simeq37.8s_{max}\left({m\over2m^*}\right)
\left({n\over4n_0}\right)^{2/3}{\rm~MeV}.\label{tmax}
\end{eqnarray}                                                   
We henceforth neglect the density dependence of $m^*$ and use $m^*\simeq0.5m$.
Notice that the estimate of $T_{max}$ in Eq. (\ref{tmax}) agrees quite well 
with our previously tabulated results.

The cooling is dominated by the $\mu$- and $\tau$-neutrinos, since $b>a$.  With
these simplifications, Eq.~(\ref{fullcool}) becomes
\begin{eqnarray}
{6a_{\ell d}n\over\pi^2(a+b)}{\partial T^2\over \partial t} = {1\over r^2}
{\partial\over\partial r}\left(r^2{\partial T^2\over\partial r}\right)\,.
\end{eqnarray}  
Once again, we may separate the resulting equation in terms of the time and
radial dependence of the temperature.  Writing $T=T_{max}\psi(r)\phi(t)$, we
find
\begin{eqnarray}
{12a_{\ell d}n\over\pi^2(a+b)}{1\over\phi}{\partial\phi\over\partial t} =
{1\over\psi^2 r^2}{\partial\over\partial r}
\left(r^2{\partial\psi^2\over\partial r}\right) = -\alpha\,,
\end{eqnarray}
where $\alpha$ is a separation constant.  The solution of the radial equation
is that of an $n=1$ Lane-Emden polytrope, for which the eigenvalue is
$\alpha=\pi^2/R^2$.  The temporal equation has the solution
\begin{eqnarray}
\phi=\exp \left(-(t-\tau_d)/\tau_c\right)
\label{simple1}
\end{eqnarray}
for $t>\tau_d$, where
\begin{eqnarray}
\tau_c={12a_{\ell d}nR^2\over\pi^4(a+b)} 
= {3s_{max}(E_{\nu o}/T_{max})R^2 \over Y_{\nu o}
\pi^4 c(\lambda_{abs}^o/2+\lambda_{\mu s}^o(E_{\nu_\mu o}/E_{\nu o})^2)}\,\,.
\end{eqnarray}
Note that
\begin{eqnarray}
{\tau_c\over\tau_d}={19s_{max}\over 3Y_{\nu o}\pi^4}
{E_{\nu o}\over T_{max}}{1\over 1/2+(\lambda_{\mu
s}^o/\lambda_{abs}^o)(E_{\nu_\mu o}/E_{\nu_ o})^2}\simeq 1.7\,.
\label{taurat}
\end{eqnarray}
This result is independent of our assumptions regarding $s_{max}$ or $T_{max}$.
 This is also of the right magnitude to match the numerical calculations, which
indicate that $\tau_c=(1-2)\tau_d$.
Thus, in spite of the fact that the mean free paths that dominate cooling
are larger than those that dominate deleptonization, the large ratio of the
matter's heat capacity to that of the neutrinos forces the cooling time to be
longer than the deleptonization time.

At late times, however, the decay of the central entropy or temperature is
roughly linear with time~\cite{birth}.  This feature can be seen to be a result
of the increasing degeneracy of the star.  For degenerate nucleons, the mean
free paths have an $E_\nu^{-3}$ dependence.  Ignoring the absorption
contributions, this energy dependence leads to a linear time decay, with a time
constant
\begin{eqnarray}
\tau_c^\prime={s_{max} R^2\over 76(\ln 2) Y_{\nu o}c\lambda_s^\prime}
\left({E_{\nu o}\over E_{\nu_\mu o}}\right)^3 \,,
\end{eqnarray}
where the fiducial scattering mean free path for degenerate nucleons
is~\cite{iwamo}
\begin{eqnarray}
\lambda_s^{\prime}=\lambda_{\mu s}^o{5\over 1+4g_A^2}\left({p_{F_n}\over 
E_{\nu_\mu o}}\right)\simeq 1.6 \lambda_{\mu s}^o \,.
\end{eqnarray}
For the values we have been assuming, one finds that $\tau_c^\prime\simeq 11$
s. Of course, since this result is applicable to the later stages of cooling,
the use of a smaller value of $E_{\nu_\mu o}$ may be appropriate, which will
lead to an increase in this timescale estimate.   In any case, one expects the
experimental behavior of Eq.~(\ref{taurat}) to alter to a linear decay after an
$e$-folding time.   

It is finally interesting to note the overall sensitivity
to the equation of state.  For a fixed mass star, employing $R \propto
n_c^{-1/3}$, where $n_c$ is the central density, we determine that $\tau_d
\propto n_c^{1/3}$ and $\tau_c \propto n_c^0$ so that for nucleons-only matter,
the central density affects the timescales rather weakly.  However, larger
changes in the opacities and radii are trigerred by the appearance of
negatively charged strongly interacting particles in any form~\cite{rp},
suggesting that their onset may significantly affect these timescales.

\section{Implications}                 
\setcounter{section}{6}
\setcounter{subsection}{0}

Our main thrust in this work has been to elucidate how the structure of a
proto-neutron star depends on its composition, which is chiefly determined
through the nature of the strong interactions at high baryon density.   During
its early evolution, a neutron star with an entropy per baryon of order unity 
contains neutrinos that are trapped in matter on dynamical timescales.  After a
time of a few tens of seconds, the star achieves its cold, catalyzed structure
with essentially zero temperature and no trapped neutrinos.  The influence of
finite temperature on the star's structure is dominated by the behavior of the
baryonic thermal pressures, which are governed  by the behavior of the baryonic
effective masses.  Baryonic thermal pressures are proportional to their Landau
effective masses, so that nuclear models that lead to  extremely small
effective masses at high density, such as Skyrme-type interactions, will
generally show substantially larger effects at finite temperature than other
models.  We have shown, however, that the gross reduction of the effective mass
in the Skyrme case leads to acausal sound speeds in dense matter and should be
discounted.  Thus, finite temperature effects upon the maximum neutron star
mass are naturally limited.                                
          
In general, however, changes in the maximum mass due to neutrino trapping are
larger than those due to finite temperatures.  These changes depend sensitively
on the composition of matter, in particular, on the question of whether or not
a new component that substantially softens matter can appear in the cold,
catalyzed star at high density.  The new components that have been discussed to
date include hyperons, a pion or kaon condensate, and a transition to quark
matter.  All these components involve negatively charged non-leptonic matter; 
hence, they appear at lower density in the cold, catalyzed star than in the hot,
neutrino-trapped star.  Since thermal pressure is always positive, the cold,
catalyzed star always has a higher density than a hot, neutrino-trapped star. 
Consequently, a cold, catalyzed star contains the softening component in a
larger proportion of the star's mass than the hot,  neutrino-trapped star, and
this leads to a {\it smaller} maximum mass.                                    

This behavior is opposite to that found for equations of state containing only
nucleons and leptons and no additional softening component.  In this case,
neutrino trapping generally reduces the maximum mass from the 
value found in neutrino 
free matter; although neutrino-trapped matter contains more leptons and more
leptonic pressure, it also contains more protons and, therefore, less baryonic
symmetry pressure.  While finite entropy provides additional pressure support,
the amount of the increase over the zero temperature case is generally small,
especially if realistic forces are employed.             

It must be emphasized that the maximum mass of the cold catalyzed star still
remains uncertain, due to the uncertainty in strong interactions at high
density.  At present, all nuclear models can only be effectively constrained
at nuclear density and by the condition of causality at high density.  
The resulting uncertainty is evident from the range of possible maximum masses
predicted by the different models considered in this work.  Despite this
uncertainty, our findings concerning the effects of finite entropy and
neutrino-trapping offer intriguing possibilities for distinguishing between the
different physical states of matter.  These possibilities include both black
hole formation in supernovae and the signature of neutrinos to be expected from
supernovae, as we now discuss.

\subsection{Black hole formation}
\setcounter{subsubsection}{0}

The gravitational collapse of the core of a massive star produces a
lepton-rich, neutrino trapped, proto-neutron star and an expanding shock wave. 
Energy losses from dissociation of heavy nuclei and neutrinos weaken the shock,
preventing a ``prompt'' explosion.  Within a few milliseconds, the shock wave
stagnates into an accretion shock at a distance of 100-200 km from the
proto-neutron star.  Gandhi and Burrows~\cite{gb} have demonstrated that the
neutrino luminosity of the stellar remnant is able to quasi-statically support
the shock against the ram pressure of infalling matter, which accretes onto the
neutron star.  Recent successful models of gravitational collapse
supernovae~\cite{snth1,snth2} invoke delayed neutrino heating, augmented by
convective motions, to power the supernova explosion. 

The explosion appears to occur within $\thalf$ to a few seconds after the core
bounce; and, once expansion occurs, accretion onto the proto-neutron star is
diminished.  Therefore, the star can be expected to accumulate nearly all of
its baryon number within a few seconds of core bounce.

As we have seen, a nucleons-only star has a maximum mass that {\em grows} as
neutrinos leak out of it.  It thus appears unlikely that such a star could form
a black hole during the longer-term deleptonization era.  In this case, a black
hole could only be produced immediately after bounce or during the short-term
accretion stage, when the shock is stagnant.

However, this is not the case for matter with non-leptonic, negatively charged
softening components.  For matter containing hyperons, kaons, or quarks, the
maximum mass of the neutrino-trapped star is larger than that of the
neutrino-free star.  Should the maximum baryon mass of the cold neutrino-free
star be close to 1.5$M_\odot$, black hole formation could occur as the 
neutrinos diffuse out of the protoneutron star, i.e., during the first 10 
seconds following bounce.  Burrows~\cite{bur} has demonstrated that black hole
formation should be accompanied by a dramatic cessation of the neutrino signal,
since the event horizon invariably forms outside the neutrinosphere.  Such
behavior would be relatively easy to observe from a galactic supernova and
would suggest that the equation of state produced a metastable protoneutron
star.              
                                                                     
In order to highlight some of the observable consequences for 
different
compositions of high density matter, the gravitational and baryonic masses, and
their differences, both for cold, catalyzed matter and for hot, neutrino-rich
matter, are shown in Fig.~29.  Two generic compositional cases are displayed:
nucleons-only matter ($npe^-\mu^-$), and matter with hyperons  ($npHe^-\mu^-$). 
Matter with condensed kaons and  with a quark-hadron transition  has a 
similar behavior, and is discussed in more detail in Refs.~\cite{tpl,pcl}.  

In all cases, the gravitational mass for the neutrino-rich cases are greater
than for the untrapped case for the same baryon mass, reflecting the binding
energy released during this stage of neutron star formation.  This is usually
less than half the total binding energy released, the remainder being emitted
during the prior stage.  The prior stage is the period between core bounce and
the production of the approximately adiabatic ($s\approx  2$), lepton-rich star 
(shown by the upper solid curves in Fig.~29), and lasts about 1-3 seconds.  The
binding energy emitted during the prior stage is difficult to show in Fig.~29, 
because during this period the neutron star is rapidly accreting mass (i.e.,
$M_B$ is increasing).  The accretion should drop substantially, and the value
of $M_B$ should approach a limiting value after the first few seconds. We note
that the binding energy emitted in the deleptonization and cooling stage
appears to be insensitive to  the composition of dense matter, just as the
total binding energy was found to be a universal function of mass.  Given its
increase with stellar mass, this means that the total energy released is
apparently not a good discriminant of composition.  
                                         
However, Fig.~29 clearly shows the consequence of different compositions for
black hole formation.  Black hole formation can be observed as an abrupt
cessation of neutrino signal, since the event horizon forms outside of the
star's neutrinosphere.  For the nucleons-only case, black hole formation is
unlikely to occur during the deleptonization and cooling stage, the one marked 
by the transition between
the upper solid and the lower dashed curves, since $M_B$ is approximately 
constant (or only
slightly increasing).  If a black hole were to form from a star with this
composition, it is much more likely to form during the post-bounce accretion 
stage.  This is not true for the
other compositional cases.  Here, the neutrino-trapped matter is always capable
of supporting more mass  than the cold, catalyzed matter.  If the 
baryon mass of the proto-neutron
star is near the maximum mass of the cold, catalyzed neutron star,
as it is, for example, for a 1.5 M$_\odot$ star in both the $npHe^-\mu^-$ and
the $npHQe^-\mu^-$  cases, then there exists a range of $0.1-0.2$ M$_\odot$
above this mass in which a hot neutrino-trapped star can be stabilized.  Thus,
with this composition, a black hole could form during either stage of neutron
star formation; although, given the relatively small value expected for the
mass of the imploding pre-bounce core, it seems more likely that a black 
hole would form during the later stage.  
     
A consequence of the potentially long delay of 10--15 seconds between core
bounce and black hole formation is that black hole formation and total binding
energy release are not necessarily correlated.  The softening in the equation
of state marked by the appearance of negatively charged hadrons is accompanied
by relatively little further binding energy release.  Thus, the large energy
release inferred by the neutrino detections from SN 1987A did not imply that
the equation of state could not yet change due to the appearance of any exotic
matter.

Black hole formation, by cutting off the neutrino luminosity from the
protoneutron star, would short-circuit the supernova mechanism if an explosion
had not already occurred.  This appears to be more likely for the nucleons-only
stars or for stars with relatively large initial core masses, i.e., very
massive stars with $M \geq 20-30~M_\odot$~\cite{wlw,ww}. 
Note also that these scenarios have different implications for
nucleosynthesis, since prompt black hole formation and a successful supernova
explosion, in which newly synthesized nuclei are ejected, may be incompatible.

Based on the calculations of Prakash and Lattimer, 
Brown and Bethe~\cite{browbet} proposed that a window exists in which neutron
stars collapse to black holes during deleptonization if hyperons, kaons or
quarks are present in neutron stars.  This proposal was meant to explain the
apparent non-existence of a neutron star in the remnant of SN1987A, although a
neutron star may have temporarily existed some 10-15~s, during which  neutrino
emission was observed.  The window naturally exists if negatively charged
hadronic matter appears after deleptonization.  The particular case of
SN1987A will be discussed below.             

\subsection{Neutrino signals from supernovae}
\setcounter{subsubsection}{0}

The composition of a neutron star will also influence the details of the star's
neutrino emission.  We focus on two possible diagnostics -- the total radiated
energy and the relative numbers of emitted neutrinos of different types.

In Fig.~30, we display the electron concentration, $Y_e$, for various dense
matter compositions.  The reference straight line is the total electron-lepton
concentration of the initial proto-neutron star, which we have assumed to be
approximately $Y_{Le}=0.4$.  The difference in any of the
other curves from the reference line shows, as a function of density, 
the total net electron-neutrino concentration (specifically, the difference 
of the $\nu_e$ and $\bar\nu_e$ concentrations)
that eventually leaks out of dense matter.  The integral of the difference over
the stellar density profile gives the net number difference for the entire
star.  This difference is always positive, although the actual value is
sensitive to composition.                          

A rough measure of the relative fluxes of escaping neutrinos can be found as
follows.  Although, in the stellar core, there are essentially only $\nu_e$'s
because $\mu_e/T\gg 1$, $\mu_e \gg 1$ and $\mu_\mu=\mu_\tau\simeq0$, the 
emerging neutrino signature consists of nearly equal proportions of all six
types of neutrinos.  This may be understood as follows.   Diffusion degrades
the high energies ($\mu_{\nu_e}\approx200$ MeV) of the core $\nu_e$'s;
therefore, the
emerging neutrinos have average energies in the range 10-20 MeV.  Pair
production in the hot matter in the outer mantle of the proto-neutron star
generates several pairs of all three neutrino flavors per core $\nu_e$.  Using
10 MeV for the emergent energy, we find about 3 pairs emerge per core $\nu_e$
emitted.   This shows that the total number of escaping neutrinos has a slight
excess of $\nu_e$'s; for this example, the ratio of $\nu_e:\nu_x$, where $x$
refers to any of the other neutrino species, is about 4:3.                     
                           
The actual situation is more complicated because, in general, different
neutrino species are emitted with different energies.  Also, the total flux of
neutrinos should be sensitive to the available binding energy change. 
Nevertheless, the basic trends are clear:  First, the smaller the value of
$Y_e$ in the cold, catalyzed neutron star, the larger the excess of $\nu_e$'s
that will be emitted.  Second, the larger the binding energy change during
deleptonization, the {\it smaller} the relative excess of $\nu_e$'s
will be.  This is because the energies of the escaping neutrinos are rather
insensitive to the structure of the proto-neutron star, including details of
the equation of state of high density matter.  The energies of the escaping
neutrinos are determined instead by the properties of the outer mantle of the
star.  Thus, higher binding energy release translates directly to  larger total
numbers of escaping neutrinos.                                      

However, referring to Fig.~29, one sees that the  change in binding energies
during deleptonization and cooling  is relatively insensitive to the dense
matter composition, although it does increase with the final neutron star mass.
Therefore, the net excess
of $\nu_e$'s from a proto-neutron star during deleptonization and cooling seems
to be a probe of the final value of $Y_e$ in dense matter, a quantity that is
quite sensitive to dense matter composition.  It remains to be seen if
differences in  the final value of $Y_e$ are large enough to be observable in
the neutrino signal from a galactic supernova.                         

\subsection{Supernova SN1987A}
\setcounter{subsubsection}{0}
 
The case of SN1987A is interesting in light of the potential metastability of
forming neutron stars.  On the one hand, on February 23 of 1987 neutrinos were
observed from the explosion of supernova SN1987A, indicating that a neutron
star, not a black hole, was initially present. On the other hand, the
ever-decreasing optical luminosity of the remnant of SN1987A suggests two
arguments~\cite{black,bb} against the presence of a neutron star.

First, accretion at the Eddington limit with the usual Thomson electron
scattering opacity onto a neutron star is already ruled out.  Chen and  Colgate
\cite{cc}, however, have recently suggested that the opacity appropriate for a
neutron star atmosphere has been underestimated by several orders of
magnitude.  Using the  opacity of iron at X-ray photon energies, they conclude
that the appropriate Eddington limit cannot yet rule out accretion onto a
neutron star. 
                           
Second, a Crab-like pulsar cannot exist in SN1987A since the emitted magnetic
dipole radiation would be too large.  The magnetic field and/or the spin rate
of a neutron star remnant must be much less than in the case of the Crab and, 
therefore, much less than is inferred for other young neutron stars.  It is
possible, however, that although the spin rate of a newly formed neutron star
is expected to be high, the timescale for the generation of a significant
magnetic field is greater than 10 years.  Unfortunately, this timescale is 
not known with certainty~\cite{muspage}.                        
                                    
The experimental measurements~\cite{exp87a} of neutrinos from SN1987A 
indicated the following:        
\begin{itemize}
\item A total binding energy of $\sim (0.1-0.2)M_{\odot}$ was 
released, indicating, from Fig.~28, a remnant gravitational mass of 
$(1.14-1.55)M_\odot$.  In addition, about
half or more of the binding energy appears to have been released during the
first 2 seconds, in agreement with the analysis of the previous section and
Fig.~29. The binding energy arguments therefore do not discriminate among the
various scenarios we have  discussed.
\item The average neutrino energy was $\sim10$ 
MeV; to lowest order, this is fixed by the mean free path
$\lambda_{\nu}(E_{\nu})$ in the outer regions of the protoneutron star,
 and also does not shed much light on the internal stellar composition. 
\item In spite of the fact that most of the binding energy is released during
the initial accretion and collapse stage in the first 2 or so seconds after
bounce, the neutrino signal continued for a period of at least 12 s.  This
latter timescale may  be significant, since it is the also about the time 
required for the neutrinos initially trapped in the star to leave.  However,
counting statistics prevented measurement of a longer duration, and this
unfortunate coincidence prevents one from distinguishing a model in which
negatively-charged hadronic matter appears and a black hole forms from a less
exotic model, in which a neutron star still exists.  As we have pointed out, the
maximum stable mass drops by $\sim 0.2M_{\odot}$ when the trapped neutrinos
depart if negatively charged hadrons are present, be they hyperons, kaons or
quarks, which could be enough to lead to continued collapse to a black hole.  
\end{itemize}

Observed neutron stars lie in a very small range of gravitational masses. The 
smallest range that is consistent with all the data~\cite{starmass} runs from 
$1.34M_{\odot}$ to $1.44M_{\odot}$, the latter value being the accurate 
measurement of PSR1913+16. Thielemann,~et.al.~\cite{thm} and Bethe and
Brown~\cite{bb} have estimated the gravitational  mass of the compact core
of SN1987A to  be in the range $\sim (1.40-1.56)M_{\odot}$, using arguments
based on the observed amounts of ejected $^{56}$Ni and/or the total explosion
energy.  This range extends above the largest accurately known value for a
neutron star mass,  $1.44~M_\odot$, so the possibility exists that the neutron
star initially produced in SN1987A could be unstable in the cold, deleptonized 
state. A possible scenario is that such a mass could be  stabilized initially
when the neutrinos were trapped, but could become unstable when the neutrino
concentration dropped to very small values (see Fig.~16, for example). In this
case, therefore, SN1987A would have become a black hole once it had
deleptonized, and no further signal would be expected. 
                                      
Note that this scenario for black hole formation in SN1987A is different from
that originally proposed by Brown, Bruenn and Wheeler~\cite{black}, who
suggested that long-term accretion on the remnant eventually resulted in the
production of a black hole.  Although recent 2-dimensional hydrodynamical
calculations of supernovae~\cite{snth1,snth2} suggest that significant accretion
ceases when the supernova shock lifts off, no more than 1 second after bounce,
it remains to be seen whether the reverse shocks generated when the shock
reaches the low-density hydrogen layers produces significant
fallback~\cite{colgate}.   

\subsection{Future detections} 
\setcounter{subsubsection}{0}

A fundamental question is whether or not future neutrino detectors will be able
to discriminate among EOS models.  The neutrino signal observed in a
terrestrial detector is the folding of the emitted neutrino spectrum with the
detector characteristics.  The latter is a combination of the appropriate
microphysical cross sections for neutrino scattering and absorption together 
with the efficiency function and fiducial (effective) volume of the detector. 
Another important parameter of the detector is its low-energy threshold, or
cutoff.  Burrows, Klein and Gandhi~\cite{bkl} list some properties of present,
under construction, and future detectors, together with a rough estimate of the
numbers of neutrinos that would be observed from a supernova located within our
Galaxy (assumed to be at a distance of 10 kpc).   Table~17 lists some of the
characteristics of the various neutrino telescopes.  
         
In an optimistic scenario, about 10,000 neutrinos will be seen from a typical
galactic supernova in a single detector.  A crucial question, which has not yet
received much attention, is whether the statistical uncertainty in a
time-dependent signal can be small enough to adequately differentiate models.
\\ 

Among the interesting features that could be sought are:

\begin{enumerate}
\item  Possible cessation of a neutrino signal due to black hole formation.
\item  Possible burst or light curve feature associated with the onset of
negatively-charged hadrons near the end of deleptonization, whether or not a
black hole is formed.
\item  Identification of the deleptonization/cooling epochs by changes in
luminosity evolution or in neutrino flavor distribution.
\item  Determination of a radius-mean-free-path correlation from the luminosity
decay time or the onset of neutrino transparency.
\item Determination of the neutron star mass from the universal-binding
energy--mass relation.
\end{enumerate}

To realize the above goals, more information about the characteristics of
neutrino telescopes must be made widely available.  This is especially
important in deciphering the time evolution of the neutrino signal (see, for
example, Lattimer and Yahil~\cite{ly}) even if a large number (10,000 or more)
of neutrinos are detected in total.   

\vskip 1in

We thank Gerry Brown, who initiated this work by raising questions about 
the extent to which neutron stars are stabilized due to thermal effects.  
We are grateful to Jason Cooke and Sanjay Reddy for helpful discussions, and for
their generous help in preparing some of the figures in this manuscript. This
work was supported in part by the U.S. Department of Energy under  grant 
numbers DOE/DE-FG02-88ER-40388 and  DOE/DE-FG02-87ER-40328 and
DOE/DE-FG02-87ER-40317, and in part by the NASA grant NAG52863. Part of this
work was done at the Institute of Theoretical Physics, Santa Barbara, during
the research program Strong Interactions at Finite Temperatures. Madappa
Prakash, Manju Prakash and Paul Ellis express gratitude for the warm
hospitality extended there and acknowledge the support of the National Science
Foundation under grant number  PHY89-04035. MP and PJE thank the Institute for 
Nuclear Theory at the University of Washington for its hospitality during their
visits  and the Department of Energy for partial support during the completion
of this work.

\newpage

\bigskip
\bigskip
\centerline{Table 1}
\centerline{Potential model parameters for nuclear matter}
\bigskip
\bigskip
\begin{center}
\begin{tabular}{cccccccc}
\hline \hline \\ 
EOS & $K_0$ & $A$ & $B$ & $B^{\prime}$ & $\sigma$ & $C_1$ & $C_2$
 \\ \hline 
BPAL1 & 120 & 75.94 & $-30.88$ & 0 & $0.498$ & $-83.84$ & $23$ \\
BPAL2 & 180 &  440.94 & $-213.41$ & 0 & $0.927$ & $-83.84$ & $23$ \\
BPAL3 & 240 & $-46.65$ & 39.45 & 0.3 & 1.663 & $-83.84$ & $23$ \\
\hline 
SL1    & 120 & 3. 706 & $-31.155$ & 0 & $0.453$ & $-41.28$ & $23$ \\
SL2    & 180 &  159.47 & $-109.04$ & 0 & $0.844$ & $-41.28$ & $23$ \\
SL3    & 240 & $-204.01$ & 72.704 & 0.3 & 1.235 & $-41.28$ & $23$
\\ \hline \hline

\end{tabular}
\begin{quote}
{Parameters in Eq.~(\ref{PAL1}) determined by fitting the equilibrium properties
of symmetric nuclear matter for some input values of the compression
modulus $K_0$~\cite{pal}. All quantities are in MeV, except for the 
dimensionless $\sigma$. The finite-range
parameters $\Lambda_1 = 1.5p_F^{(0)}$ and $\Lambda_2=3p_F^{(0)}$. }
\end{quote}
\label{palpar}
\end{center}
\newpage

\bigskip
\bigskip
\centerline{Table 2}
\centerline{Potential model parameters for neutron-rich matter}
\bigskip
\bigskip
\begin{center}
\begin{tabular}{cccccc}
\hline \hline \\ 
EOS & $K_0$ & $x_0$ & $x_3$ & $Z_1$ & $Z_2$ 
 \\ \hline 
BPAL11 & {} & $-0.689$ & $0.577$ & $-14.00$ & $16.69$ \\
BPAL12 & {120}  & $-1.361$ & $-0.244$ & $-13.91$ & $16.69$ \\
BPAL13 & {}  & $-1.903$ & $-1.056$ & $-1.83$ & $5.09$ \\ \hline
BPAL21 & {} & $0.086$ & $0.561$ & $-18.40$ & $46.27$ \\
BPAL22 & {180}  & $-0.410$ & $-0.105$ & $-9.38$ & $24.05$ \\
BPAL23 & {}  & $-1.256$ & $-1.358$ & $-11.67$ & $-10.90$ \\  \hline 
BPAL31 & {} & $0.376$ & $0.246$ & $-12.23$ & $-2.98$ \\
BPAL32 & {240} & $0.927$ & $-0.227$ & $-11.51$ & $8.38$ \\
BPAL33 & {}  & $1.654$ & $-1.112$ & $3.81$ & $13.16$ \\
\hline \hline 
SL12 & {120}  & $-3.548$ & $-0.5$ & $-13.355$ & $2.789$ \\
SL22 & {180}  & $-0.410$ & $-0.105$ & $9.38$ & $-4.421$ \\
SL32 & {240} & $-0.442$ & $-0.5$ & $-13.387$ & $2.917$ \\     
\hline \hline

\end{tabular}
\begin{quote}
{Parameters in Eq.~(\ref{enem}); $x_0$ and $x_3$  are dimensionless, the
remaining quantities are in MeV.  For each
compression modulus $K_0$, the three different choices of the constants yield
the potential part of the symmetry energy that varies approximately as ${\sqrt
u}~,u$ and $2u^2/(1+u)$, respectively, as in the parameterization of
Ref.~\cite{pal}.  In all cases, the symmetry energy at the nuclear matter
equilibrium density is taken to be 30 MeV. The notations BPAL$n_1n_2$ and
SL$n_1n_2$ are used used to denote different EOSs; $n_1$ refers to different
values of $K_0$, and $n_2=1,2$ and 3 indicate, respectively, a
${\sqrt u}~,u$, and $2u^2/(1+u)$ dependence of the nuclear symmetry
potential energy on the density.}
\end{quote}
\label{expalpar}
\end{center}
\newpage

\centerline{Table 3}
\centerline{Pure neutron star properties at finite entropy in the potential 
models.}
\begin{center}
\begin{tabular}{ccccccccc}\hline\hline
 EOS  & $S$ &
${\displaystyle\frac{\strut M_{\rm max}}{M_{\odot}}}$ & $R$&
${\displaystyle \frac{n_c}{n_0}}$ & $P_c$ & $T_c$ & $\lambda \cdot 10^2$ & $I$\\
&&&(km)& & ${\rm MeV~fm}^{-3}$ & ${\rm MeV}$ & & $M_\odot~{\rm km}^2$ \\ \hline
 {}       & 0 & 1.896 & 10.509 &  7.344 & 546.6 &   0.0 & {}   &  87.60\\
BPAL 22   & 1 & 1.941 & 10.987 &  6.812 & 493.0 &  72.9 & 2.61 &  95.73\\
 {}       & 2 & 2.093 & 12.331 &  5.392 & 346.4 & 138.4 & {}   & 126.60\\ \hline
 {}       & 0 & 2.020 & 10.55  &  7.05  & 645.2 &   0.0 & {}   &  98.97\\
SL22      & 1 & 2.109 & 11.12  &  6.37  & 565.6 & 117.0 & 4.31 & 113.53\\
 {}       & 2 & 2.369 & 12.59  &  4.90  & 429.2 & 208.2 & {}   & 163.24\\ 
\hline \hline
\end{tabular}
\begin{quote}
$R$, $n_c,~P_c$, $T_c$, and $I$ refer to the radius, central 
density, pressure, temperature, and moment of inertia of the maximum mass star.  
The coefficient
$\lambda$ (see Eq.~(\ref{lambda})) shows the increase in the maximum mass due
to thermal effects.                                          
\end{quote}
\label{bpalnem}
\end{center}
\newpage

\centerline {Table 4}
\centerline{Star properties for matter in beta equilibrium  at finite entropy}
\centerline{in the BPAL potential  model.}
\bigskip
\begin{center}
\begin{tabular}{ccccccccc}\hline\hline
 EOS  & $S$ &
${\displaystyle\frac{\strut M_{\rm max}}{M_{\odot}}}$ & $R$&
${\displaystyle \frac{n_c}{n_0}}$ & $P_c$ & $T_c$ & $\lambda \cdot 10^2$ & $I$\\
&&& km & & ${\rm MeV~fm}^{-3}$ & ${\rm MeV}$ & & $M_\odot~{\rm km}^2$ \\ \hline
 {}       & 0 & 1.393 & 8.219 & 12.656 & 843.1  & 0.0 & {}   & 36.49 \\
BPAL 11   & 1 & 1.415 & 8.540 & 11.875 & 786.6  &55.9 & 1.73 & 39.11 \\
 {}       & 2 & 1.489 & 9.571 &  9.687 & 552.4  &97.6 & {}   & 48.84 \\ \hline
 {}       & 0 & 1.454 & 8.943 & 10.938 & 659.5  & 0.0 & {}   & 43.49 \\
BPAL 12   & 1 & 1.475 & 9.250 & 10.156 & 621.5  &43.5 & 1.48 & 46.35 \\
 {}       & 2 & 1.540 &10.219 &  8.594 & 451.6  &82.4 & {}   & 56.38 \\ \hline
 {}       & 0 & 1.473 & 9.446 & 10.156 & 566.6  & 0.0 & {}   & 47.16 \\
BPAL 13   & 1 & 1.495 & 9.872 &  9.375 & 449.7  &39.3 & 1.60 & 51.00 \\
 {}       & 2 & 1.567 &10.840 &  7.812 & 376.4  &74.1 & {}   & 62.67  \\ \hline
 {}       & 0 & 1.672 & 9.172 &  9.687 & 766.2  & 0.0 & {}   & 58.83 \\
BPAL 21   & 1 & 1.689 & 9.437 &  9.219 & 712.4  &50.5 & 1.00 & 61.73 \\
 {}       & 2 & 1.739 &10.239 &  7.969 & 550.9  &88.4 & {}   & 71.07 \\ \hline
 {}       & 0 & 1.722 & 9.721 &  8.750 & 638.3  & 0.0 & {}   & 66.36 \\
BPAL 22   & 1 & 1.735 & 9.943 &  8.437 & 564.2  &40.5 & 0.79 & 68.78 \\
 {}       & 2 & 1.776 &10.630 &  7.500 & 500.7  &77.1 & {}   & 77.64 \\ \hline
 {}       & 0 & 1.737 &10.104 &  8.281 & 566.5  & 0.0 & {}   & 70.13 \\
BPAL 23   & 1 & 1.752 &10.350 &  7.969 & 536.4  &36.1 & 0.88 & 73.26  \\
 {}       & 2 & 1.798 &11.120 &  6.875 & 409.9  &69.5 & {}   & 83.63  \\ \hline
 {}       & 0 & 1.905 &10.107 &  7.734 & 652.1  & 0.0 & {}   & 84.79 \\
BPAL 31   & 1 & 1.917 &10.280 &  7.500 & 629.0  &41.3 & 0.62 & 87.31 \\
 {}       & 2 & 1.952 &10.878 &  6.878 & 543.5  &78.4 & {}   & 95.29 \\ \hline
 {}       & 0 & 1.933 &10.420 &  7.343 & 590.2  & 0.0 & {}   & 90.14 \\
BPAL 32   & 1 & 1.943 &10.589 &  7.138 & 577.7  &36.7 & 0.53 & 92.52 \\
 {}       & 2 & 1.974 &11.136 &  6.506 & 482.8  &71.5 & {}   &100.17 \\ \hline
 {}       & 0 & 1.955 &10.797 &  7.000 & 532.0  & 0.0 & {}   & 95.35  \\
BPAL 33   & 1 & 1.966 &11.020 &  6.719 & 507.0  &33.3 & 0.49 & 98.57  \\
 {}       & 2 & 1.994 &11.518 &  6.198 & 454.3  &66.0 & {}   &105.72  \\ \hline 
\hline
\end{tabular}
\label{bpalnpem}
\end{center}
\newpage

\centerline {Table 5}
\centerline{Star properties for matter in beta equilibrium  at finite entropy}
\centerline{in the SL potential  model.}
\bigskip
\begin{center}
\begin{tabular}{ccccccccc}\hline\hline
 EOS  & $S$ &
${\displaystyle\frac{\strut M_{\rm max}}{M_{\odot}}}$ & $R$&
${\displaystyle \frac{n_c}{n_0}}$ & $P_c$ & $T_c$ & $\lambda \cdot 10^2$ & $I$\\
&&&km& & ${\rm MeV~fm}^{-3}$ & ${\rm MeV}$& & $M_\odot~{\rm km}^2$ \\ \hline
 {}       & 0 & 1.7423 & 9.145  & 9.53 & 923.3 &   0.0 & {}   &  62.66 \\
SL12   	  & 1 & 1.7711 & 9.526  & 8.91 & 816.8 &  71.4 & 1.86 &  67.46 \\
 {}       & 2 & 1.8710 & 10.61  & 7.27 & 558.2 & 121.0 & {}   &  85.53 \\ \hline
 {}       & 0 & 1.890  & 9.840  & 8.13 & 773.5 &   0.0 & {}   &  80.26 \\
SL22      & 1 & 1.920  & 10.132 & 7.73 & 724.9 &  64.0 & 1.59 &  85.15 \\
 {}       & 2 & 2.010  & 11.110 & 6.56 & 500.7 & 113.5 & {}   & 102.60 \\ \hline
 {}       & 0 & 2.0971 & 10.572 & 6.81 & 689.9 &   0.0 & {}   & 107.14 \\
SL32      & 1 & 2.1211 & 10.790 & 6.60 & 651.6 &  56.6 & 1.11 & 111.60 \\
 {}       & 2 & 2.1901 & 11.549 & 5.83 & 532.2 & 103.2 & {}   & 127.13 \\ \hline
\hline
\end{tabular}
\label{snpem}
\end{center}
\newpage

\centerline {Table 6}
\centerline{Star properties for matter with trapped neutrinos ($Y_{Le}=0.4$) 
in beta equilibrium}
\centerline{at finite entropy in the BPAL potential  model.}
\bigskip
\begin{center}
\begin{tabular}{ccccccccc}\hline\hline
 EOS  & $S$ &
${\displaystyle\frac{\strut M_{\rm max}}{M_{\odot}}}$ & $R$&
${\displaystyle \frac{n_c}{n_0}}$ & $P_c$ & $T_c$ & $\lambda \cdot 10^2$ & $I$\\
&&& km & & ${\rm MeV~fm}^{-3}$ & ${\rm MeV}$ & & $M_\odot~{\rm km}^2$ \\ \hline
 {}       & 0 & 1.376 & 8.459 & 11.875 & 766.3 &  0.0 & {}   & 36.48 \\
BPAL 11   & 1 & 1.396 & 8.705 & 11.250 & 710.6 & 43.3 & 1.63 & 38.78 \\
 {}       & 2 & 1.465 & 9.750 &  9.219 & 505.3 & 78.9 & {}   & 48.30 \\ \hline
 {}       & 0 & 1.394 & 8.635 & 11.406 & 719.0 &  0.0 & {}   & 38.37 \\
BPAL 12   & 1 & 1.413 & 8.884 & 10.781 & 663.8 & 41.2 & 1.59 & 40.77 \\
 {}       & 2 & 1.482 & 9.919 &  8.906 & 478.1 & 75.7 & {}   & 50.47 \\ \hline
 {}       & 0 & 1.403 & 8.782 & 11.094 & 681.6 &  0.0 & {}   & 39.59 \\
BPAL 13   & 1 & 1.422 & 9.035 & 10.469 & 620.5 & 39.7 & 1.60 & 42.11 \\
 {}       & 2 & 1.492 &10.099 &  8.594 & 447.3 & 73.0 & {}   & 52.37 \\ \hline
 {}       & 0 & 1.641 & 9.300 &  9.219 & 705.4 &  0.0 & {}   & 57.38 \\
BPAL 21   & 1 & 1.656 & 9.465 &  8.958 & 684.7 & 38.7 & 0.96 & 59.40 \\
 {}       & 2 & 1.704 &10.199 &  7.917 & 549.0 & 72.9 & {}   & 67.80 \\ \hline
 {}       & 0 & 1.655 & 9.403 &  9.062 & 688.6 &  0.0 & {}   & 59.03 \\
BPAL 22   & 1 & 1.669 & 9.580 &  8.750 & 664.7 & 37.0 & 0.93 & 61.24 \\
 {}       & 2 & 1.716 &10.330 &  7.708 & 519.5 & 70.3 & {}   & 69.79 \\ \hline
 {}       & 0 & 1.662 & 9.505 &  8.906 & 665.1 &  0.0 & {}   & 60.21 \\
BPAL 23   & 1 & 1.676 & 9.677 &  8.594 & 636.5 & 35.9 & 0.92 & 62.45 \\
 {}       & 2 & 1.723 &10.427 &  7.604 & 509.2 & 68.5 & {}   & 71.06 \\ \hline
 {}       & 0 & 1.855 &10.036 &  7.656 & 644.6 &  0.0 & {}   & 79.65 \\
BPAL 31   & 1 & 1.867 &10.198 &  7.422 & 618.4 & 34.1 & 0.66 & 82.01 \\
 {}       & 2 & 1.904 &10.747 &  6.812 & 523.9 & 66.4 & {}   & 89.78 \\ \hline
 {}       & 0 & 1.862 &10.092 &  7.578 & 633.2 &  0.0 & {}   & 80.74 \\
BPAL 32   & 1 & 1.874 &10.249 &  7.344 & 610.7 & 33.7 & 0.64 & 83.17 \\
 {}       & 2 & 1.910 &10.820 &  6.719 & 514.1 & 65.3 & {}   & 90.82 \\ \hline
 {}       & 0 & 1.871 &10.203 &  7.437 & 609.6 &  0.0 & {}   & 82.42 \\
BPAL 33   & 1 & 1.883 &10.330 &  7.266 & 604.6 & 32.8 & 0.63 & 84.44 \\
 {}       & 2 & 1.918 &10.922 &  6.625 & 503.7 & 63.7 & {}   & 92.35 \\ \hline
\hline
\end{tabular}
\label{bpaltnpem}
\end{center}
\newpage

\centerline{Table 7}
\centerline{MRHA coupling constants and saturation properties}
\begin{center}
\begin{tabular}{cccccc}\hline \hline
${\displaystyle \frac{\mu_r}{M}}$&
${\displaystyle\frac{\strut M^*_{N{\rm sat}}}{M}}$&$K_0$&$C^2_{\omega}$&
$C^2_{\sigma}$&$C^2_{\rho}$\\
&&(MeV)&&&\\ \hline
0.79&0.66&354&180.6&317.5&73.5\\
1.00&0.73&461&137.7&215.0&81.6\\
1.25&0.82&264&\phantom{1}78.6&178.6&90.8\\ \hline \hline
\end{tabular}
\end{center}
\newpage

\vskip1cm
\centerline{Table 8}
\centerline{Star properties for matter in beta equilibrium  
at finite entropy}
\centerline{using relativistic models.} 
\begin{center}
\begin{tabular}{ccccccccc}\hline \hline
${\displaystyle \frac{\mu_r}{M}}$&$S$&
${\displaystyle\frac{\strut M_{\rm max}}{M_{\odot}}}$&$R$&
${\displaystyle \frac{n_c}{n_0}}$& $P_c$& $T_c$ & $\lambda \cdot 10^2$ & $I$\\
&&&(km)& & ${\rm MeV~fm}^{-3}$ & ${\rm MeV}$& & $M_\odot~{\rm km}^2$ \\ \hline

{}   & 0 & 2.532& 12.62& 4.73& 432.2 &  0.0 & {}   & 191.18 \\
0.79 & 1 & 2.535& 12.75& 4.63& 416.3 & 28.9 & 0.10 & 193.37 \\
{}   & 2 & 2.542& 13.00& 4.46& 394.9 & 58.3 & {}   & 197.24 
\\ \hline
{}   & 0 & 2.305& 12.02& 5.34& 428.2 &  0.0 & {}   & 151.05 \\
1.00 & 1 & 2.311& 12.14& 5.24& 416.6 & 27.9 & 0.25 & 153.19  \\
{}   & 2 & 2.328& 12.45& 4.98& 387.2 & 56.7 & {}   & 158.82 
\\ \hline 
{}   & 0 & 1.857& 10.60& 7.29& 484.9 &  0.0 & {}   & 85.63 \\
1.25 & 1 & 1.868& 10.72& 7.19& 466.1 & 29.1 & 0.56 & 87.38  \\
{}   & 2 & 1.899& 11.19& 6.60& 419.6 & 58.8 & {}   & 94.33
\\ \hline \hline
{}   & 0 & 2.005& 10.92& 7.14& 545.8 &  0.0 & {}   & 100.6 \\
GM   & 1 & 2.014& 11.08& 6.95& 521.9 & 31.6 & 0.47 & 103.1 \\
{}   & 2 & 2.044& 11.56& 6.43& 458.2 & 62.6 & {}   & 110.6 
\\ \hline \hline
\end{tabular}
\begin{quote}
The symbol GM refers to the EOS of
Ref.~\cite{glenmos}.  The other three sets are for  the MRHA model.
\end{quote}
\end{center}

\newpage
\vskip1cm
\centerline{Table 9}
\centerline{Star properties for matter with trapped neutrinos ($Y_{Le}=0.4$) 
in beta 
equilibrium }
\centerline{at finite entropy using relativistic models.} 
\begin{center}
\begin{tabular}{ccccccccc}\hline \hline
${\displaystyle \frac{\mu_r}{M}}$&$S$&
${\displaystyle\frac{\strut M_{\rm max}}{M_{\odot}}}$&$R$&
${\displaystyle \frac{n_c}{n_0}}$& $P_c$& $T_c$ & $\lambda \cdot 10^2$ & $I$\\
&&&(km)& & ${\rm MeV~fm}^{-3}$ & ${\rm MeV}$& & $M_\odot~{\rm km}^2$ \\ \hline

{}   & 0 & 2.447& 12.28& 4.86& 460.1 &  0.0 & {}   & 172.77 \\
0.79 & 1 & 2.452& 12.29& 4.78& 448.9 & 26.4 &0.25  & 174.73 \\
{}   & 2 & 2.471& 12.70& 4.55& 415.7 & 52.6 & {}   & 181.38 
\\ \hline
{}   & 0 & 2.222& 11.68& 5.47& 451.2 &  0.0 & {}   & 135.90 \\
1.00 & 1 & 2.228& 11.72& 5.42& 448.5 & 25.9 &0.40  & 137.27  \\
{}   & 2 & 2.257& 12.16& 5.09& 406.4 & 51.5 & {}   & 145.39 
\\ \hline 
{}   & 0 & 1.784& 10.28& 7.50& 514.1 &  0.0 & {}   & 76.29 \\
1.25 & 1 & 1.791& 10.23& 7.42& 511.4 & 27.6 &0.75  & 77.16  \\
{}   & 2 & 1.836& 10.89& 6.81& 448.3 & 54.6 & {}   & 85.25
\\ \hline \hline
{}   & 0 & 1.935& 10.53& 7.41& 595.8 &  0.0 & {}   & 90.13 \\
GM   & 1 & 1.946& 10.69& 7.22& 568.3 & 30.2 &0.58  & 92.48 \\
{}   & 2 & 1.980& 11.19& 6.67& 496.6 & 59.0 & {}   & 100.1 
\\ \hline \hline
\end{tabular}

\end{center}
\vskip.25cm
\newpage

\centerline{Table 10}
\centerline{MRHA coupling constants and saturation properties with hyperons.}
\begin{center}
\begin{tabular}{ccccccc}\hline \hline
${\displaystyle \frac{\mu_r}{M}}$&
${\displaystyle\frac{\strut M^*_{N{\rm sat}}}{M}}$&$K_0$&$C^2_{\omega}$&
$C^2_{\sigma}$&$C^2_{\rho}$&$x_\omega$\\
&&(MeV)&&&\\ \hline
0.79&0.76&177&118.7&258.1&84.8&0.660\\
1.00&0.73&455&133.1&210.3&82.4&0.658\\
1.25&0.84&228&\phantom{1}64.9&174.0&92.6&0.679\\ \hline \hline
\end{tabular}
\end{center}
\vskip.25cm
\begin{quote}
In all cases, the ratio of hyperon to nucleon sigma and rho couplings are
taken to be equal: $x_\sigma=g_{\sigma H/}g_{\sigma n} = 
x_\rho=g_{\rho H}/g_{\rho n}= 0.6$. 
\end{quote}
\newpage

\newpage
\vskip1cm
\centerline{Table 11}
\centerline{Star properties for matter, including hyperons, in beta
equilibrium}
\centerline{at finite entropy using relativistic models.}
\begin{center}
\begin{tabular}{ccccccccc}\hline \hline
${\displaystyle \frac{\mu_r}{M}}$&$S$&
${\displaystyle\frac{\strut M_{\rm max}}{M_{\odot}}}$&$R$&
${\displaystyle \frac{n_c}{n_0}}$& $P_c$& $T_c$ & $\lambda \cdot 10^2$ & $I$\\
&&&(km)& & ${\rm MeV~fm}^{-3}$ & ${\rm MeV}$& & $M_\odot~{\rm km}^2$ \\ \hline

{}       & 0 & 1.638& 10.62& 7.81& 376.7 &  0.0  & {}   & 67.8 \\
0.79     & 1 & 1.639& 10.73& 7.66& 365.6 &  19.7 & 0.08 & 68.6 \\
{}       & 2 & 1.643& 10.99& 7.39& 353.5 &  42.2 & {}   & 70.0  
\\ \hline
{}       & 0 & 1.886& 12.12& 5.64& 248.1 &  0.0 & {}   & 107.4 \\
1.00     & 1 & 1.884& 12.16& 5.64& 250.1 &  16.9   & $-0.11$   & 106.9 \\
{}       & 2 & 1.878& 12.32& 5.54& 247.9 &  35.8   & {}   & 106.6 
\\ \hline
{}       & 0 & 1.407& 10.43& 8.44& 310.3 &  0.0  & {}   & 51.0 \\
1.25     & 1 & 1.414& 10.59& 8.13& 293.1 &  18.4 & 0.36    & 52.5 \\
{}       & 2 & 1.428& 10.86& 7.81& 283.9 &  39.2 & {}   & 54.5 
\\ \hline \hline

{}       & 0 & 1.544& 10.78& 7.66& 311.4 &  0.0  & {}   & 63.2 \\
{\rm GM} & 1 & 1.551& 10.95& 7.34& 290.5 &  19.7 & 0.47 & 65.2 \\
{}       & 2 & 1.573& 11.32& 6.88& 269.5 &  41.4 & {}   & 69.2  
\\ \hline \hline
\end{tabular}

\end{center}
\vskip.25cm

\newpage
\vskip1cm
\centerline{Table 12}
\centerline{Star properties for matter, including hyperons and trapped
neutrinos ($Y_{Le}=0.4$),}
\centerline{in beta equilibrium  at finite entropy using relativistic models.}
\begin{center}
\begin{tabular}{ccccccccc}\hline \hline
${\displaystyle \frac{\mu_r}{M}}$&$S$&
${\displaystyle\frac{\strut M_{\rm max}}{M_{\odot}}}$&$R$&
${\displaystyle \frac{n_c}{n_0}}$& $P_c$& $T_c$ & $\lambda \cdot 10^2$ & $I$\\
&&&(km)& & ${\rm MeV~fm}^{-3}$ & ${\rm MeV}$& & $M_\odot~{\rm km}^2$ \\ \hline

{}       & 0 & 1.843& 11.03& 6.60 & 368.5 &  0.0 & {}   & 89.5 \\
0.79     & 1 & 1.837& 11.15& 6.46 & 355.4 &  17.0   & $-0.15$    & 89.8 \\
{}       & 2 & 1.831& 11.49& 6.25 & 337.7 &  36.3   & {}   & 91.2 
\\ \hline
{}       & 0 & 2.066& 12.15& 5.27 & 290.2 &  0.0 & {}   & 126.5 \\
1.00     & 1 & 2.063& 12.27& 5.19 & 281.5 & 15.4 & $-0.11$    & 127.1 \\
{}       & 2 & 2.057& 12.56& 5.03 & 266.9 & 32.2 & {}   & 128.8 
\\ \hline
{}       & 0 & 1.580& 10.48& 7.66& 355.9 &  0.0  &  {}   & 63.1 \\
1.25     & 1 & 1.585& 10.61& 7.50& 344.6 &  18.0 & 0.30  & 64.3 \\
{}       & 2 & 1.599& 11.11& 6.88& 299.7 &  36.9 & {}    & 64.9  
\\ \hline \hline
{}       & 0 & 1.768& 11.11& 6.63& 334.8&  0.0 & {}   & 83.6 \\
{\rm GM} & 1 & 1.772& 11.21& 6.56& 332.0 &17.5 & 0.10 & 84.5 \\
{}       & 2 & 1.776& 11.66& 6.15& 296.7 &37.0 & {}   &    88.5
\\ \hline \hline
\end{tabular}

\end{center}
                           
\newpage
\vskip1cm
\centerline{Table 13}
\centerline{Critical density ratio, $u_{crit}=n_{crit}/n_0$, 
for kaon condensation in the relativistic}
\centerline{mean field model, GM, for the neutrino-free 
and trapped neutrino cases ($Y_{Le}=0.4$),}
\centerline{with and without hyperons.}
\begin{center}
\begin{tabular}{ccc|cc|cc}\hline \hline
&&&\multicolumn{2}{c|}{Without hyperons}&\multicolumn{2}{c}{With hyperons}
\\ \cline{4-7}
$a_3m_s$(MeV) & model &$S$&$Y_\nu=0$ & $Y_{Le}=0.4$&$Y_\nu=0$ & $Y_{Le}=0.4$\\
\hline
$-134$&chiral&0& 4.15 & 6.38 & 9.46 &  ** \\ 
&mes. exch.&0  & 4.54 & 7.29  &  **  & ** \\ \hline
$-222$&chiral&0& 3.15 & 4.35 & 4.22 & ? \\ 
&mes. exch.&0  & 3.59 & 5.46 &   **  & ** \\ \hline 
$-310$&chiral&0& 2.49 & 3.15 & 2.73 & ? \\ 
&mes. exch.&0  & 2.86 & 4.03 & 3.76 & ? \\ \hline
\end{tabular}
\begin{quote}
The symbol ** indicates that no condensation takes place for this set of 
couplings.
\end{quote}
\end{center}

\newpage
\vskip1cm
\centerline{Table 14}
\centerline{Properties of a star, both without
and with trapped neutrinos ($Y_{Le}=0.4$), which contains neutrons, protons,}
\centerline{and kaon condensates in beta equilibrium in the 
relativistic mean field GM model.}
\begin{center}
\begin{tabular}{ccc|cccc|cccc}\hline \hline
& & & \multicolumn{4}{c|}{$Y_\nu=0$} & \multicolumn{4}{c}{$Y_{Le}=0.4$}\\ 
\cline{4-11}
Model& $a_3m_s$ &$S$&
${\displaystyle\frac{\strut M_{\rm max}}{M_{\odot}}}$&$R$&
${\displaystyle \frac{n_c}{n_0}}$& $I$ &
${\displaystyle\frac{\strut M_{\rm max}}{M_{\odot}}}$&$R$&
${\displaystyle \frac{n_c}{n_0}}$& $I$ \\
&  MeV & &&(km)& & $M_\odot~{\rm km}^2$ 
& &(km)& & $M_\odot~{\rm km}^2$ \\ \hline 
   & $-134$  & 0& 1.911 & 11.39 & 6.38 & 98.8 & 1.934 & 10.56 & 7.05 & 90.6 \\
chiral& $-222$&0& 1.781 &  9.89 & 8.62 & 69.4 & 1.902 & 10.60 & 7.05 & 88.1 \\
   & $-310$&0& 1.779 &  9.02 & 9.78 & 63.4 & 1.838 &  9.94 & 8.05 & 75.2 \\
\hline 
mes.&$-134$& 0& 1.950 & 11.36 & 6.38 & 102.2 & 1.935 & 10.53 & 7.05 & 90.3 \\
exch.&&1      & 1.971 & 11.50 & 6.25 & 105.1 & 1.945 & 10.74 & 6.84 & 92.9 \\
&&2           & 2.015 & 11.79 & 5.99 & 111.3 & 1.977 & 11.22 & 6.37 & 100.1 \\
\hline
mes.&$-222$& 0& 1.832 & 10.65 & 7.50 &  81.6 & 1.928 & 10.65 & 6.88 & 91.4 \\
exch.&&1     & 1.866 & 11.22 & 6.72 &  90.9 & 1.939 & 10.78 & 6.84 & 92.9 \\
&&2          & 1.946 & 12.01 & 5.80 & 107.8 & 1.970 & 11.23 & 6.38 & 99.8 \\
\hline \hline
\end{tabular}

\end{center}

\newpage
\centerline{Table 15}
\begin{quote}
Ratios of hyperon-meson to nucleon-meson coupling constants,
$x_{iH}=g_{iH}/g_{iN}$, where $i=\sigma,\ \omega$ or $\rho$, and $H$ is a
hyperon species.
\end{quote}
\begin{center}
\begin{tabular}{c|ccc|ccc|ccc} \hline \hline
Case & $x_{\sigma \Lambda}$ & $x_{\omega \Lambda}$ & $x_{\rho \Lambda}$ & 
$x_{\sigma \Sigma}$ & $x_{\omega \Sigma}$ & $x_{\rho \Sigma}$ & 
$x_{\sigma \Xi}$ & $x_{\omega \Xi}$ & $x_{\rho \Xi}$ \\ \hline
1 & 0.60 & 0.65 & 0.60 & 0.54 & 0.67 & 0.67 & 0.60 & 0.65 & 0.60\\
2 & 0.60 & 0.65 & 0.60 & 0.77 & 1.00 & 0.67 & 0.60 & 0.65 & 0.60\\
3 & 0.60 & 0.65 & 0.60 & 0.77 & 1.00 & 0.67 & 0.77 & 1.00 & 0.67\\
\hline \hline
\end{tabular}
\end{center}

\newpage
\centerline {Table 16}
\begin{quote} 
{Maximum masses of stars, $M_{\rm max}/M_{\odot}$, with  baryonic matter that 
undergoes a phase
transition to quark matter without ($Y_\nu=0$) and with ($Y_{Le}=0.4$) trapped
neutrinos.  Results are for a mean field model of baryons and a bag model of
quarks. $B$ denotes the bag pressure in the quark EOS. }
\end{quote}
\begin{center}
\begin{tabular}{c|c|c|c|c} \hline \hline
$ \begin{array}{c} \, ~ \, \\ \, ~ \, \end{array}$ 
& \multicolumn{2}{c|}{Without hyperons} 
& \multicolumn{2}{c}{With hyperons} \\ \hline
$ \begin{array}{c} B \\ ({\rm MeV~fm}^{-3}) \end{array} $ 
& $Y_{\nu}=0$ & $Y_{Le}=0.4 $ & $Y_{\nu}=0$ & $Y_{Le}=0.4$ \\ \hline
136.6    & 1.440 & 1.610 & 1.434 & 1.595 \\
150      & 1.444 & 1.616 & 1.436 & 1.597 \\
200      & 1.493 & 1.632 & 1.471 & 1.597 \\
250      & 1.562 & 1.640 & 1.506 & 1.597 \\ \hline
${\rm No~~quarks}$ & 1.711 & 1.645 & 1.516 & 1.597 \\ \hline \hline
\end{tabular}
\end{center}

\newpage

\newpage
\centerline {Table 17}
\vskip 2cm
\centerline {\bf SUPERNOVA NEUTRINO TELESCOPE CHARACTERISTICS }
\begin{center}
\begin{tabular}{lcccc} \hline \hline
Detector & Total~mass~(tonnes) & Composition & Threshold & \#~Events \\
         & (Fiducial~Mass~(tonnes)) & & (MeV) & at 10~kpc \\ \hline
{\bf \v{C}ERENKOV:}     &   &   &   &   \\
~~{\bf KIII} &  3000    &   H$_2$O  &  5  & 370 \\
           &  (2140)  &           &     &      \\
~~Super Kamiokande & 40,000  & H$_2$O  &  5  & 5500 \\
                 & (32,000)  &           &     &      \\
~~SNO        & 1600/1000  & H$_2$O/D$_2$O  & 5  &  780  \\ \hline
{\bf SCINTILLATION}: & & & & \\
~~{\bf LVD}   & 1800  &   Kerosene  &  5--7  &    375  \\
            & (1200)  &  &  &  \\
~~{\bf MACRO}  &  1000  &  ``CH$_2"$   &  10  &  240 \\
~~Baksan   &  330   &  ``White~Spirits" &  10 & 70 \\
           & (200)  &  ``CH$_2"$   &  &  \\
~~LSND & 200  & ``CH$_2"$   &  5  &  70  \\
~~Borexino &  300  &   (BO)$_3$(OCH$_3)_3$  &$\sim 0.2$ &  200 \\
~~Caltech  &  1000  & -- &  2.8  &  290 \\ \hline
{\bf DRIFT CHAMBER:}  &   &   &   &   \\
~~ICARUS  &  3600  &  $^{40}$Ar  &  5  &  120 \\ \hline
{\bf RADIOCHEMICAL:}  &   &   &   &   \\
~~Homestake~ $^{37}$Cl   &   610  & C$_2$Cl$_4$  &  0.814  &  4 \\
~~Homestake~ $^{127}$I &   -- & NaI &  0.664  &  25 \\
~~Baksan$^{37}$Cl   &   3000 & C$_2$Cl$_4$  &  0.814  &  22 \\ \hline
{\bf EXTRAGALACTIC:}  &   &   &   &   \\
~~SNBO  &  100,000  & CaCO$_3$  &  --  & 10,000 \\
~~JULIA &   40,000  & H$_2$O  &  --  & 10,000 \\ \hline \hline
\end{tabular}
\end{center}

\newpage

\section{Figure Captions}

\noindent Fig. 1. The main stages of evolution of a neutron star. Numbers in
parentheses refer to the stages discussed in the text. \\ 

\noindent Fig. 2. Results (left panels for BPAL EOS and right panels for
SL interactions) for pure neutron matter. Top panels show the neutron 
effective mass ratio from Eq.~(\ref{emasbpals}) and Eq.~(\ref{skmstar}) versus
the density ratio  $u=n/n_0$, where $n_0=0.16~{\rm fm}^{-3}$ is the equilibrium
nuclear density. The center panels show isentropic pressures, and the bottom
panels show star masses versus central density ratio at fixed entropy per 
baryon.  \\

\noindent Fig. 3.  Results (left panels for BPAL EOS and right panels for
SL interactions) for matter in beta equilibrium among $n,~p,~e^-$, 
and $\mu^-$, at an entropy per baryon $S=1$.  Top panels:  Individual
concentrations $Y_i=n_i/n_b$, where $i=n,~p,~e^-~{\rm and}~\mu^-$. Center
panels: The electron chemical potential  $\mu_e=\mu_\mu=\mu_n-\mu_p$.   Bottom
panels: Individual contributions to the entropy per baryon. \\

\noindent Fig. 4. Results (left panels for BPAL EOS and right panels for
SL interactions) for matter in beta equilibrium among $n,~p,~e^-$
and $\mu^-$. (See caption to Fig.~2 for
further details. Proton effective masses are also shown here.) \\

\noindent Fig. 5. The moment of inertia, $I$, as a function of density
(left panel) and baryonic mass, $M_B$ (right panel). The BPAL22 equation of 
state is employed for fixed values of the entropy per baryon. The full dots 
on the curves indicate the maximum gravitational mass. \\

\noindent Fig. 6. Results for the BPAL model with trapped neutrinos at 
entropy per baryon $S=1$. The upper panel shows individual concentrations, 
the center panel gives the leptonic chemical potentials, with 
$\mu=\mu_e-\mu_{\nu_e}$, and the lower panel separates the nucleon and lepton 
contributions to the entropy per baryon.\\ 

\noindent Fig. 7. Stellar temperature, $T$, as a function of the density 
ratio $u$ for the MRHA model with \mum=1.25. The full curves are for
neutrino free matter, and the dotted curves refer to matter with trapped 
neutrinos. In the upper (lower) panel, the baryons are nucleons without 
(with) hyperons. \\

\noindent Fig. 8.  Results for matter in beta equilibrium among $n,~p,~e^-$, 
and $\mu^-$ in the MRHA model, with \mum=1.25, at an entropy per baryon 
$S=1$. (See caption to Fig.~3 for further details.) \\  

\noindent Fig. 9.  Top panel: Nucleon Landau effective mass ratios versus 
density
ratio for the MRHA model (with \mum=1.25).  Middle panel: Isentropic
pressures.  Bottom panel: Star mass versus central density ratio at fixed
entropy per baryon. \\                                 

\noindent Fig. 10. Results for matter in beta equilibrium among $n,~p,~e^-$,
$\mu^-$, and trapped neutrinos, in the MRHA model, with \mum=1.25, at an 
entropy per baryon $S=1$. Shown are the individual concentrations (top panel), 
the leptonic chemical potentials (middle panel), and the baryonic and leptonic
contributions to the entropy as a function of density. \\ 
                  
\noindent Fig. 11. Relative fractions and the electron chemical potential for
beta-equilibrated matter containing nucleons, hyperons, electrons, and muons
in the MRHA model (\mum=1.25) at zero temperature.\\

\noindent Fig. 12. Results for matter in beta equilibrium among nucleons, 
hyperons, electrons, and muons in the MRHA model with \mum=1.25 at an 
entropy per baryon $S=1$. Shown are the individual concentrations (top panel), 
the electron chemical potential (middle panel), and the baryonic and leptonic
contributions to the entropy as a function of density. \\ 

\noindent Fig. 13.  Relative fractions and leptonic chemical potentials for
beta-equilibrated matter containing nucleons, hyperons, electrons, muons, 
and trapped neutrinos in the MRHA model (\mum=1.25) at zero temperature.
Here $\mu=\mu_e-\mu_{\nu_e}$.\\ 
                                                        
\noindent Fig. 14. Results for beta-equilibrated matter containing nucleons, 
hyperons, electrons, muons and trapped neutrinos in the MRHA model 
(\mum=1.25) at an entropy per baryon $S=1$.
Shown are the individual concentrations (top panel), the
leptonic chemical potentials (middle panel), and the baryonic and leptonic
contributions to the entropy as a function of density. \\ 

\noindent Fig. 15. Panel (1): Ratio of gravitational mass $M_G$ to baryonic mass
$M_B$ as a function of $M_B$ for the GM model. Solid lines are for lepton-rich 
matter, dashed lines for neutrino-poor matter. A dot at the end of a curve 
indicates matter with hyperons, a star indicates matter without hyperons. For 
the neutrino-poor cases, the entropy per baryon is given next to the curves. 
Panel (2): Gravitational mass as a function of baryonic mass. The symbols
are the same as in panel (1). \\

\noindent Fig. 16. Maximum neutron star mass as a function of 
electron-neutrino fraction $Y_{\nu_e}$ in the GM model for
matter with and without hyperons, labeled by npH and np, respectively. \\ 

\noindent Fig. 17. Illustrative plot of the kaon energies $\omega^{\pm}$ 
in the meson exchange model as a function of the density ratio $u$. Here 
the baryons are nucleons. The chemical potential $\mu$ is also shown;
the dashed portion of the curve indicates the behavior when kaons are absent. \\

\noindent Fig. 18.  Neutrino-free  matter in beta equilibrium among
nucleons, (thermal) kaons, electrons and muons in the GM model, with the 
meson-exchange formalism as a function of
temperature, $T$. Results are shown for three different values of the 
kaon-nucleon sigma term $\Sigma^{KN}$. Bottom panel: Critical nucleon density 
ratio for the onset of
kaon condensation. Next to bottom panel: The electron chemical potential
$\mu_e=\mu_\mu=\mu_n-\mu_p=\mu_K$. Next to top panel: Thermal kaon to baryon
ratio at threshold, for kaon condensation. Top panel: Proton fraction 
at threshold.  \\ 

\noindent Fig. 19.  Results for neutrino-free  matter in beta equilibrium among
nucleons, kaons, electrons and muons  in  the GM model  for an entropy per
baryon $S=1$. The top panel shows the relative  concentrations. The center
panel  shows the electron chemical potential, and the bottom panel shows the 
contributions to the total entropy from the strongly interacting particles  and
the leptons, respectively.\\

\noindent Fig. 20.  Results for neutrino-trapped 
matter in beta equilibrium among
nucleons, (thermal) kaons, electrons, and muons in the GM model as a function of
temperature, $T$, with three different values of the kaon-nucleon sigma term
$\Sigma^{KN}$.   Bottom panel: Critical nucleon density ratio for the onset of
kaon condensation. Next to bottom panel: The electron chemical potential
$\mu_e=\mu_\mu=\mu_n-\mu_p=\mu_K$. Next to top panel: Thermal kaon to baryon
ratio. Top panel: Proton fraction at threshold, for kaon condensation.  \\ 

\noindent Fig. 21.  Results for neutrino-trapped 
matter in beta equilibrium among
nucleons, kaons, electrons, and muons  in  the GM model  for an entropy per
baryon $S=1$. The top panel shows the relative  concentrations. 
The center panel gives the leptonic chemical potentials, with 
$\mu=\mu_e-\mu_{\nu_e}$, 
and the bottom panel shows the 
contributions to the total entropy from the strongly interacting particles  and
the leptons, respectively.\\

\noindent Fig. 22.  Results for zero-temperature matter in beta equilibrium 
among nucleons, hyperons, kaons, electrons, and muons.
The chiral model with $a_3m_s = -222$ MeV is used in conjunction with
the mean field GM description of the baryons.
(a) Relative fractions $Y_i = n_i/(\sum_Bn_B)$.
(b) Baryon Dirac effective masses, the kaon chemical potential $\mu = \mu_e$, 
and the scalar field $\sigma$. (c) Baryon scalar densities. (d) Condensate
amplitude, $\theta$, in degrees. \\

\noindent Fig. 23. Particle fractions for model HS81 in conjunction with 
the kaon meson-exchange formalism for different choices of the
$\Sigma$ and $\Xi$ coupling constants. Panels (1), (2), and (3) correspond to
parameter sets 1, 2, and 3 of Table 15, respectively. \\

\noindent Fig. 24. Gravitational mass vs. baryonic mass for matter with and
without kaons in the lepton-rich ($Y_{L_e}=0.4$) 
and neutrino-poor stages ($Y_{\nu}=0$).
The chiral model with $a_3m_s = -222$ MeV is used in conjunction with
the mean field GM description for the nucleons.
The range in neutron star masses that is metastable during deleptonization
is indicated. \\

\noindent Fig. 25. Maximum neutron star mass as a function of 
electron-neutrino fraction $Y_{\nu_e}$ for
matter with and without kaons, labeled by np and npK, respectively. 
(See caption to Fig. 21 for further details.)\\

\noindent Fig. 26.  Individual concentrations for matter in beta equilibrium 
among nucleons, hyperons, quarks, electrons, and muons, 
 employing the mean field 
GM model in the baryon sector and a bag model for the quarks. 
The top panel shows the neutrino-free case and the bottom panel 
the results with trapped neutrinos.  The quark phase cavity pressure
$B=200~{\rm MeV~fm^{-3}}$. \\

\noindent Fig. 27.  Quark-hadron phase transition boundaries in
beta-equilibrated matter as a function of the bag pressure, $B$.
In the top panel, the hadrons are nucleons and, in the lower panel, 
nucleons and hyperons. The onset of a quark-hadron mixed phase occurs at a 
density ratio $u_1$, and a pure quark phase begins at $u_2$. Also shown is the 
central density ratio, $u_c$, of the maximum mass star.\\

\noindent Fig. 28. Binding energy versus baryon mass for nucleons-only matter
(np), matter with nucleons and hyperons (npH), and matter with nucleons and
kaons (npK).  The stars, dots, and triangles mark the maximum mass
configurations. The lower envelope is for an EOS that is causal above a
transition density of $n_t=0.3~{\rm fm}^{-3}$ and for the GM EOS below $n_t$. 
\\ 

\noindent Fig. 29.  Enclosed gravitational mass versus baryon mass.  Two 
generic compositional cases are displayed: nucleons-only matter and matter with
hyperons.  Solid curves are for neutrino-rich matter with $Y_{Le}=0.4$ at an
entropy per baryon $S=2$.   Dashed curves refer to cold catalyzed neutrino-free
matter.                           
\\               

\noindent Fig. 30.  Electron concentrations as a function of density for 
neutrino-free matter with various assumptions for the stellar composition, 
as indicated. The quark phase cavity pressure $B=200~{\rm
MeV~fm^{-3}}$.  Arrows indicate central densities of $1.44M_\odot$  stars.  
Differences of each curve from  $Y_{Le}=0.4$ show the net $Y_{\nu_e}$ lost at
each density during cooling.  \\

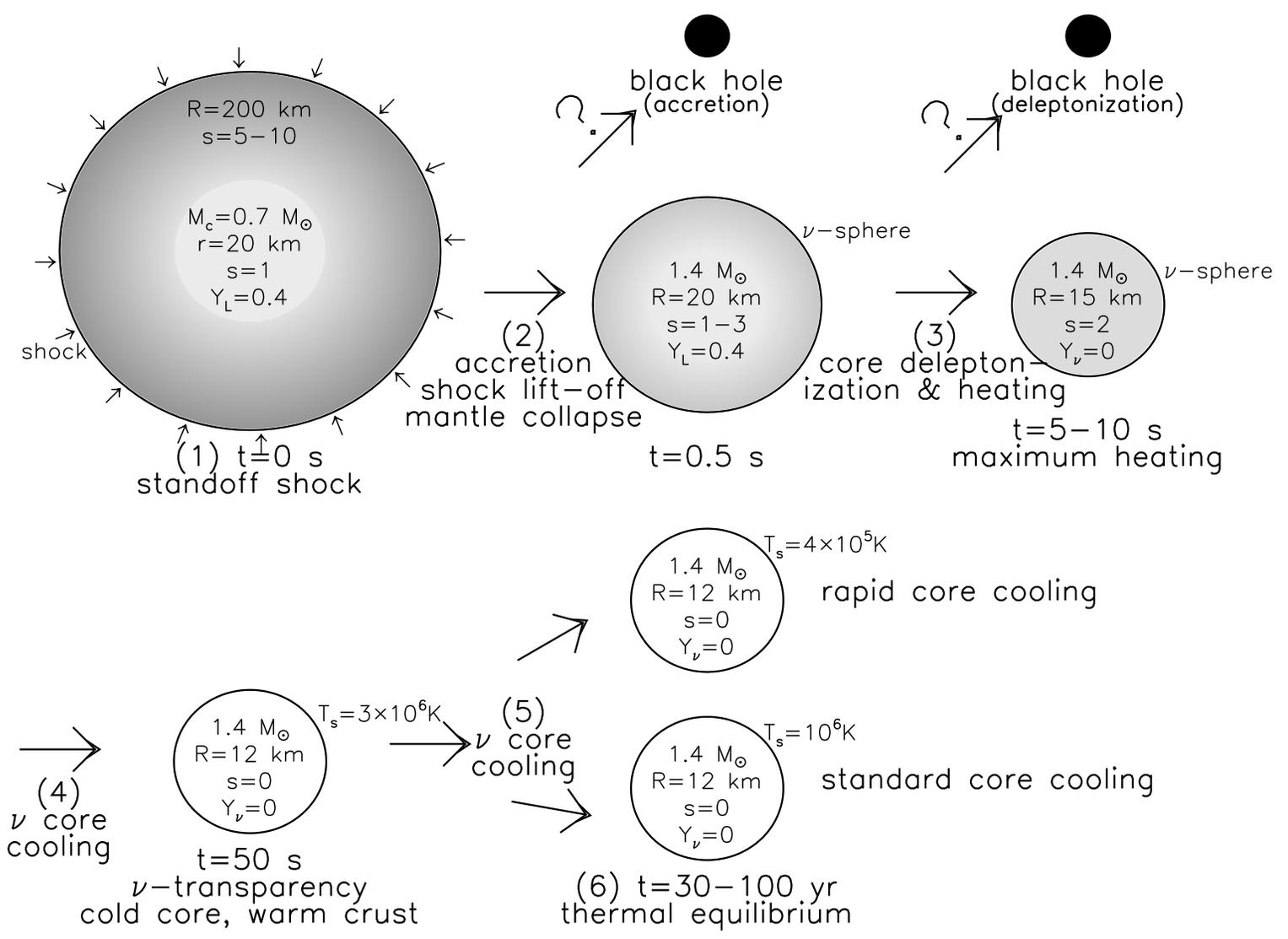
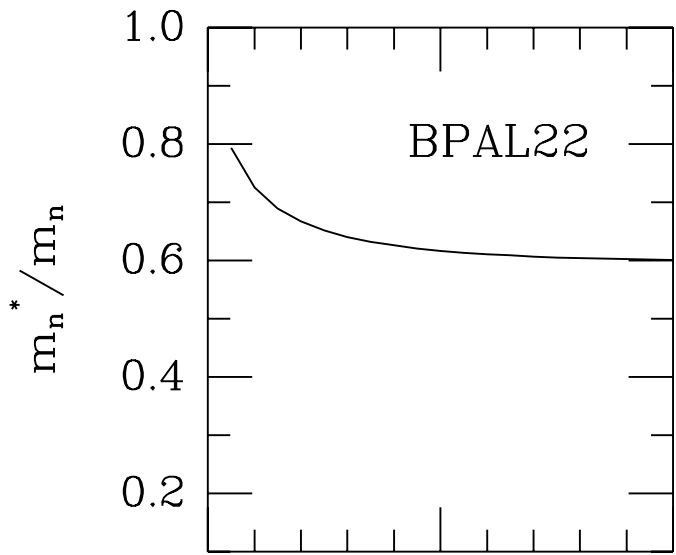
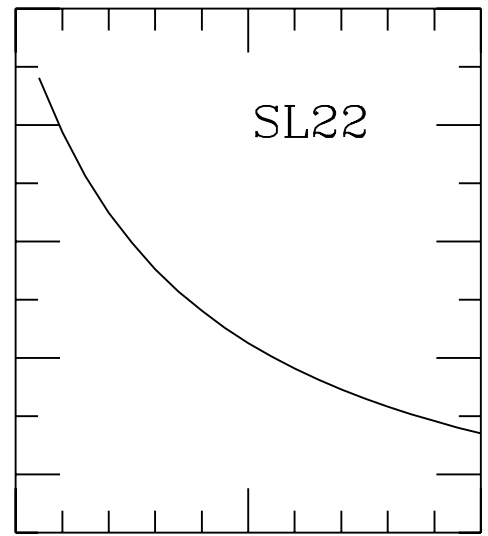
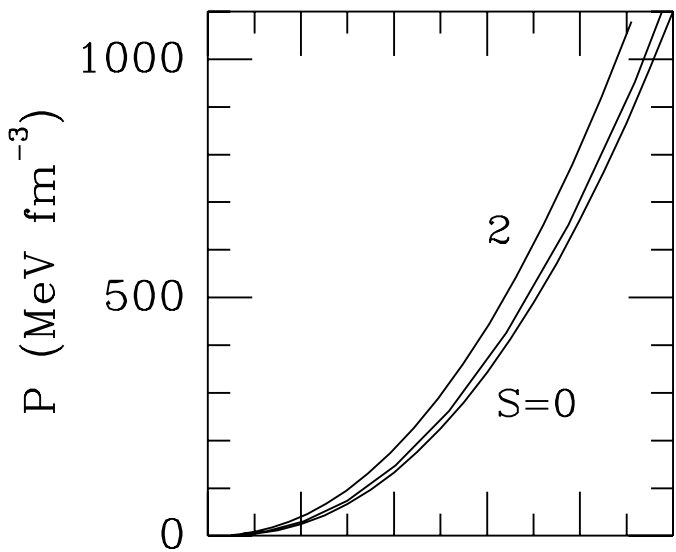
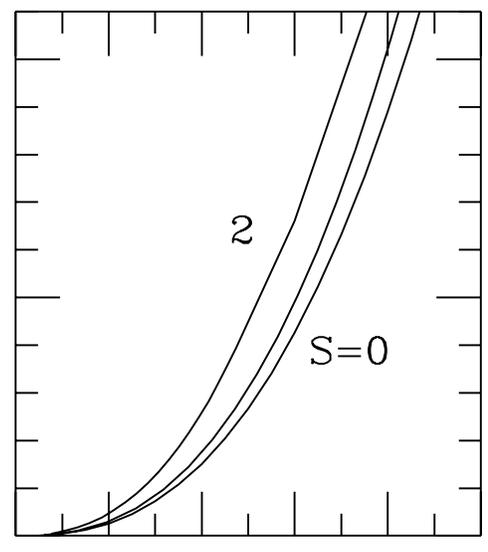
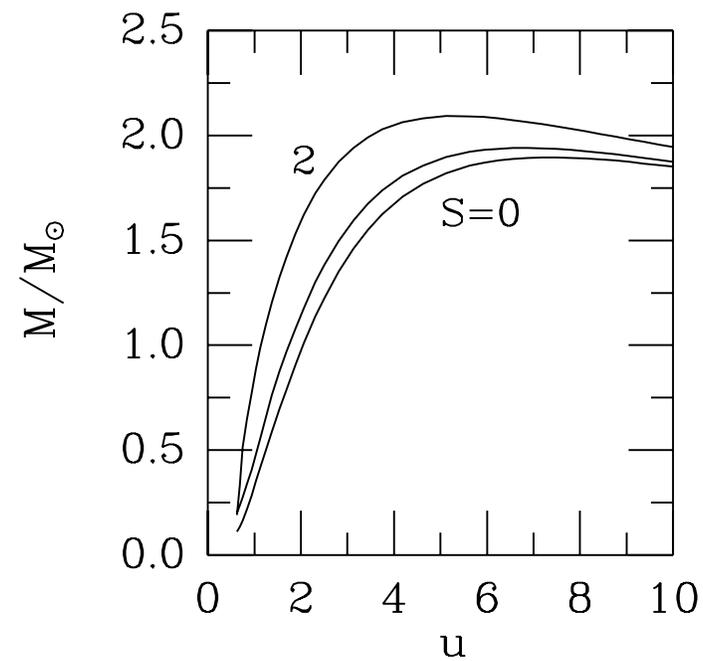
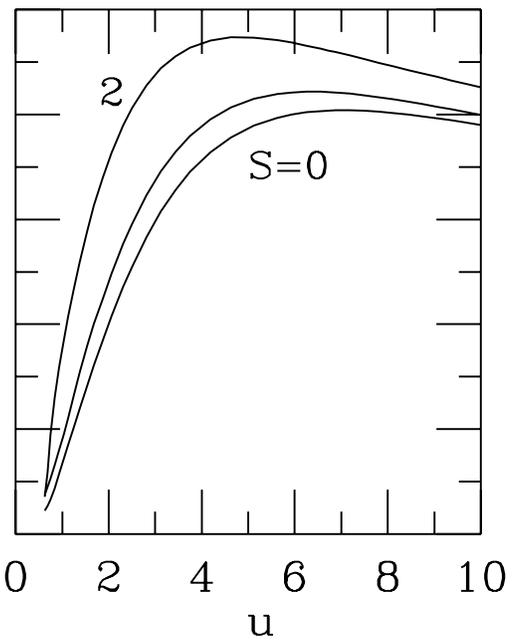
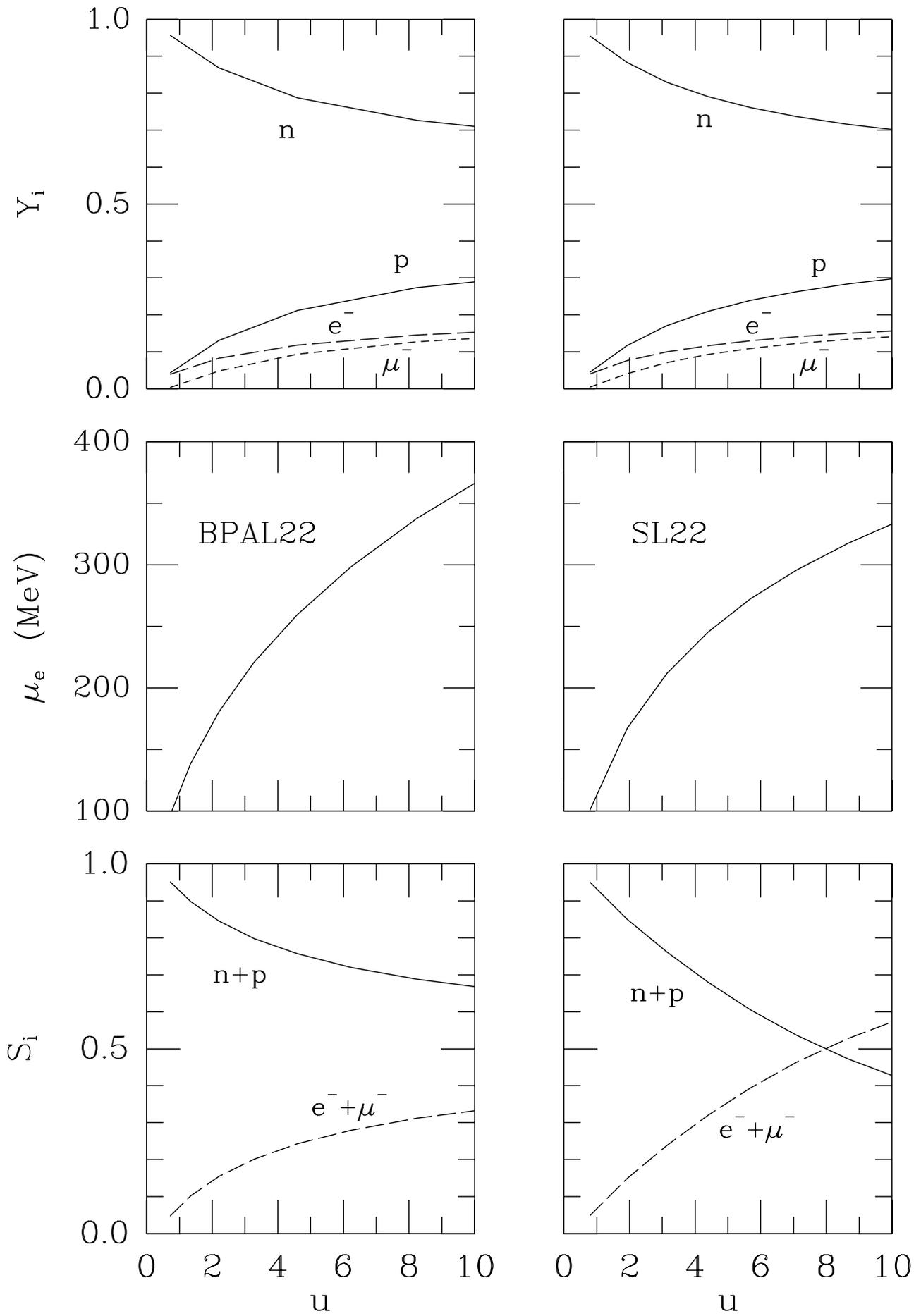
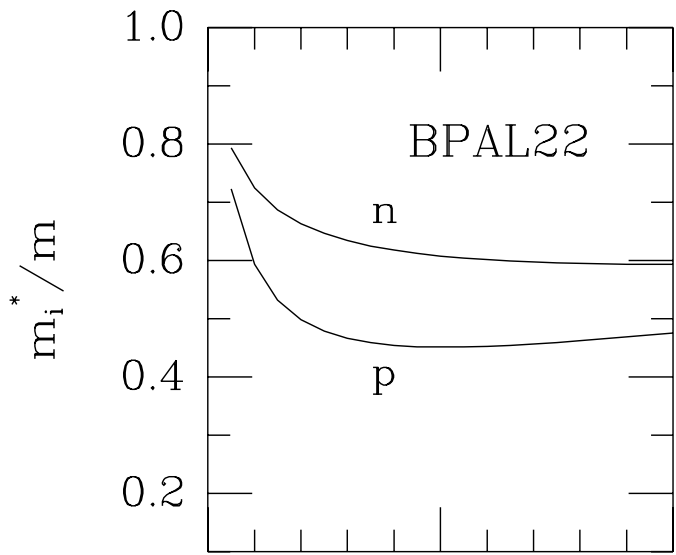
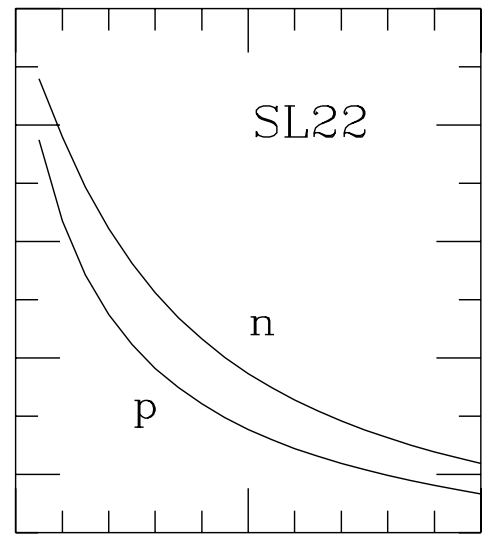
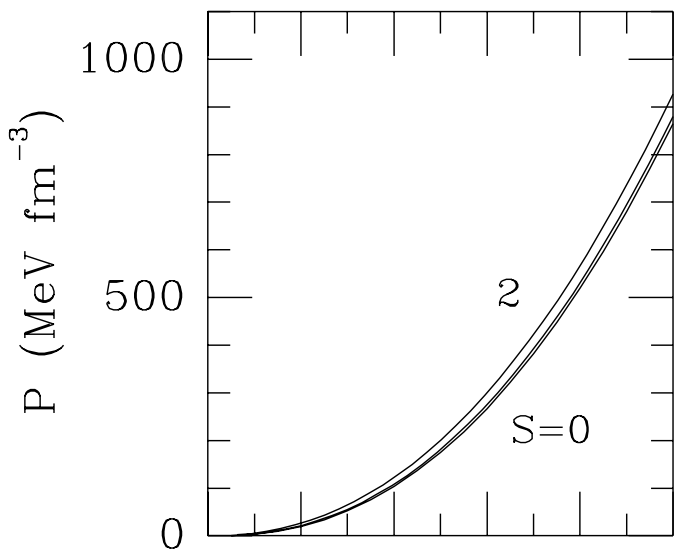
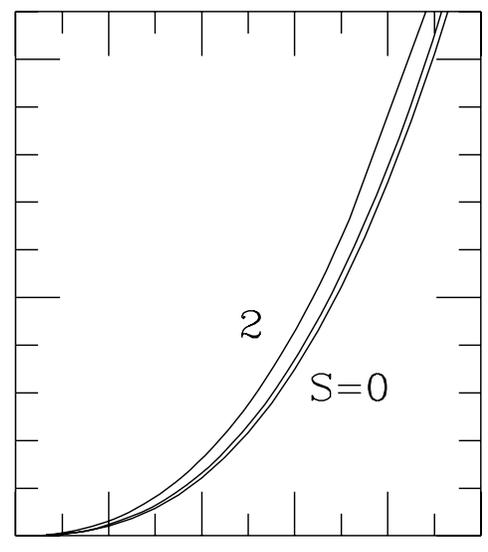
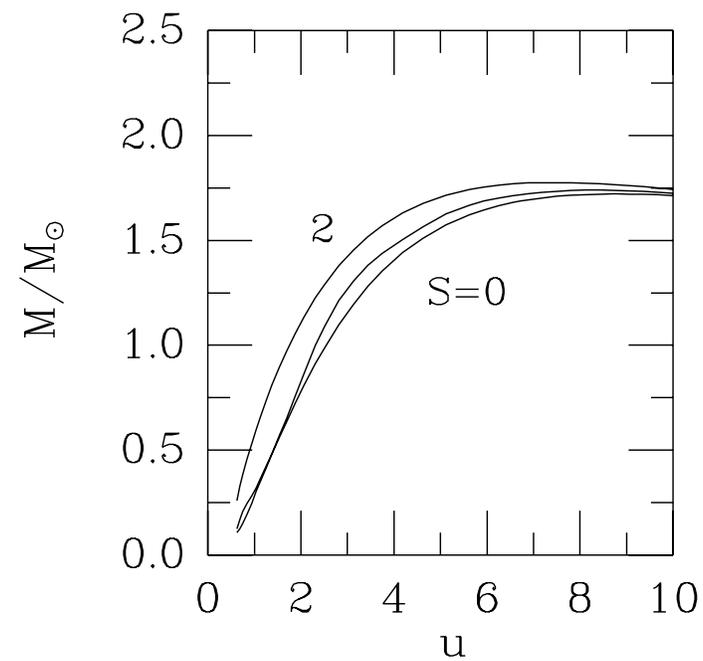
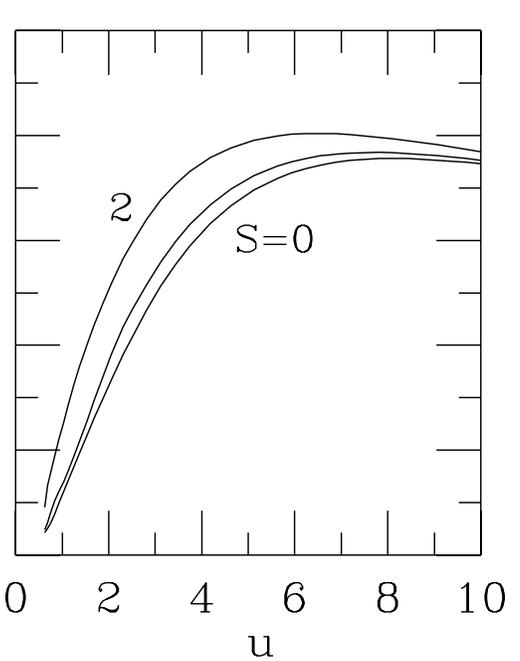
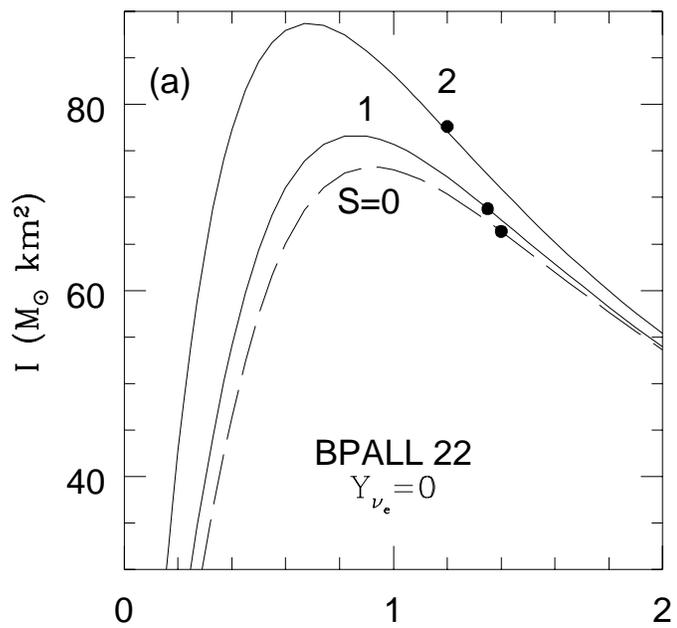
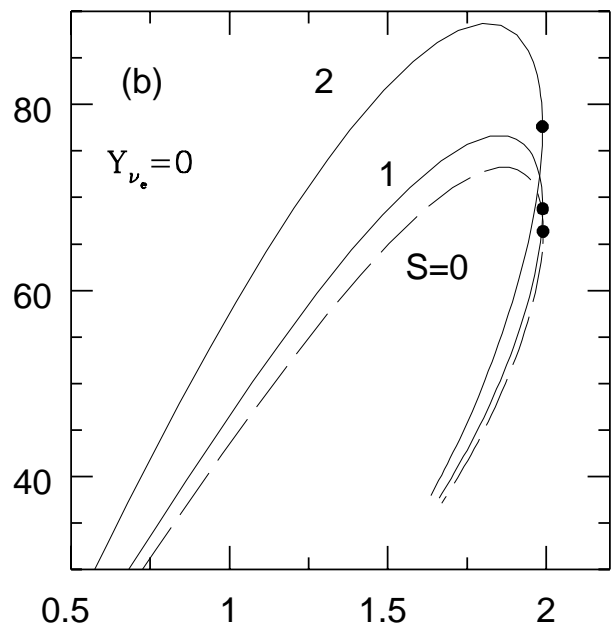
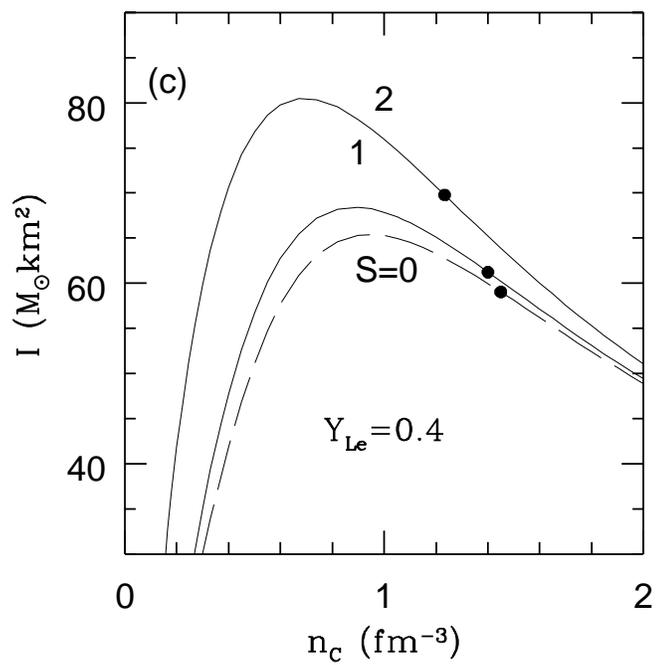
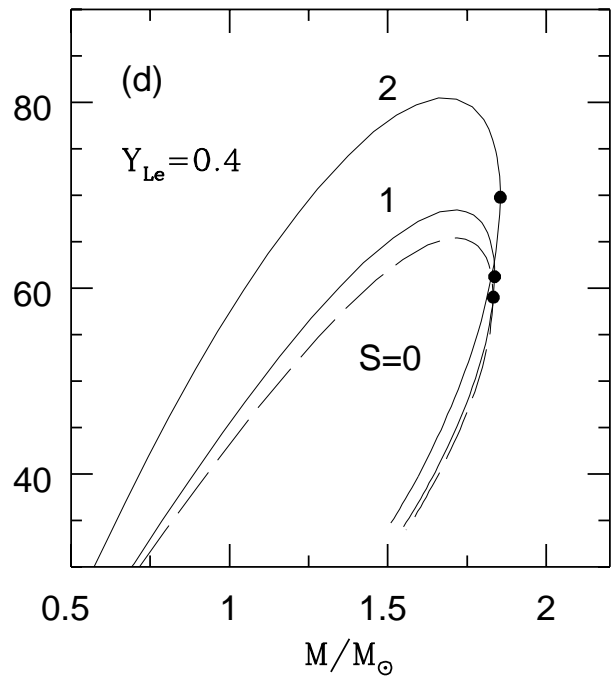
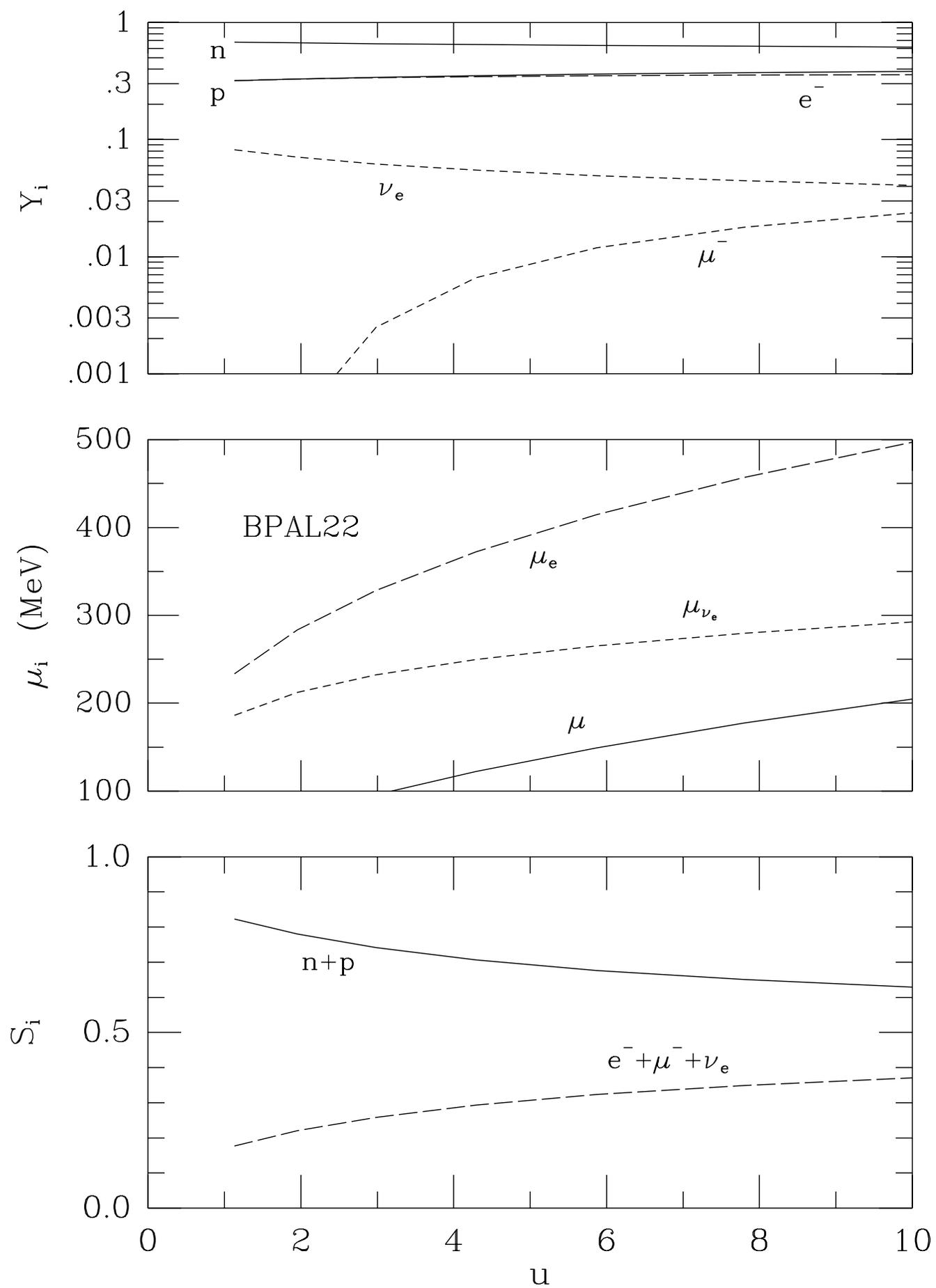
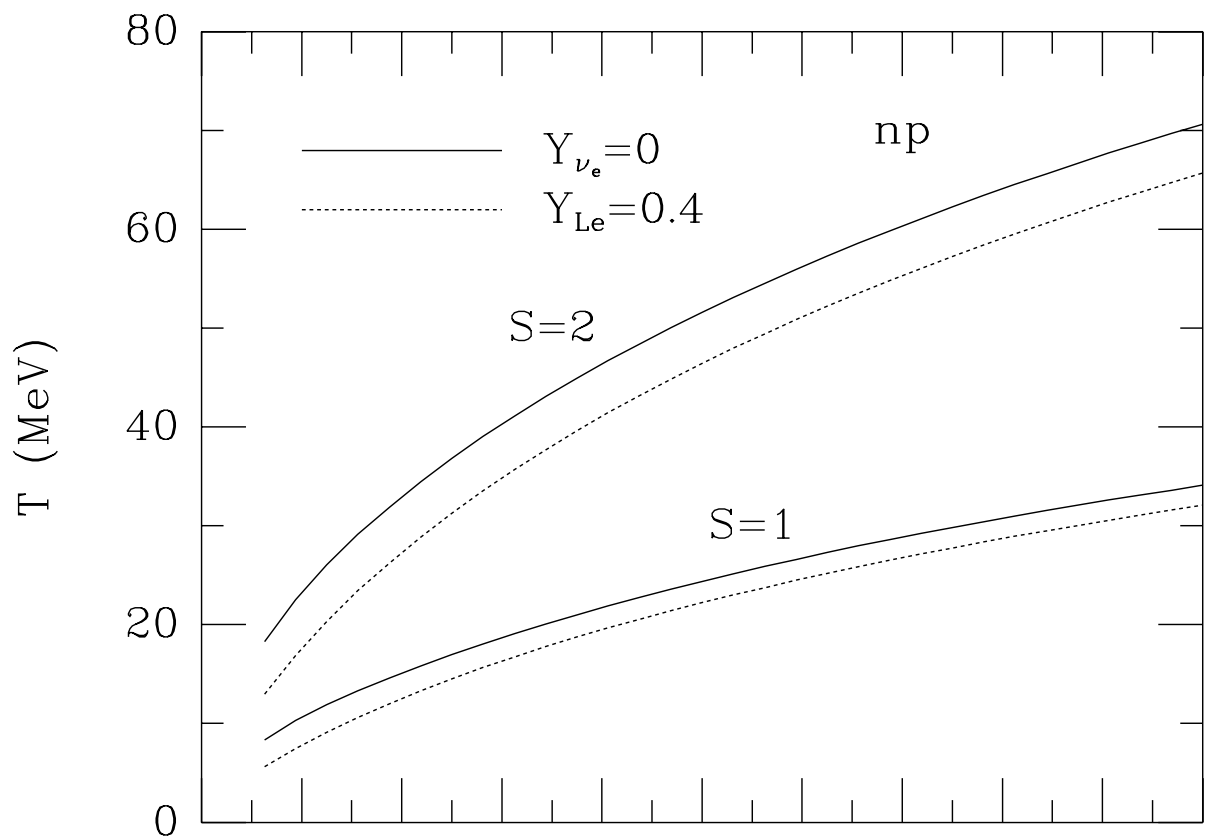
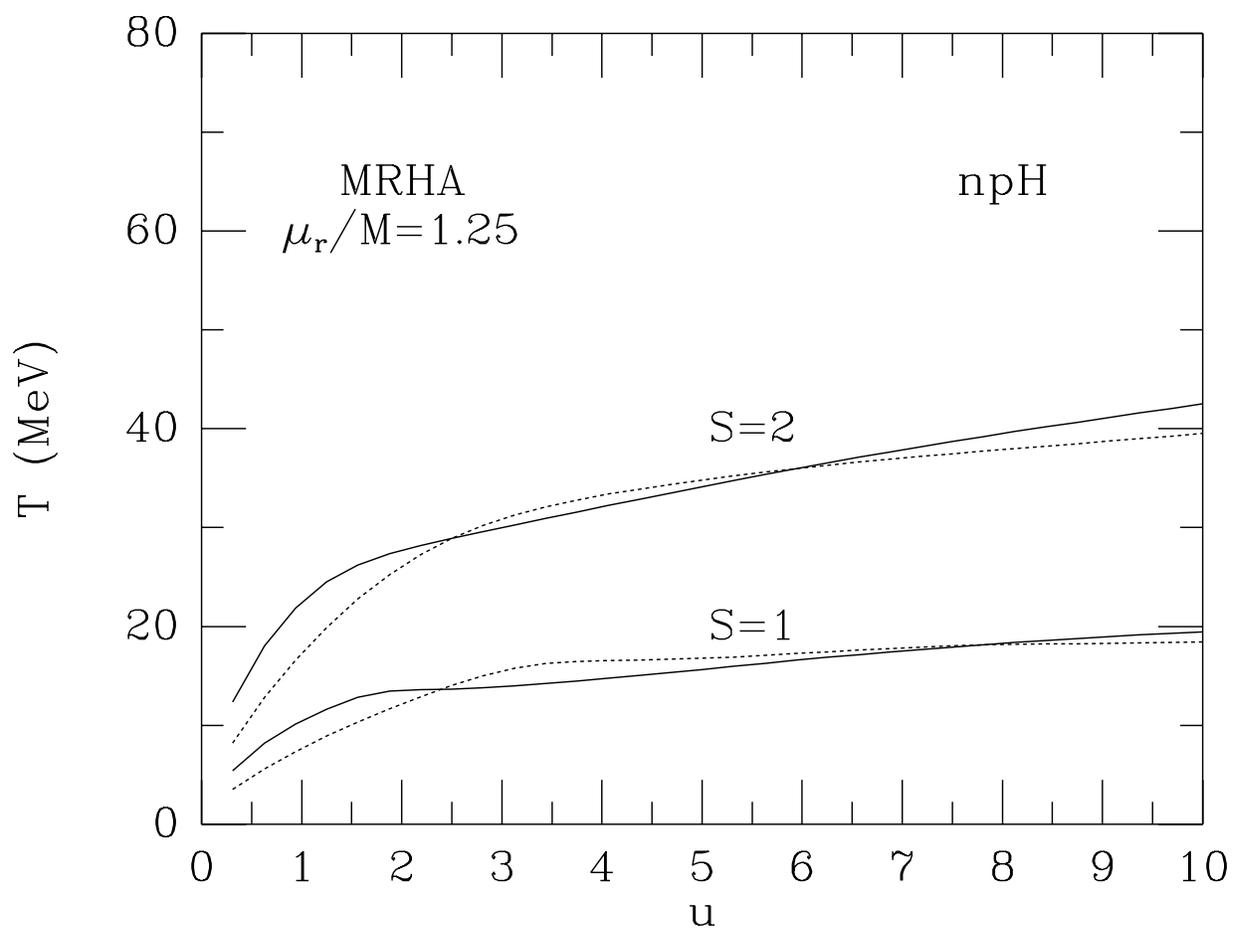
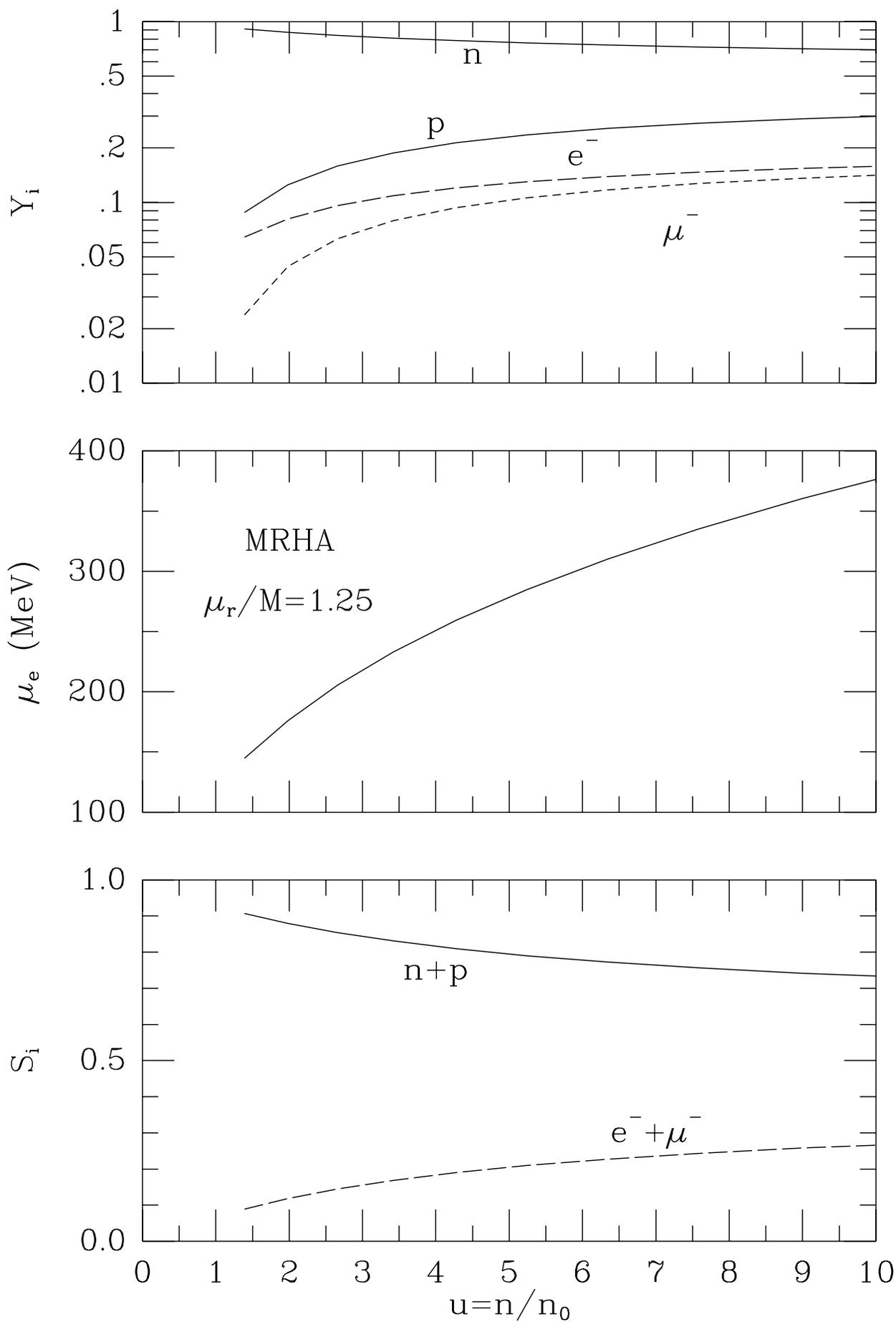
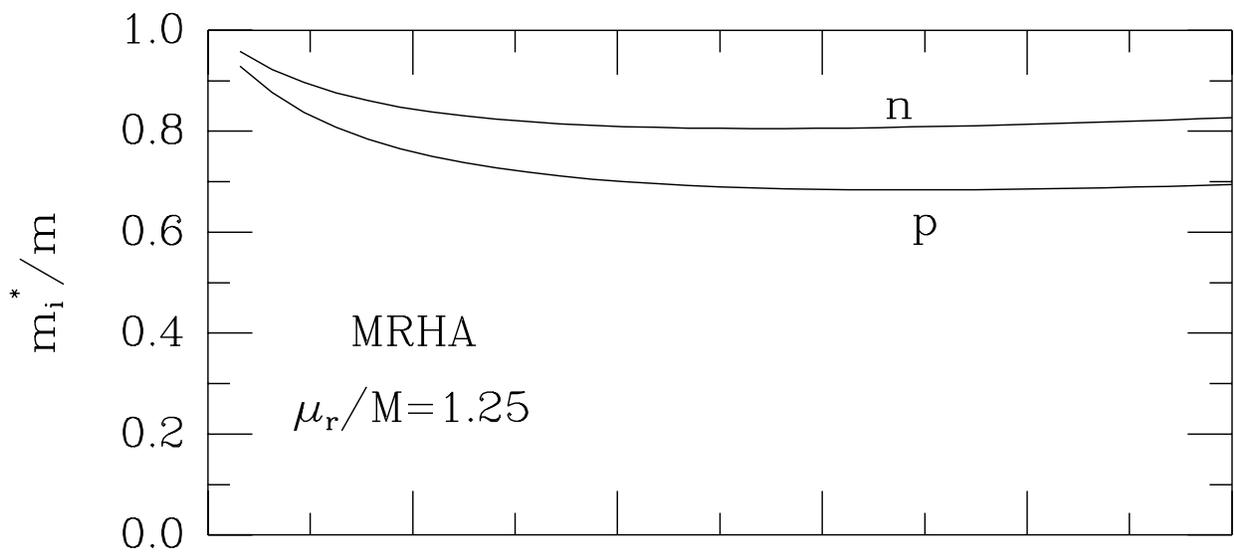
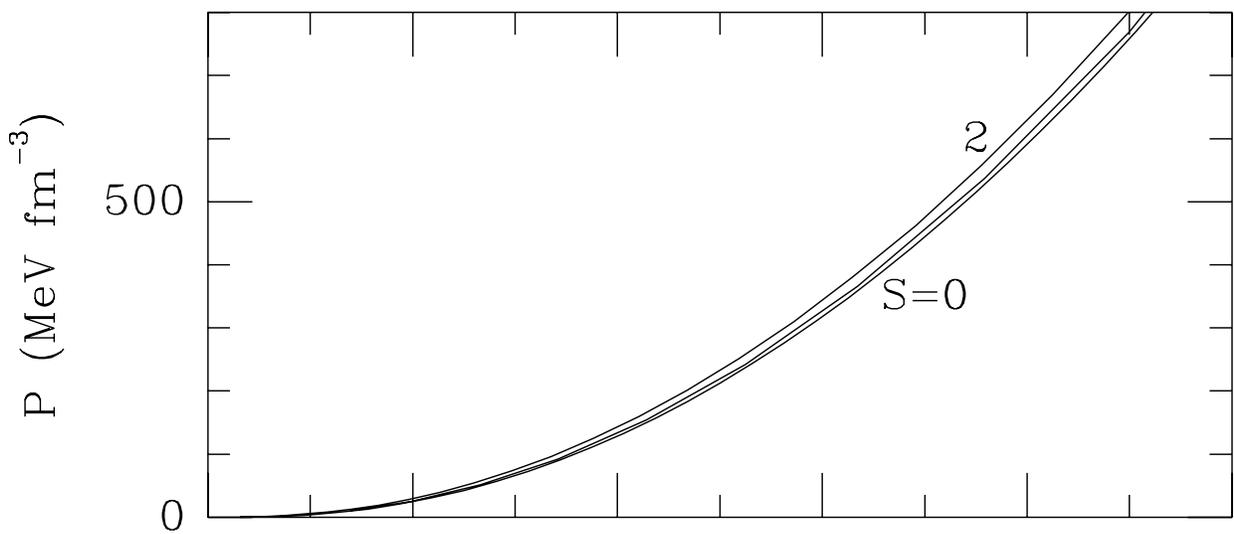
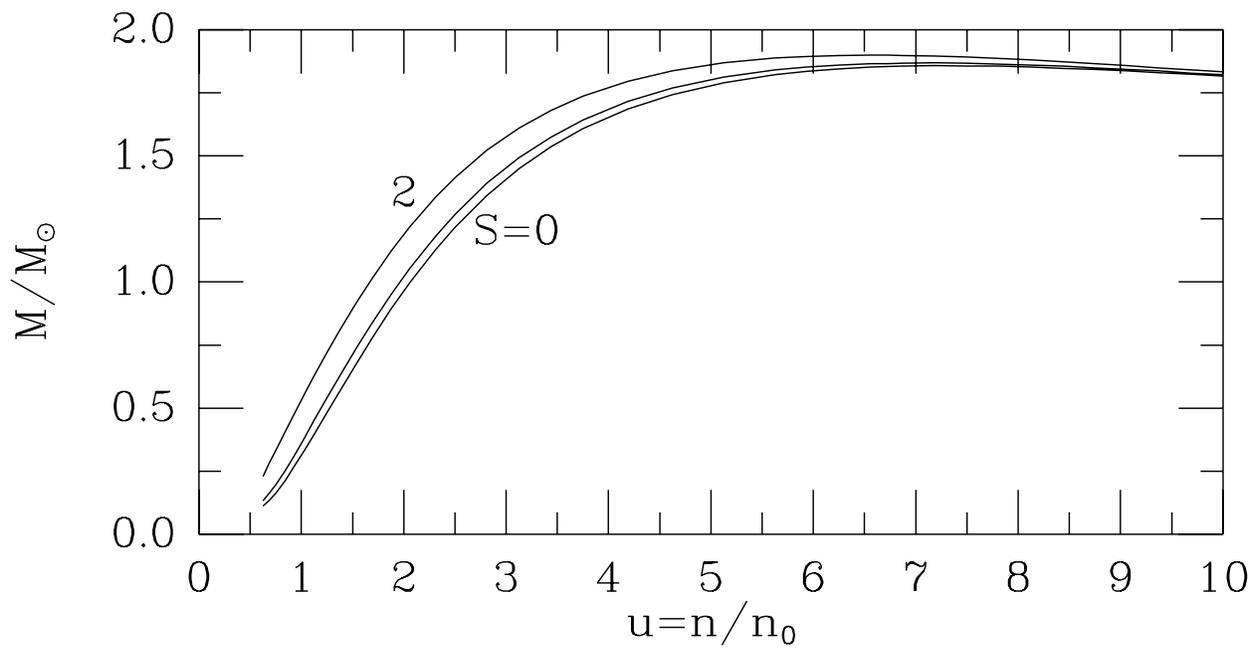
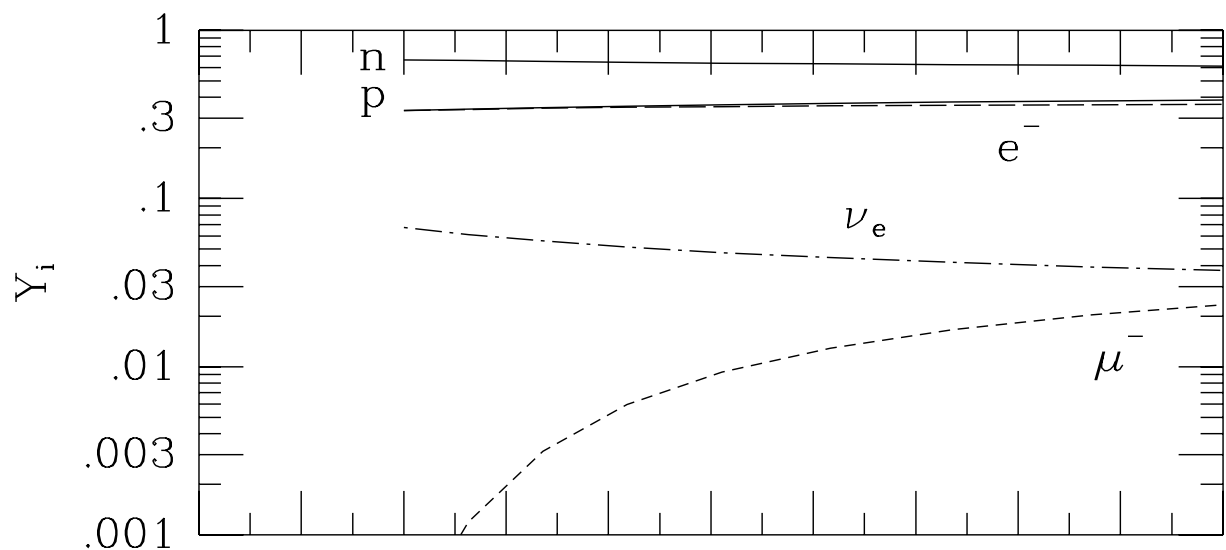
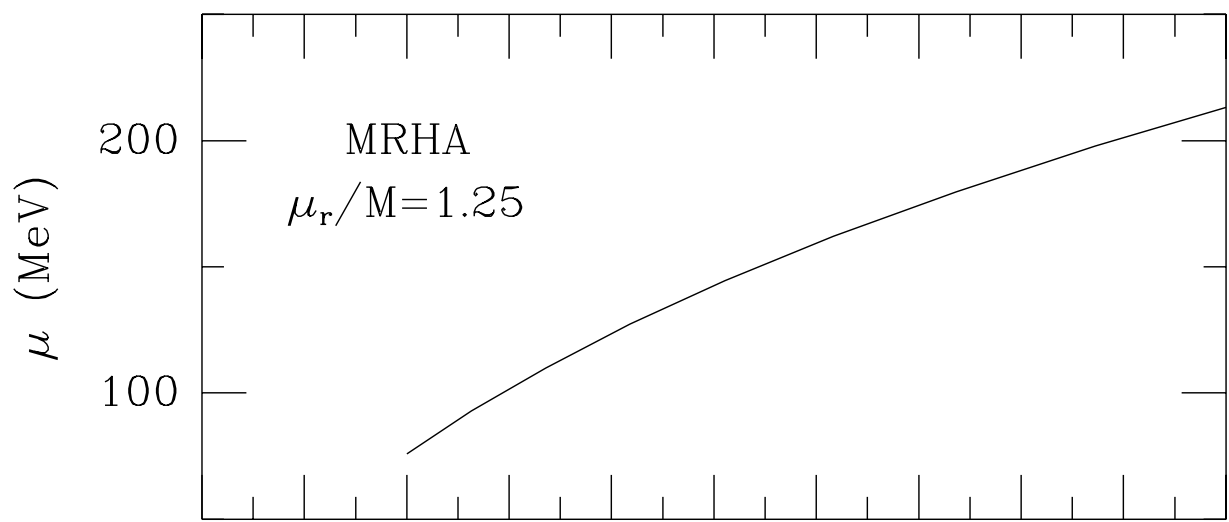
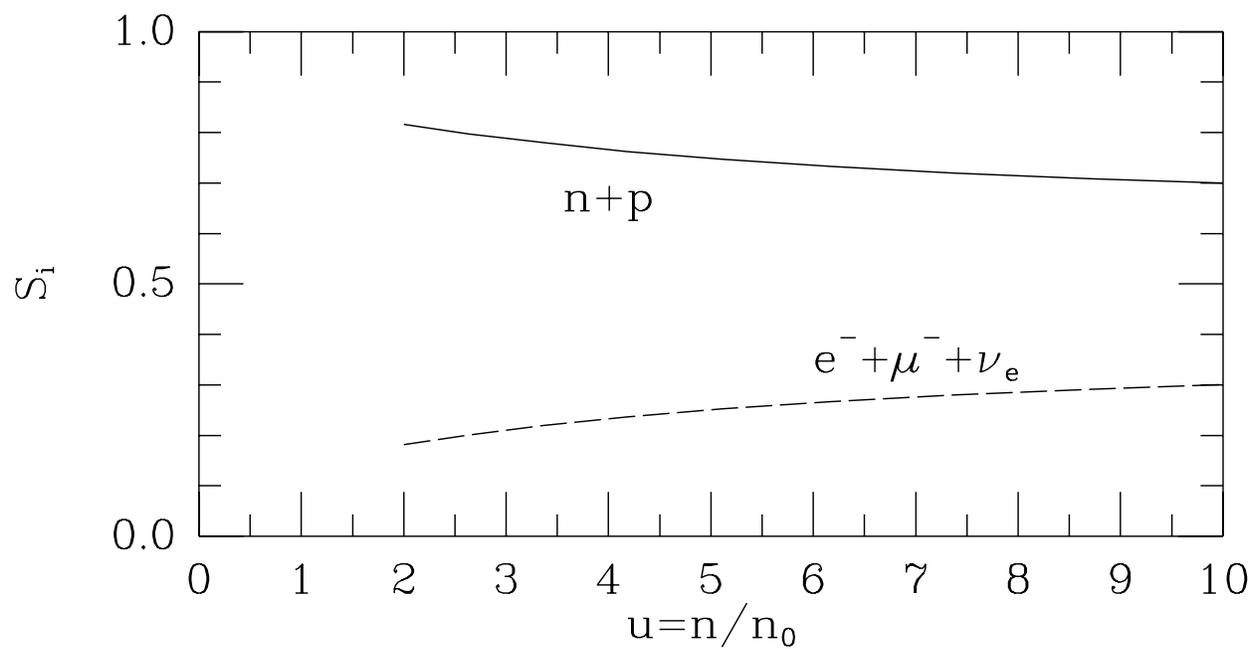
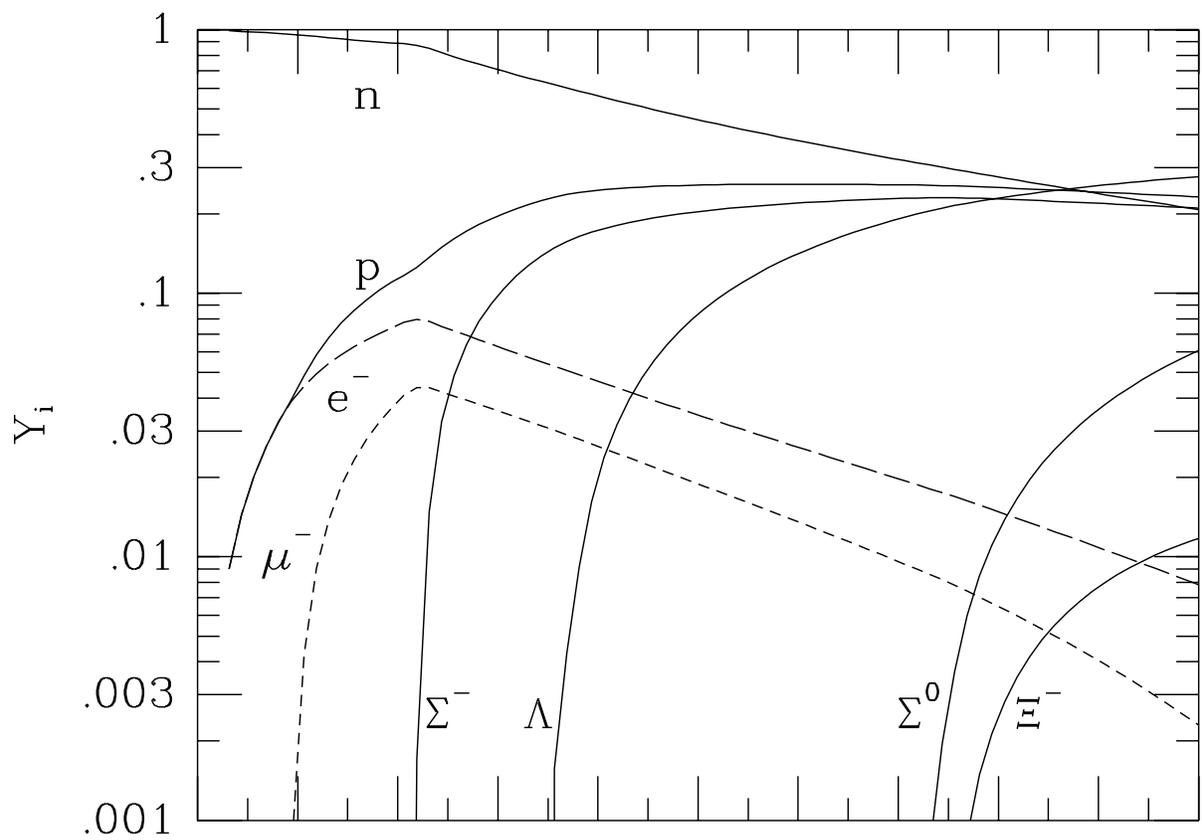
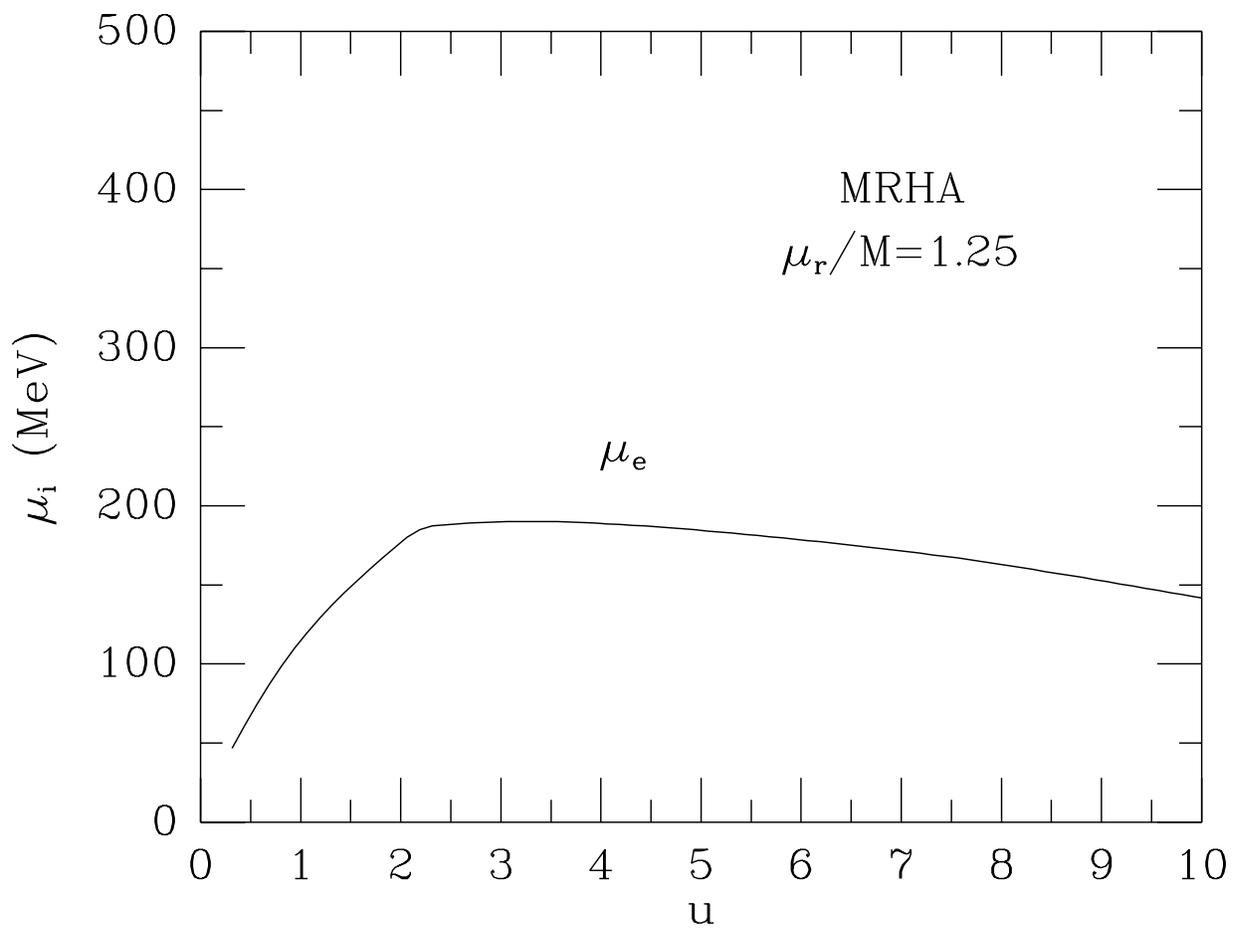
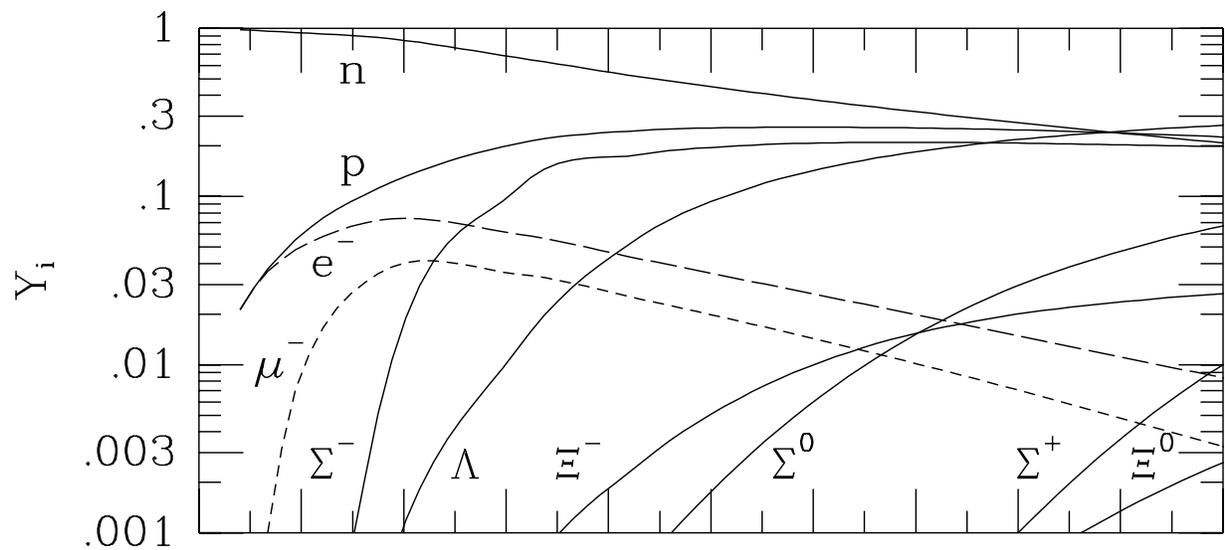
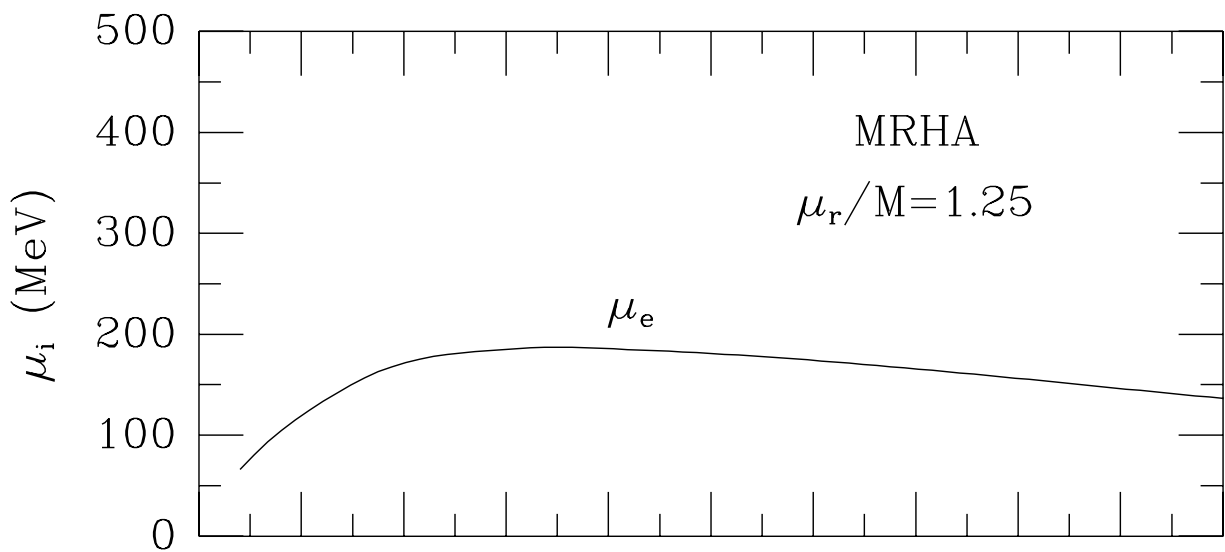
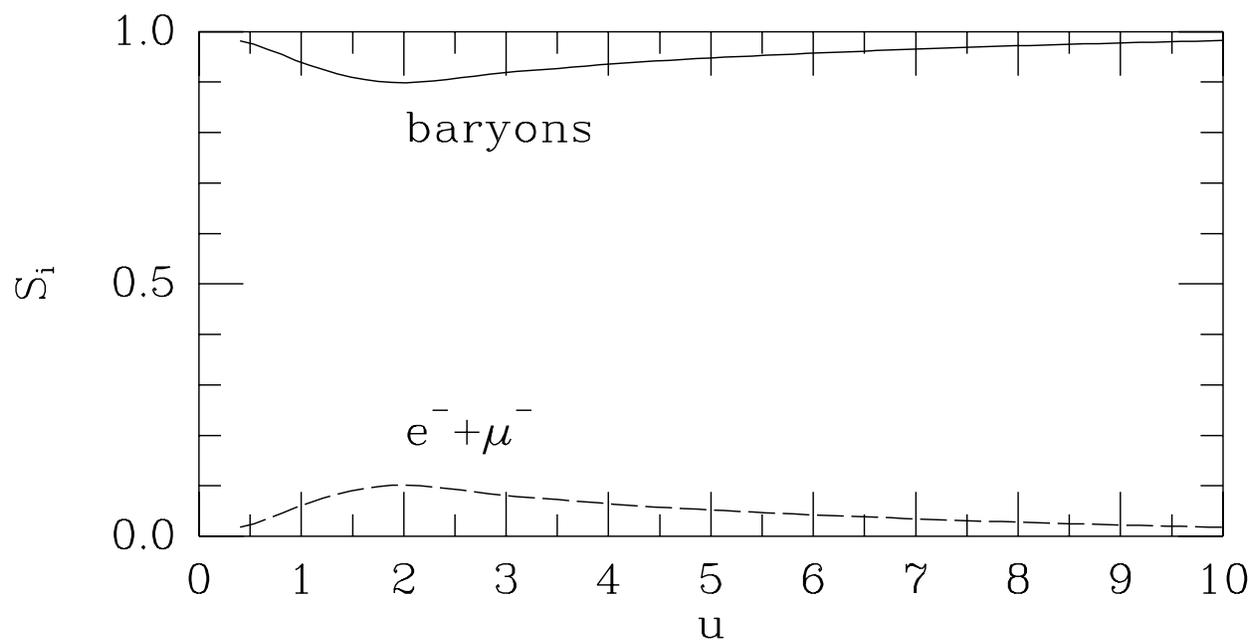
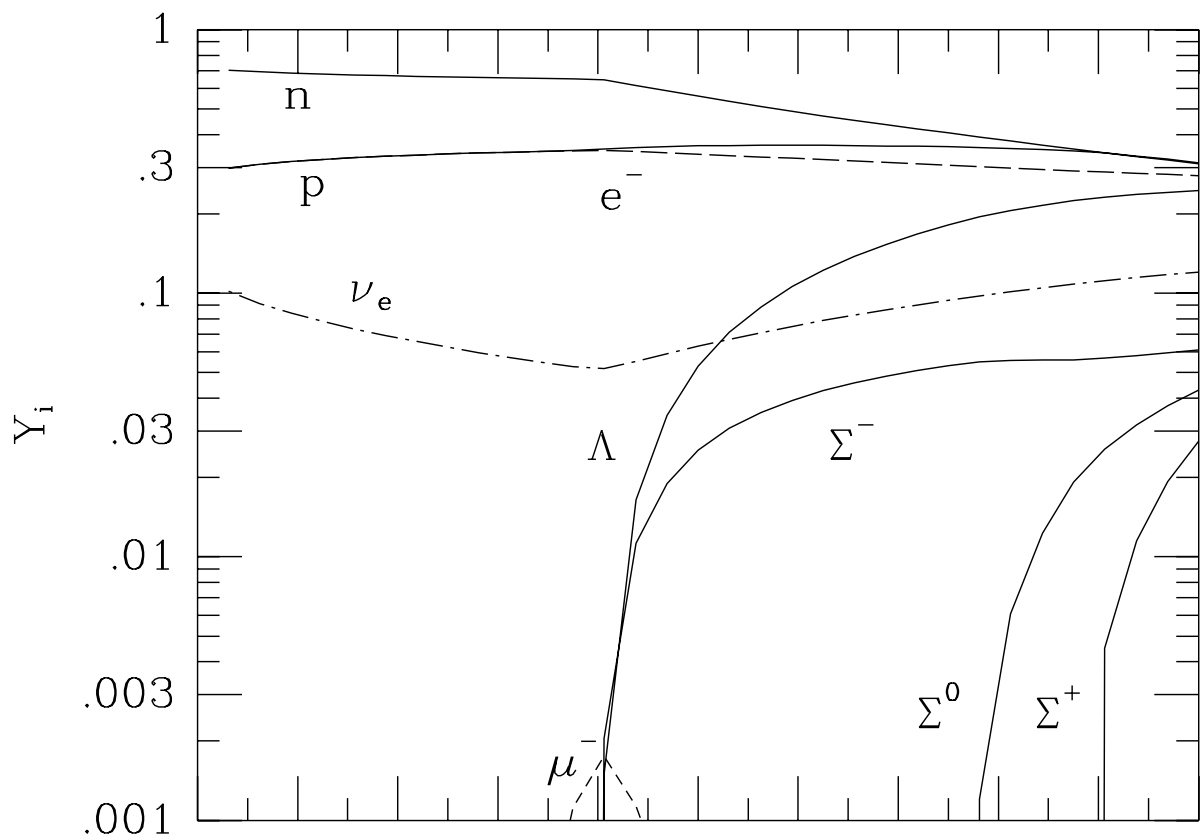
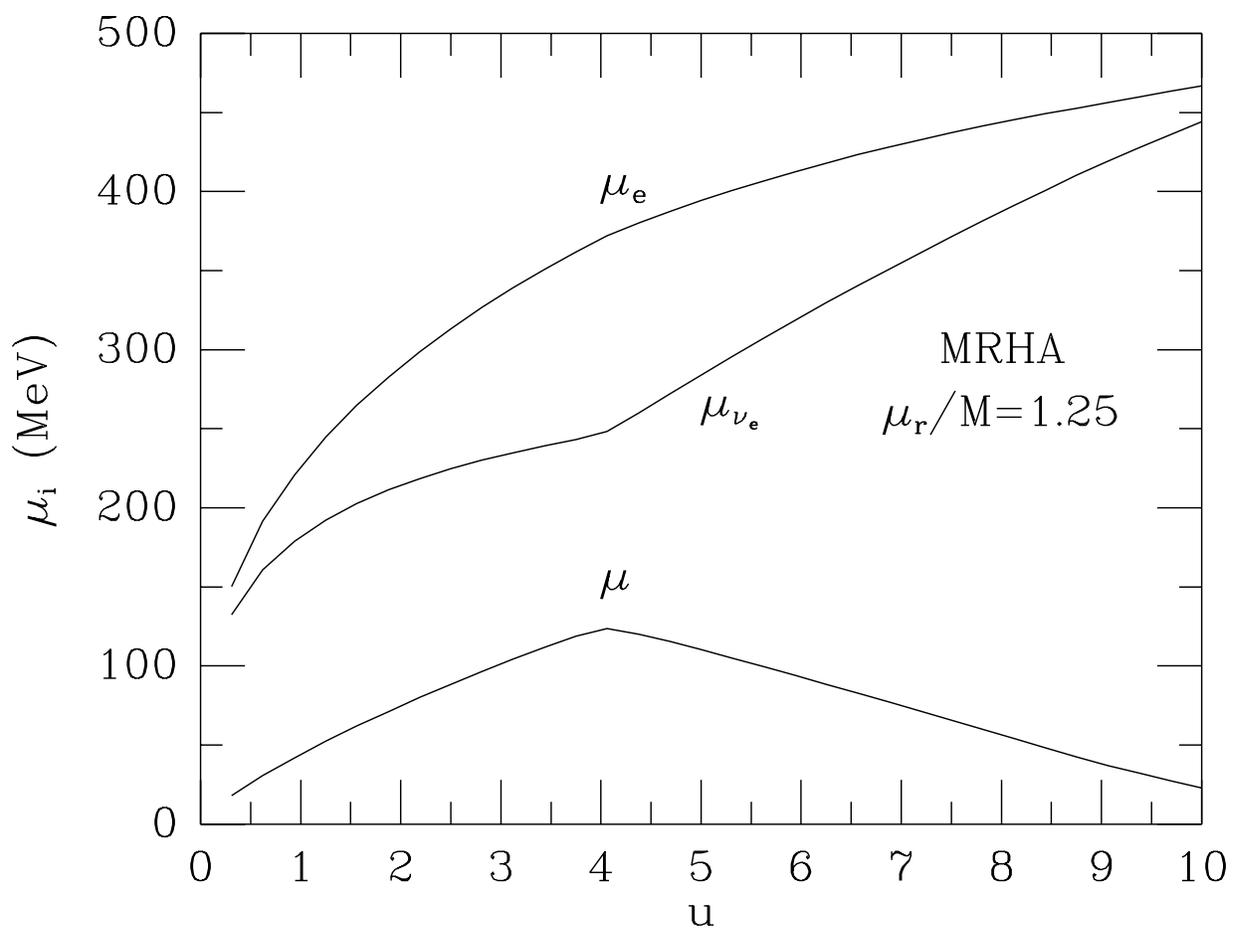
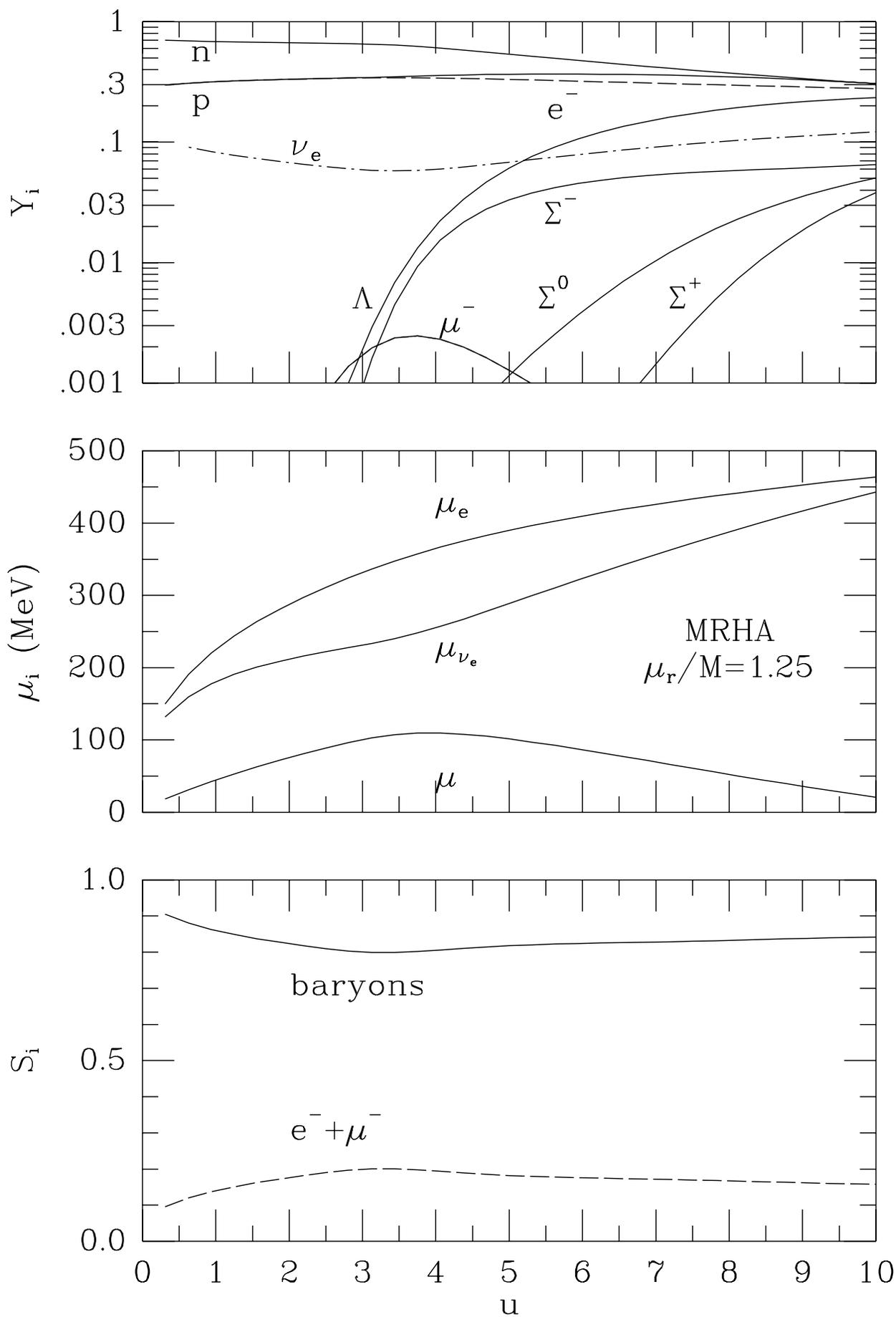
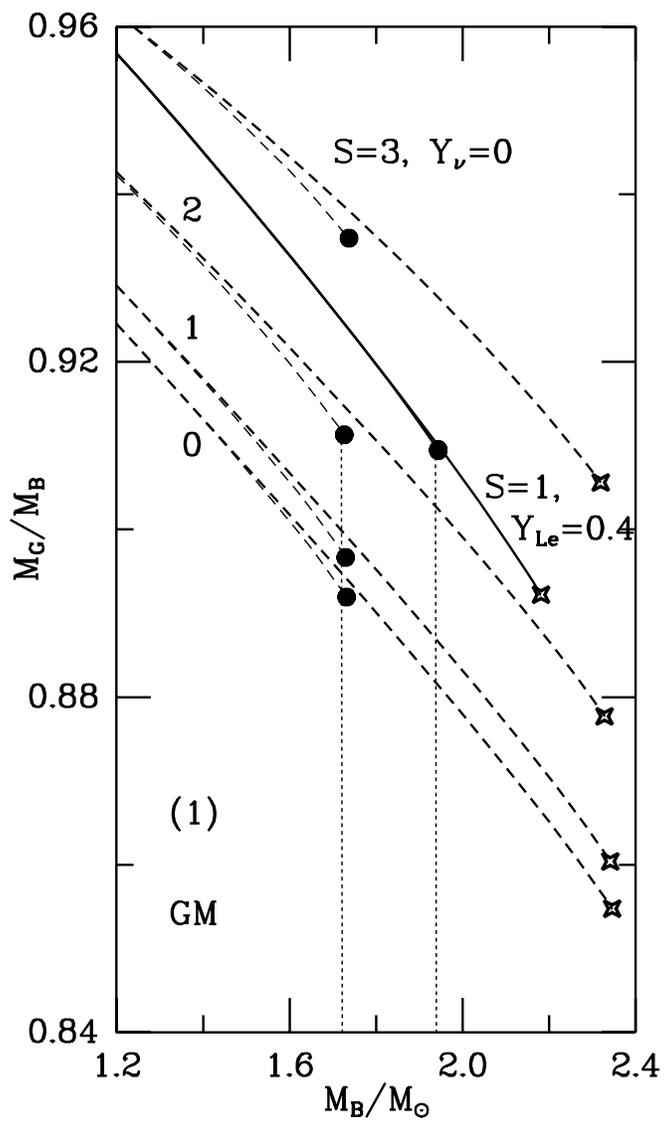
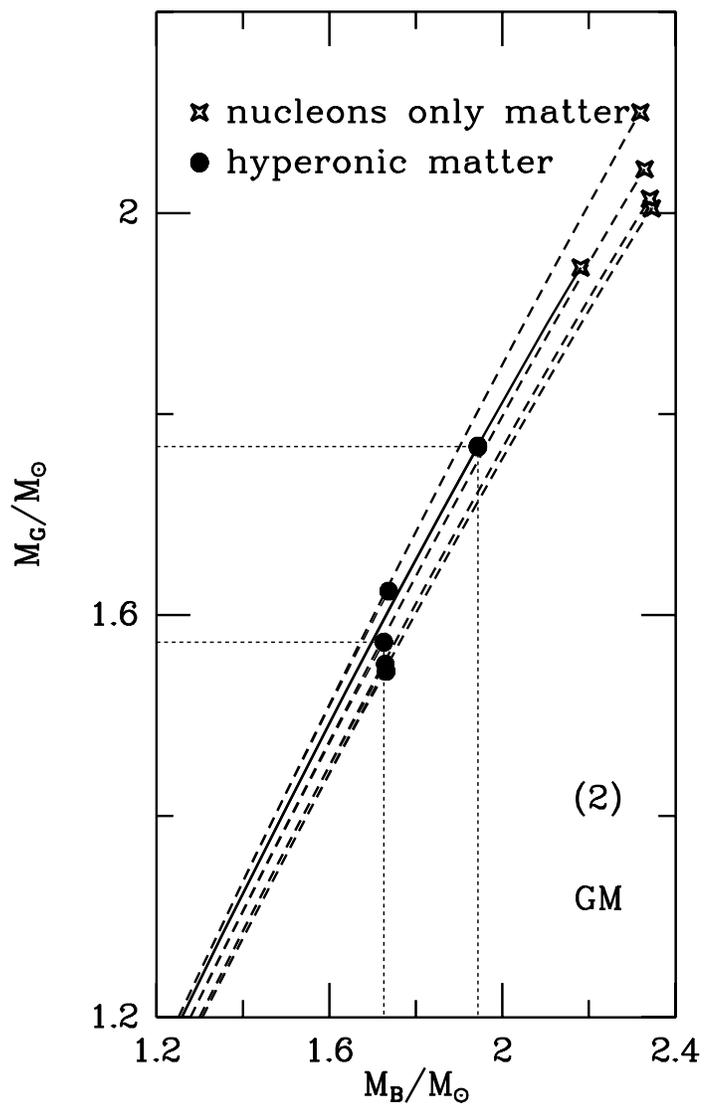
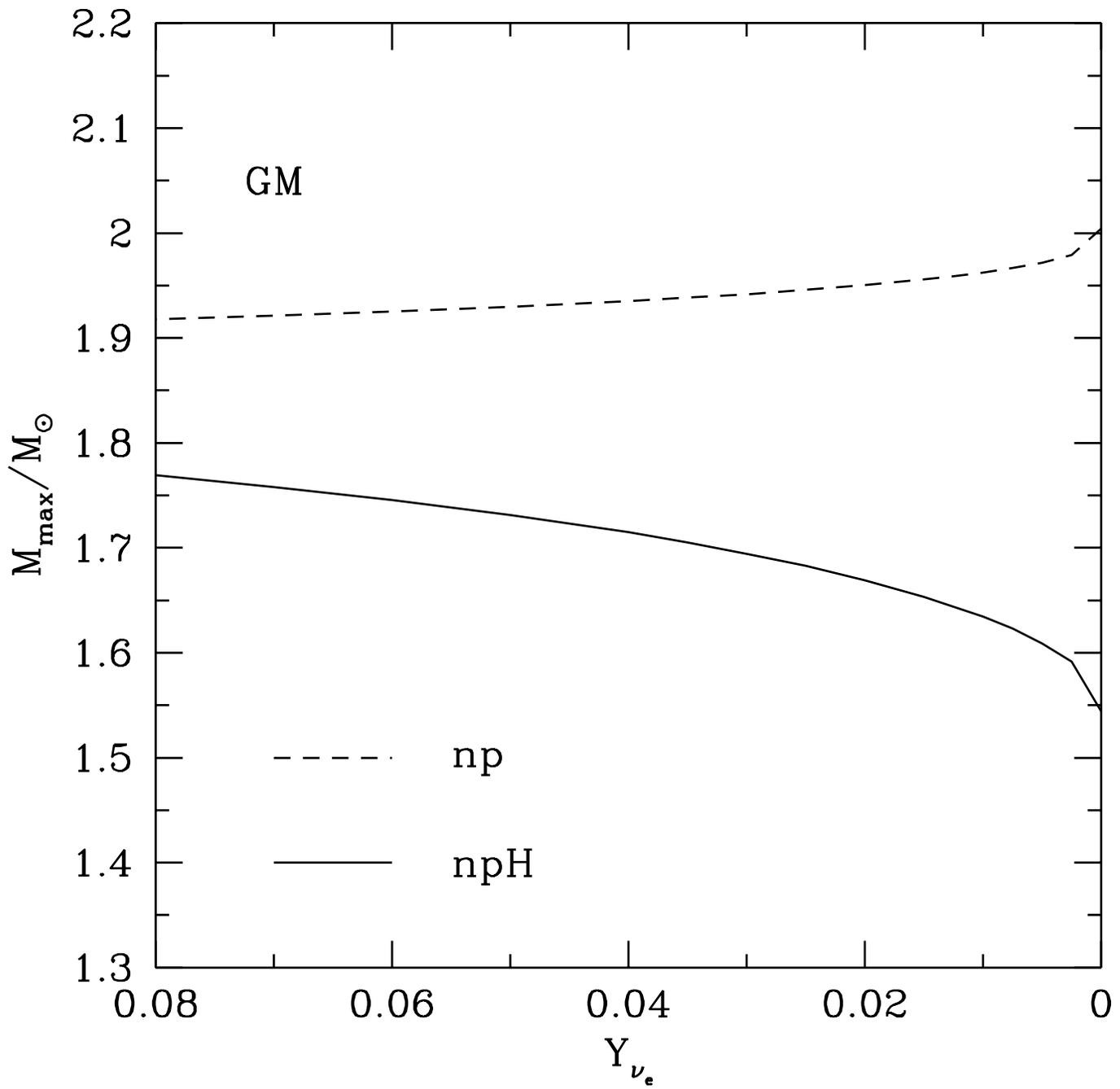
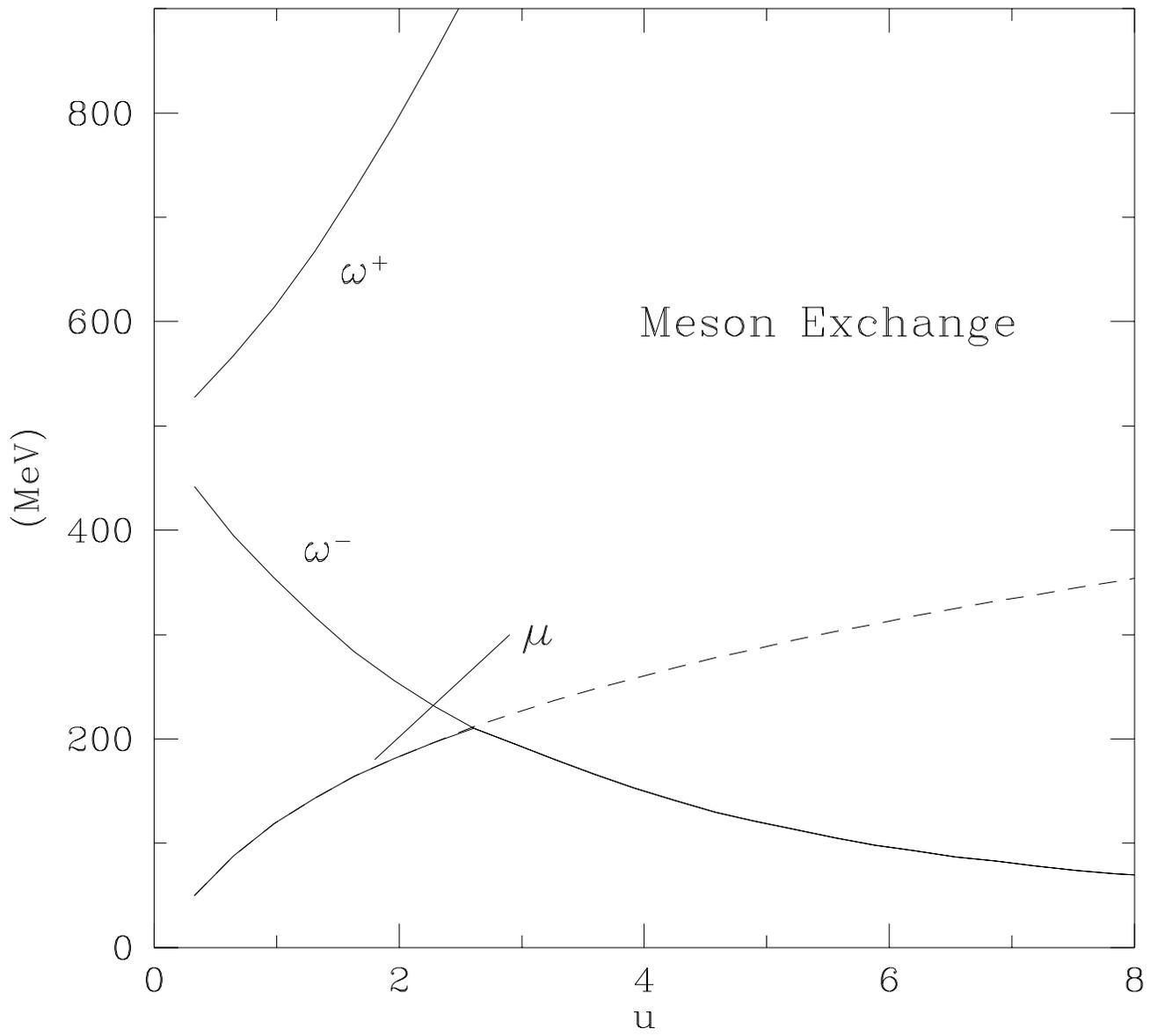
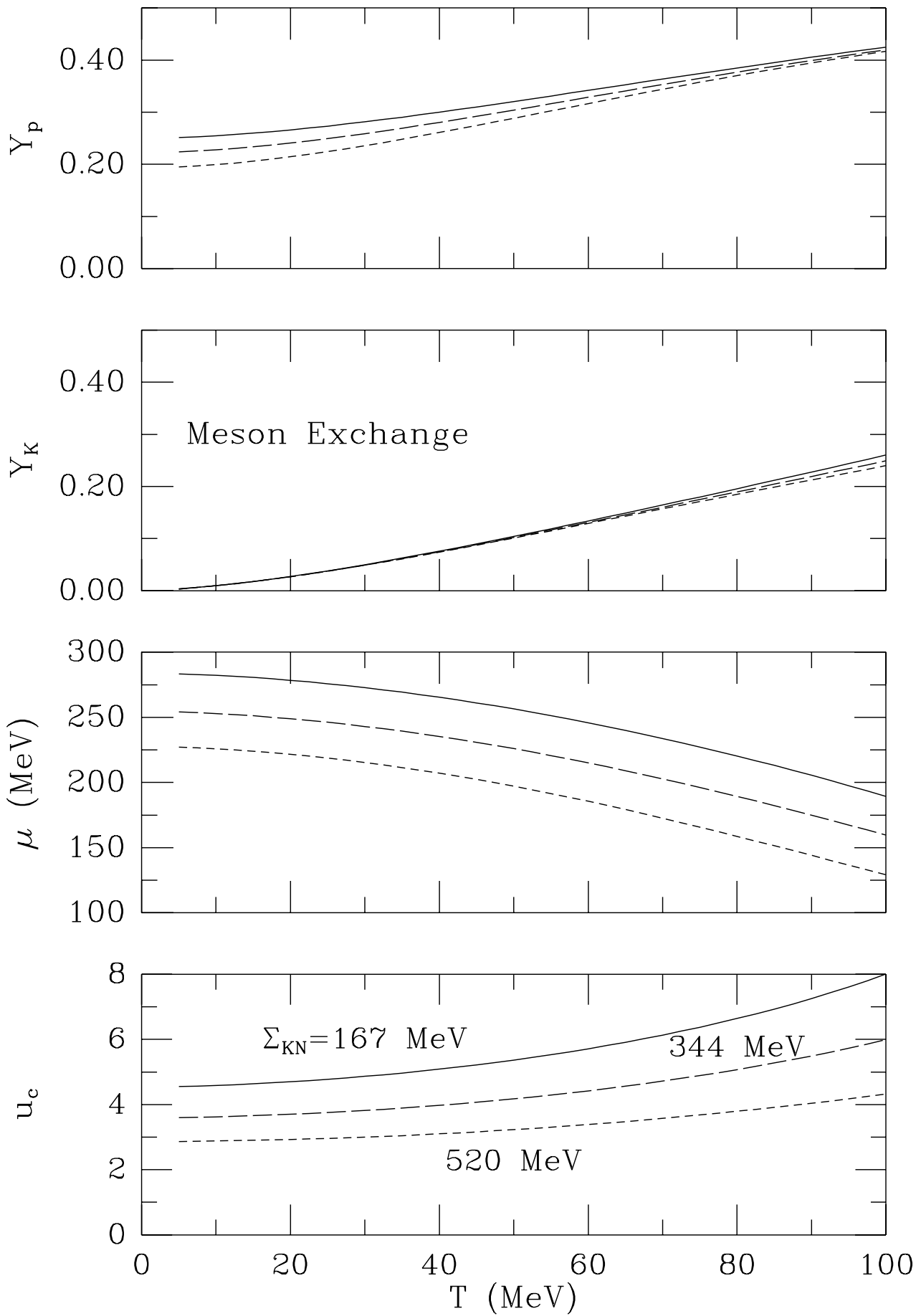
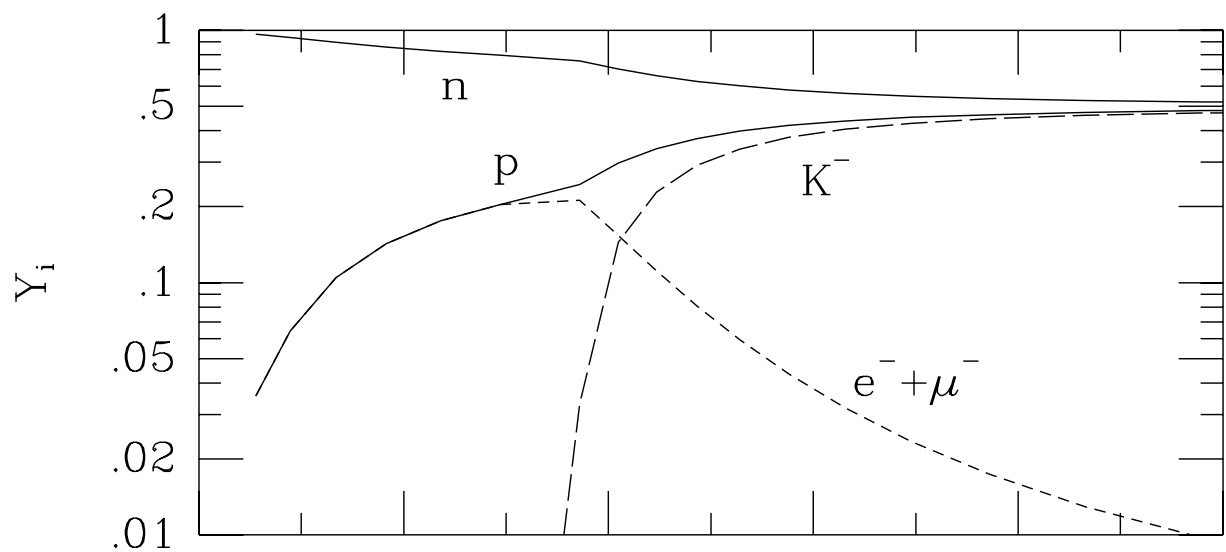
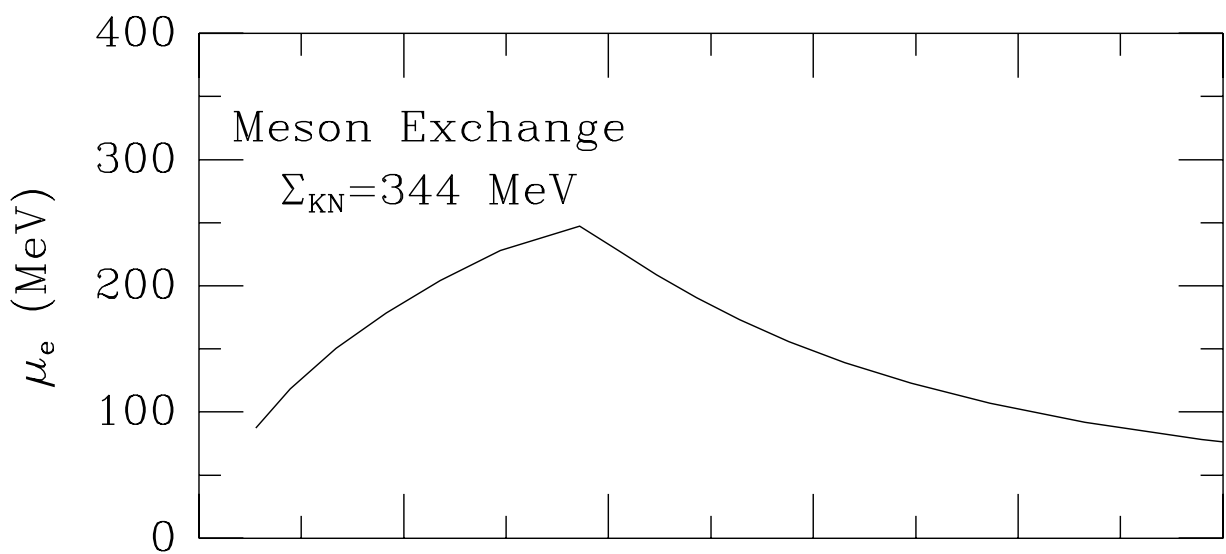
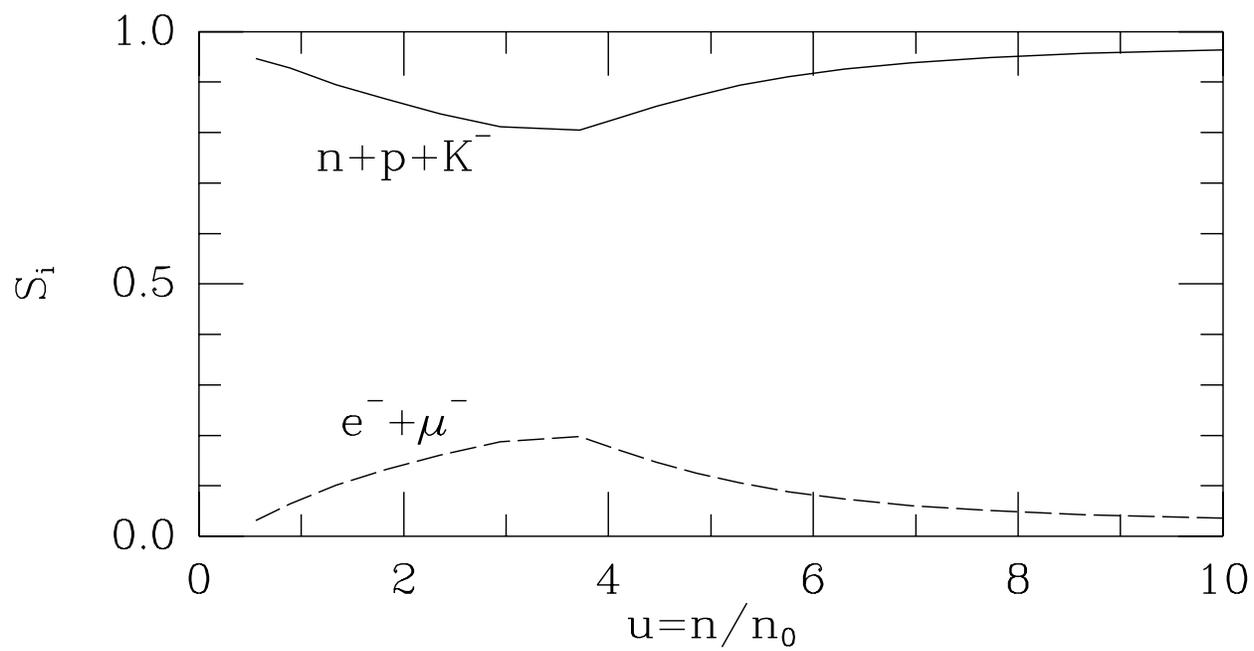
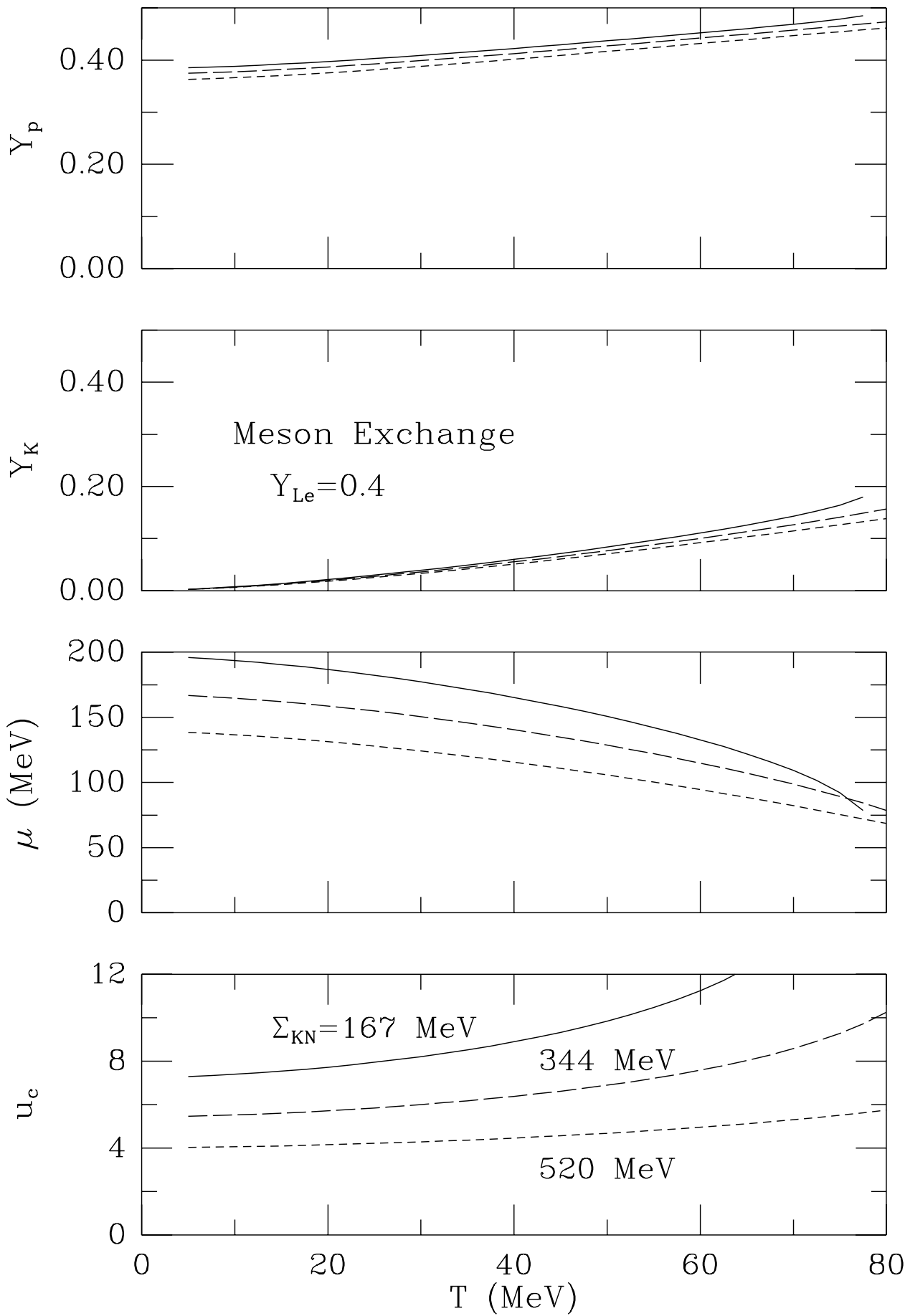
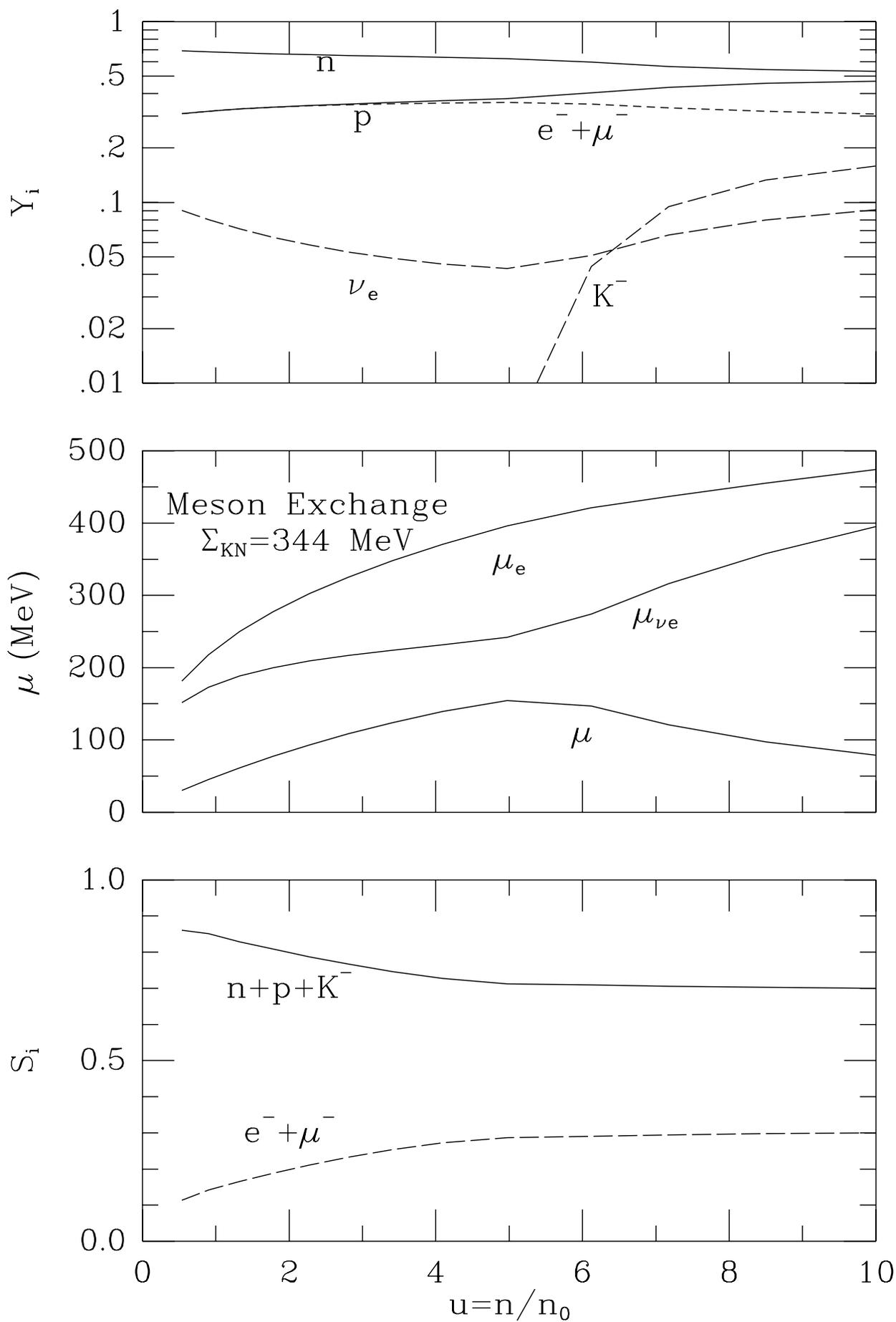
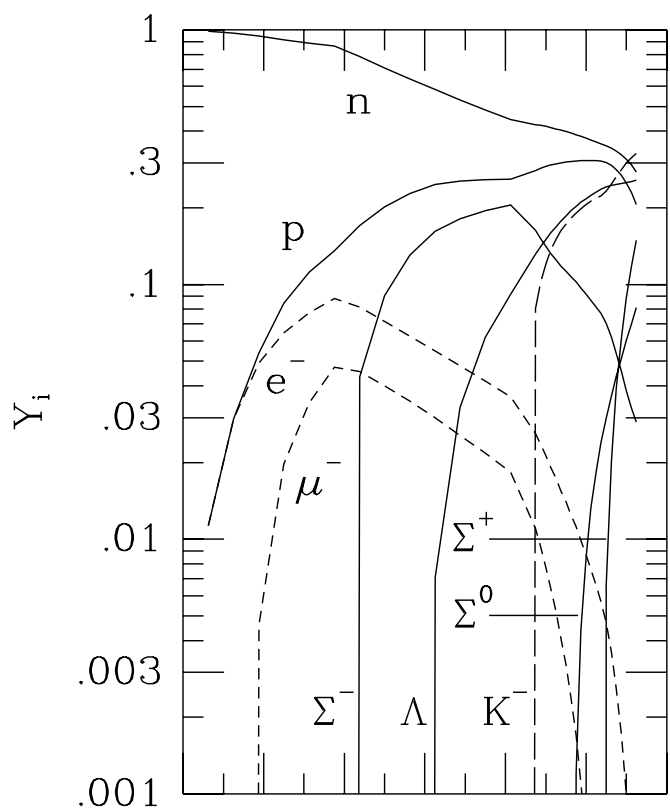
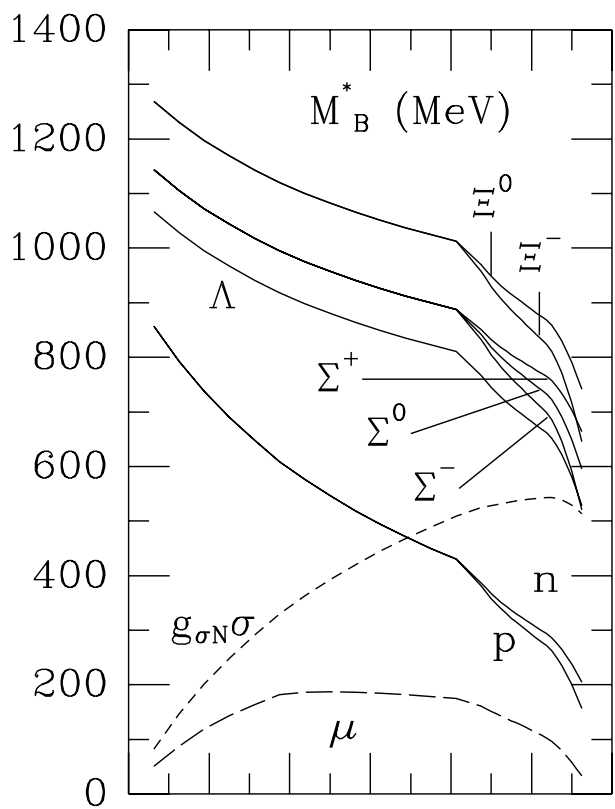
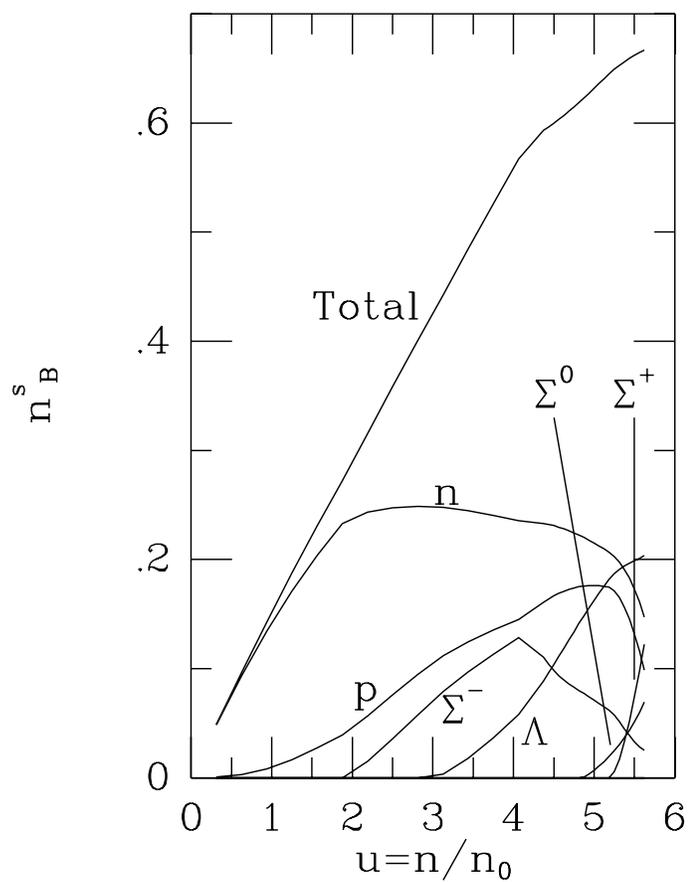
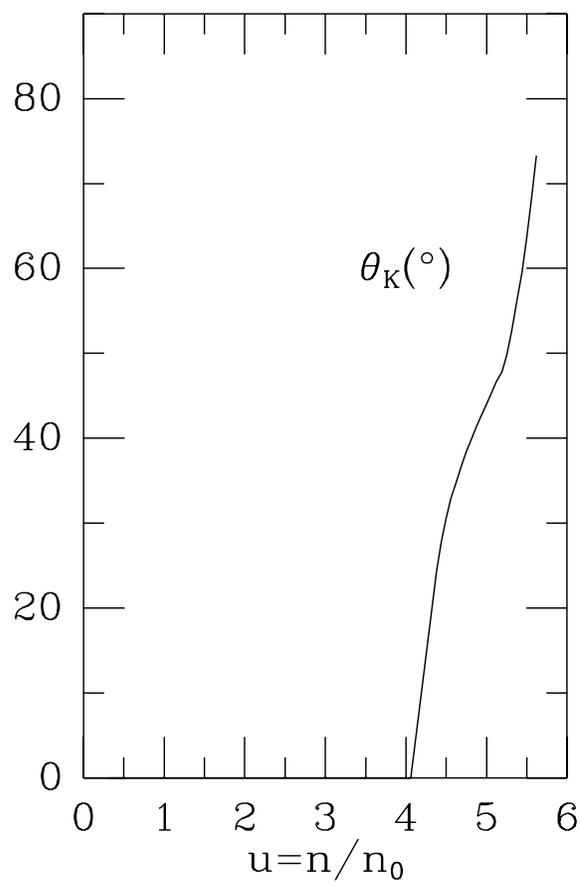
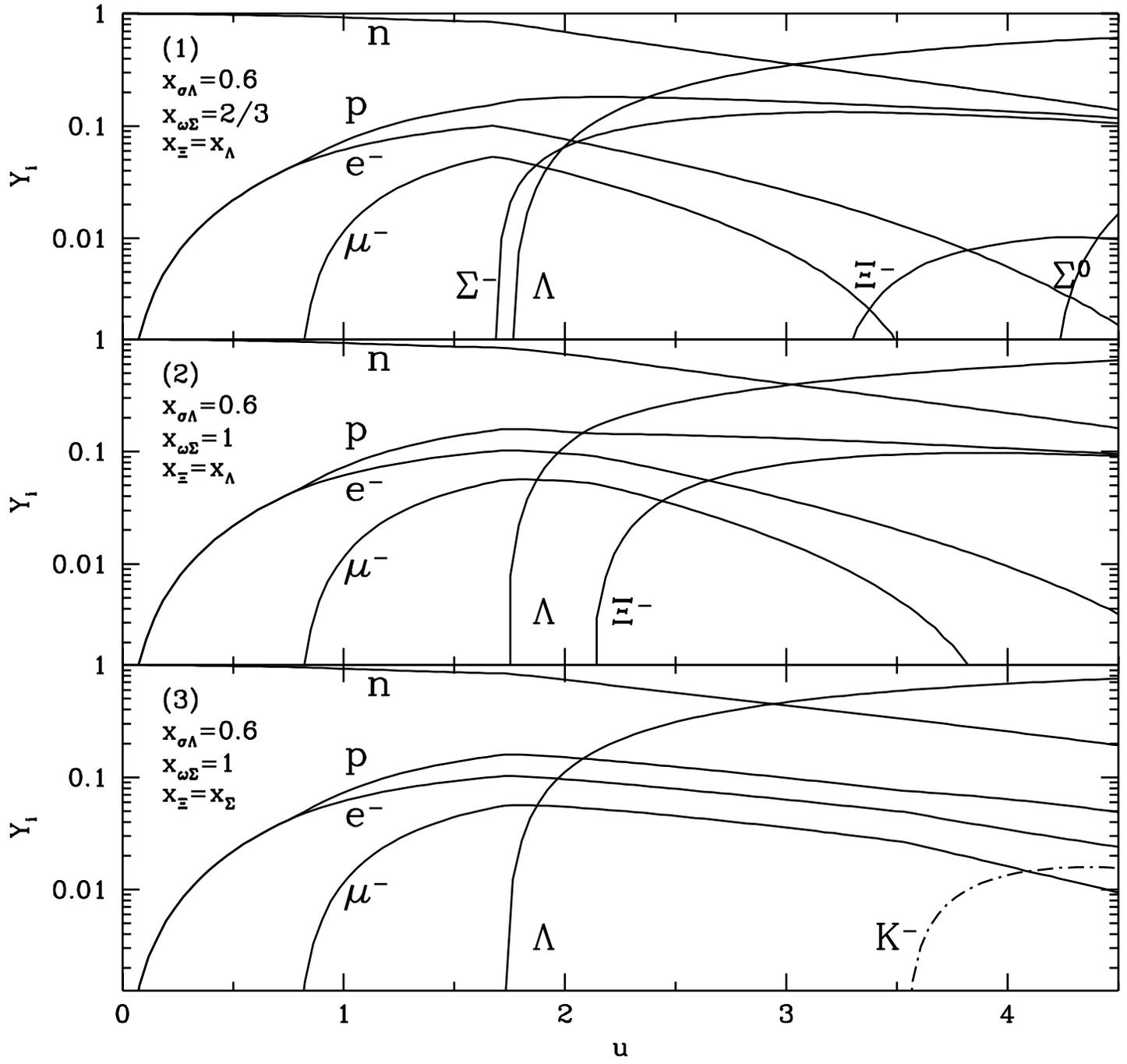
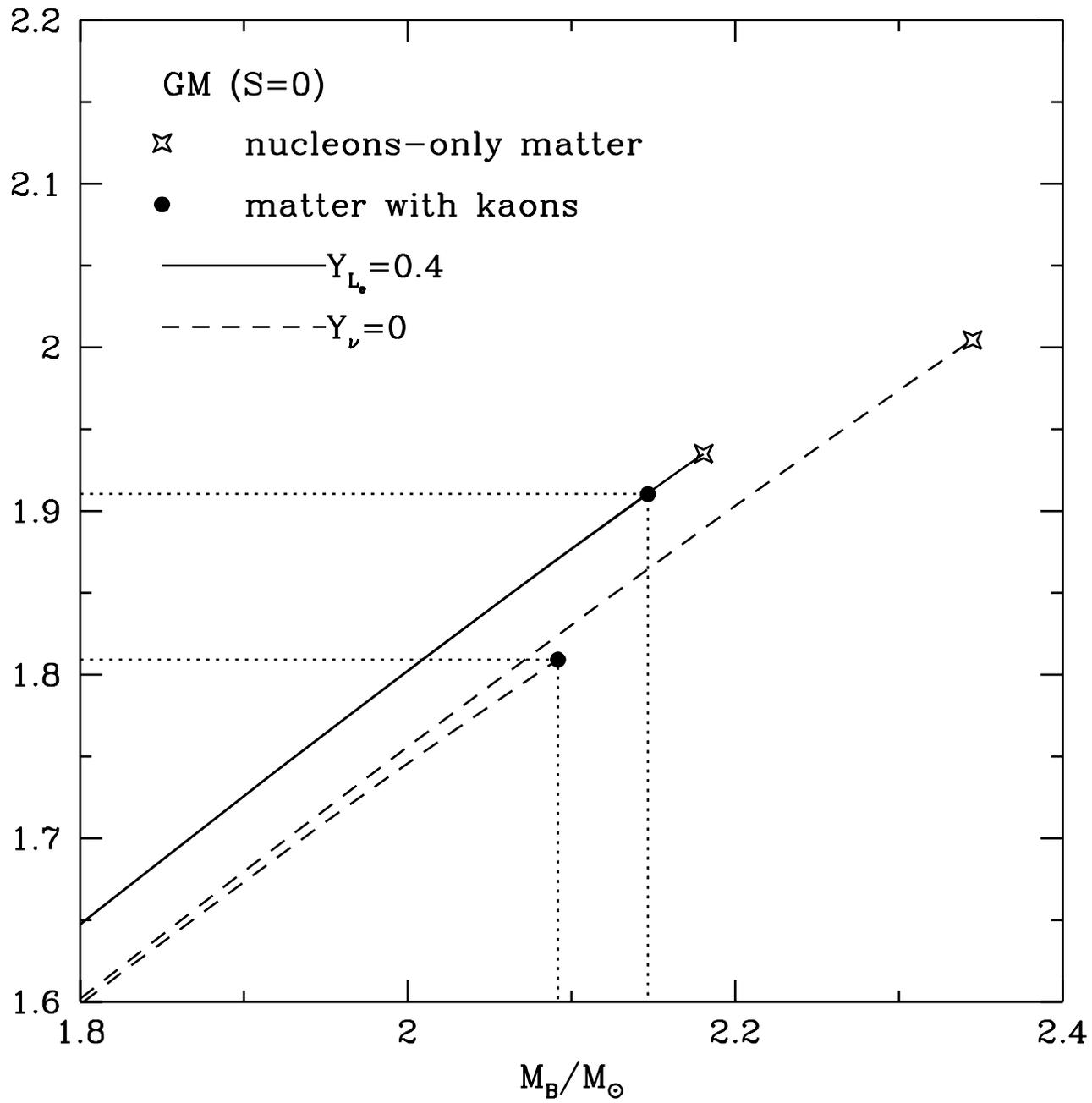
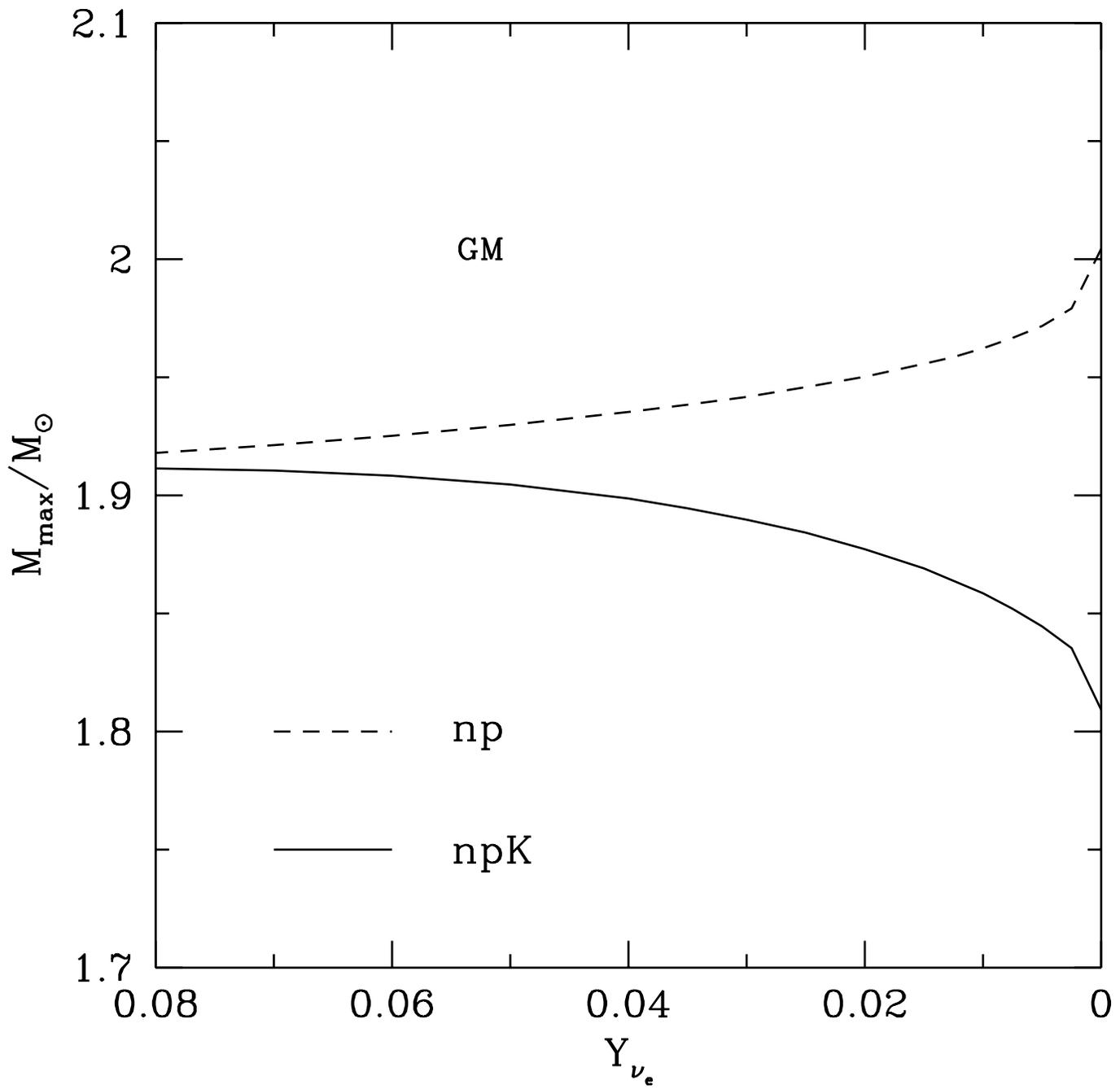
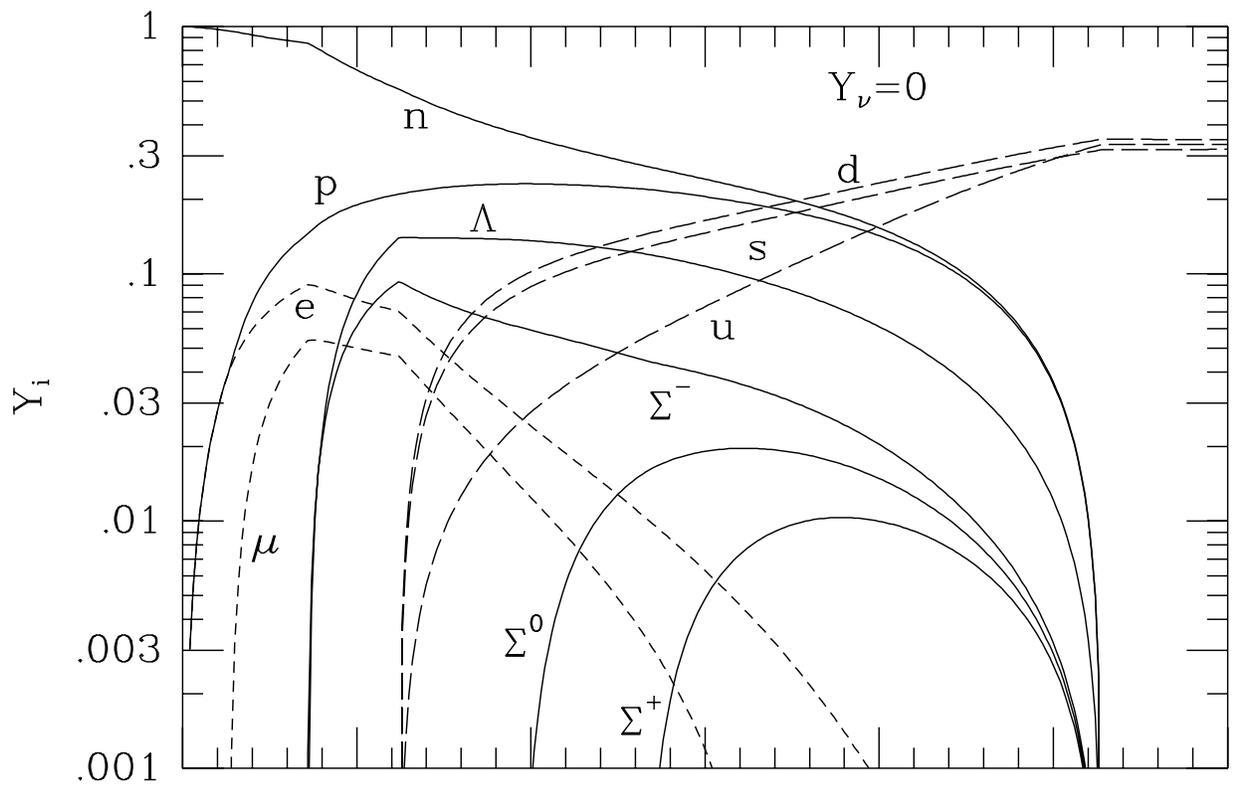
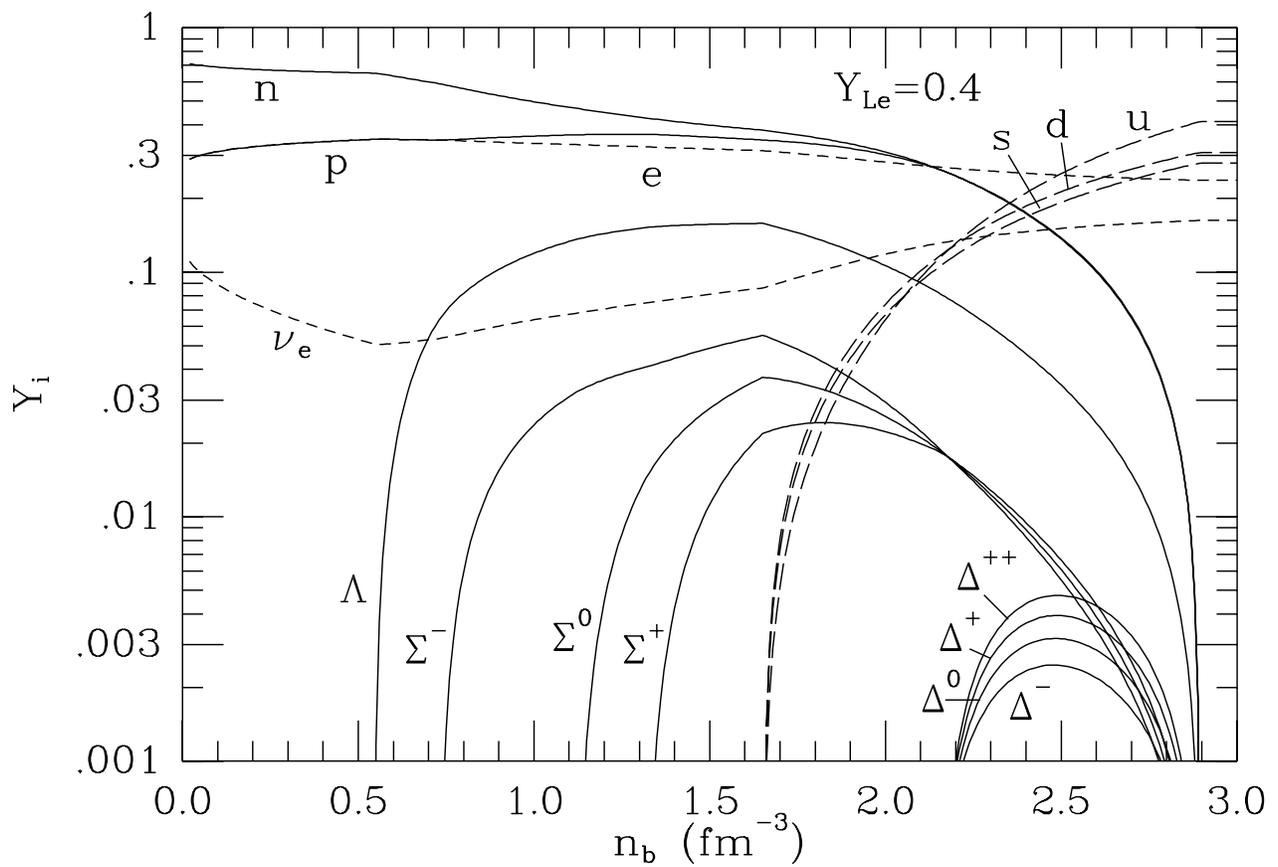
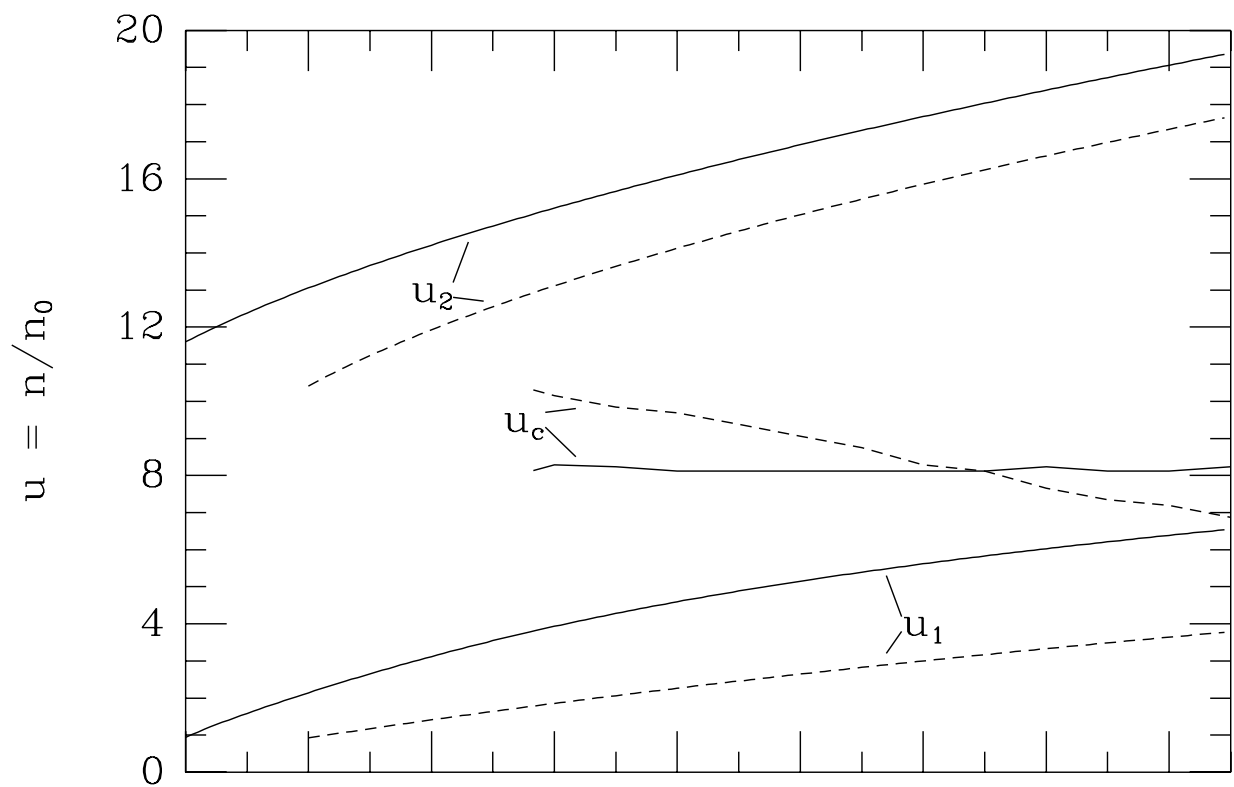
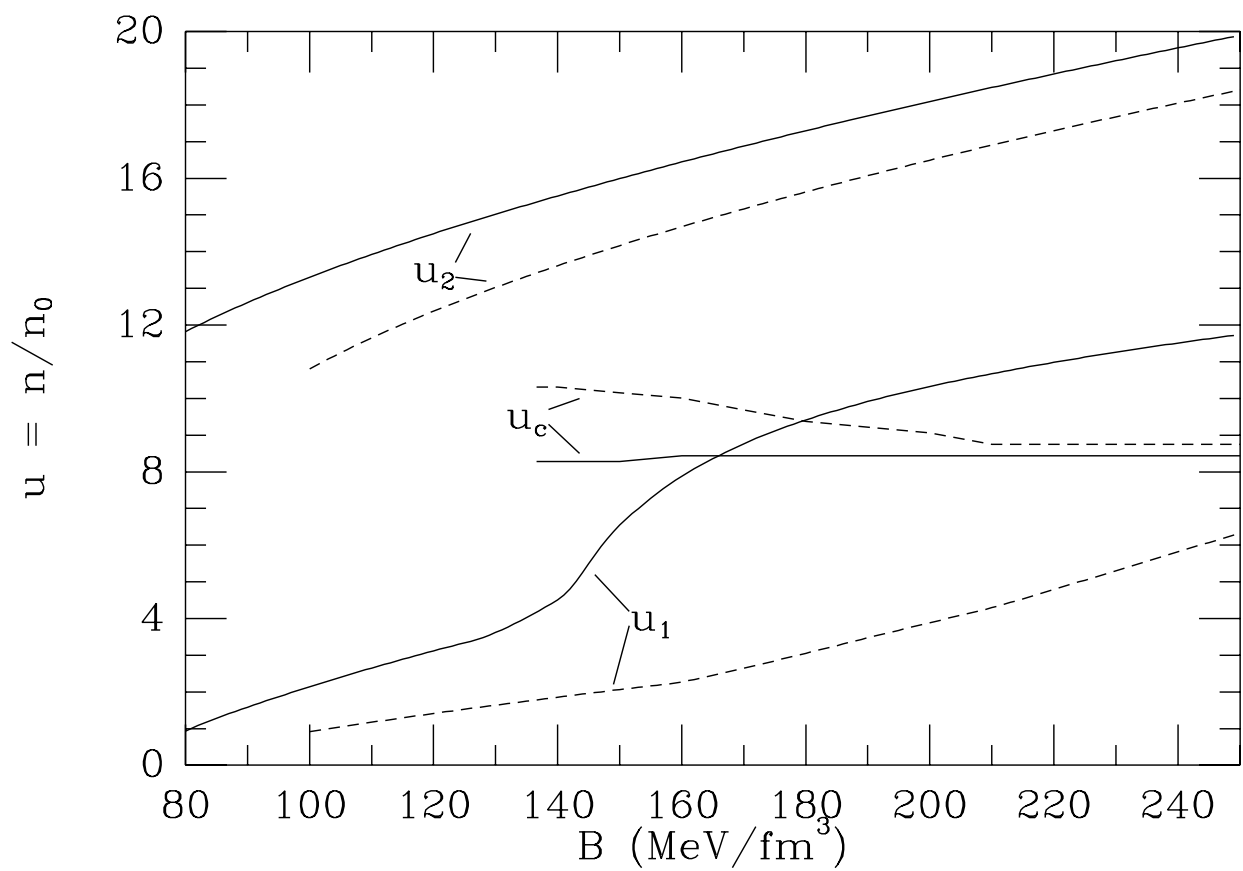
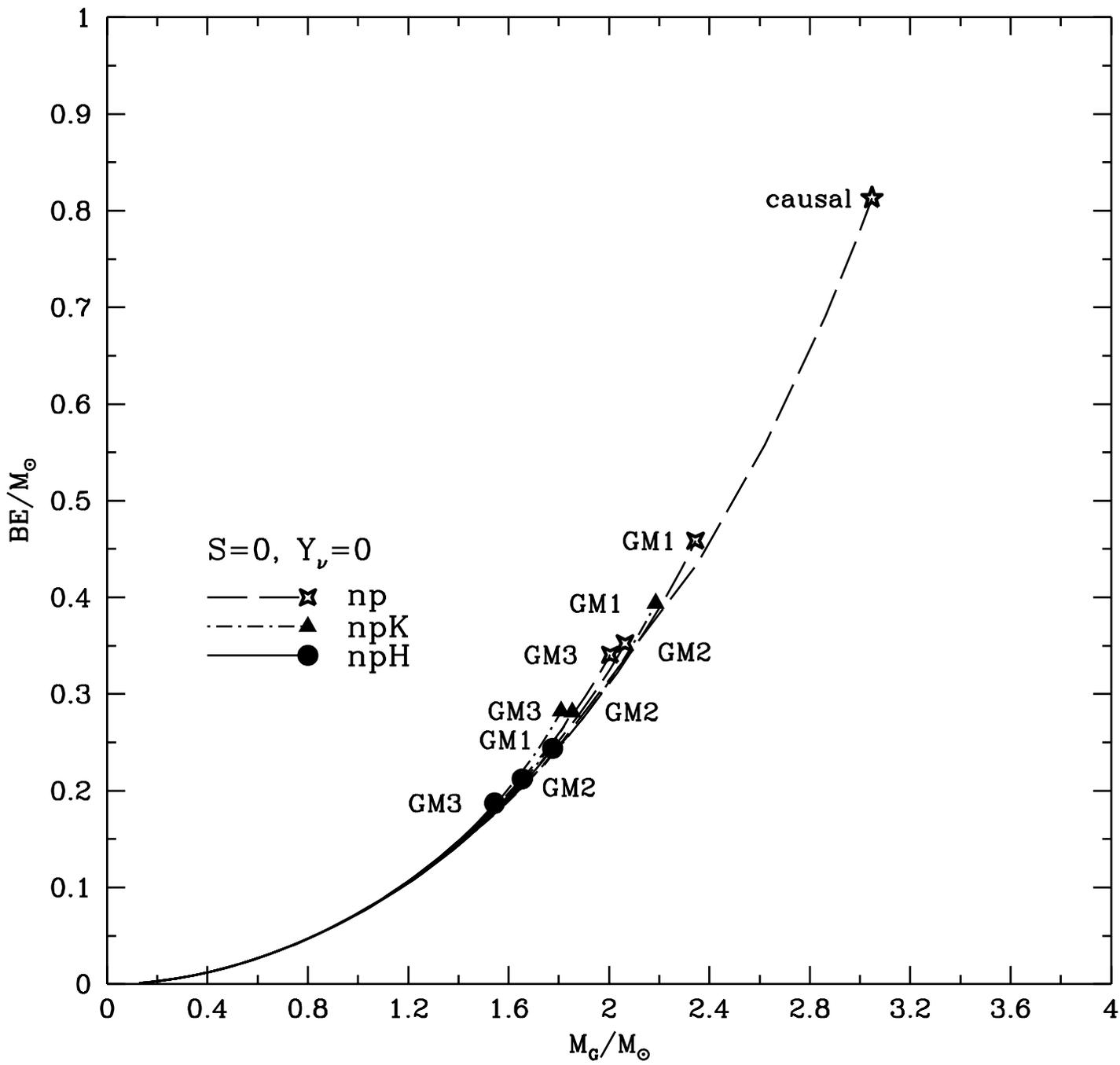
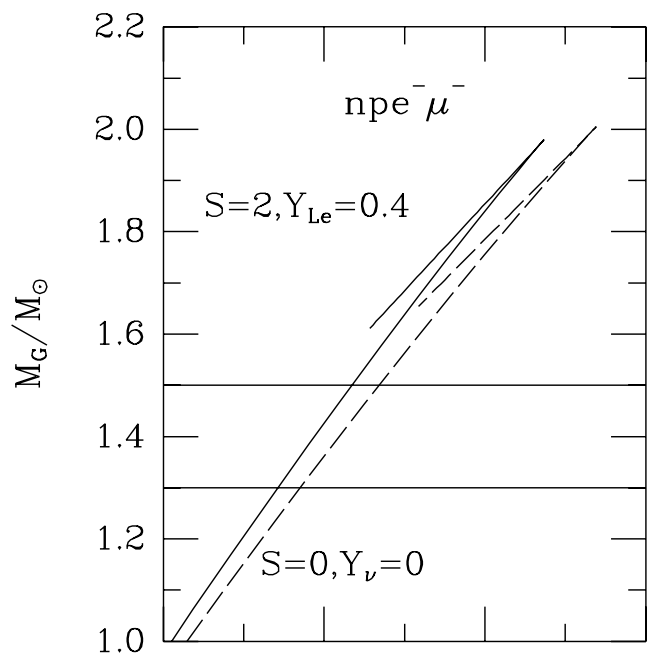
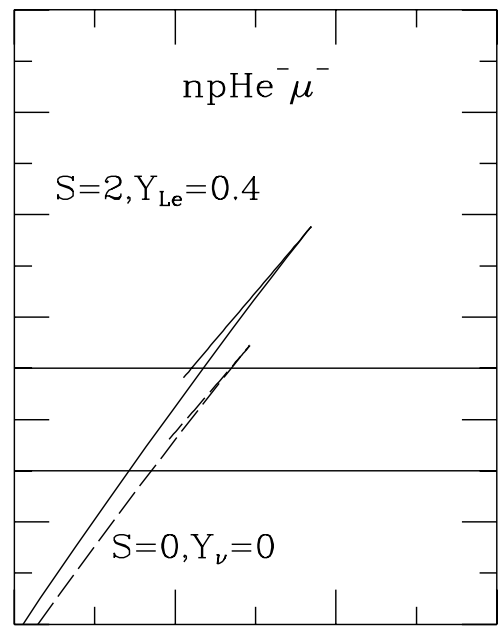
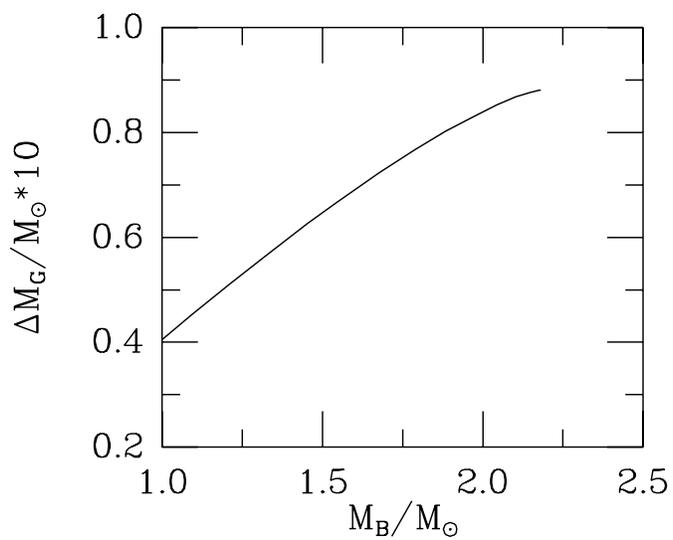
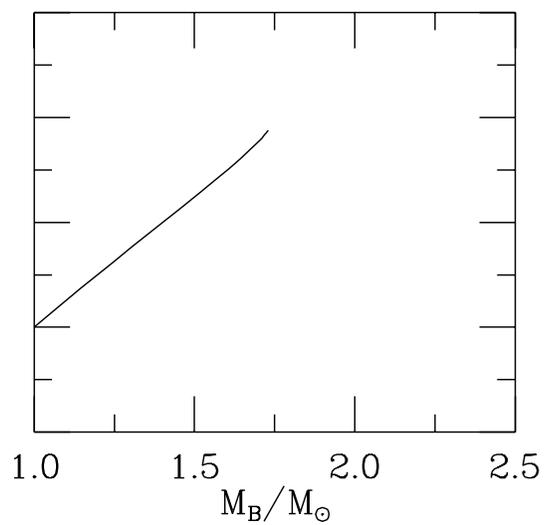
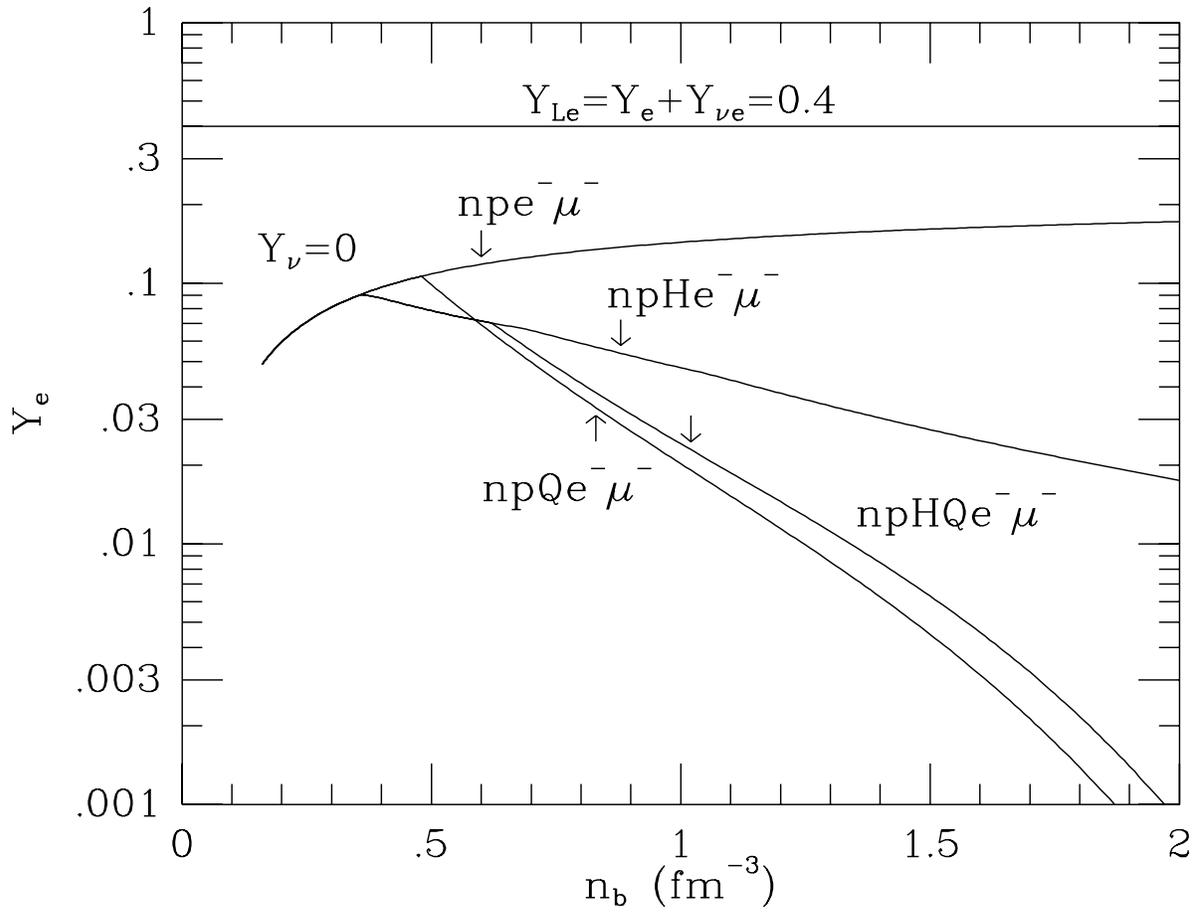
\end{document}